\documentclass[reqno]{amsart}

\usepackage{amsfonts,amsmath,amssymb,bm,enumitem,mathrsfs,mleftright,parskip,soul,wasysym}
\usepackage[margin = 1in]{geometry}
\usepackage[colorlinks = true]{hyperref}

\setlist[itemize]{leftmargin = *}
\setlist[enumerate]{leftmargin = *}

\allowdisplaybreaks

\newtheorem{Thm}{Theorem}
\newtheorem{Lem}{Lemma}
\newtheorem{Prop}{Proposition}
\newtheorem{Cor}{Corollary}

\theoremstyle{definition}

\newtheorem{Def}{Definition}

\newcommand{\C}{\mathbb{C}}
\newcommand{\D}{\mathscr{D}}
\newcommand{\HH}{\mathbb{H}}
\newcommand{\M}{\mathsf{M}}
\newcommand{\N}{\mathbb{N}}
\newcommand{\R}{\mathbb{R}}
\newcommand{\U}{\mathsf{U}}
\newcommand{\V}{\mathsf{V}}
\newcommand{\X}{\mathsf{X}}
\newcommand{\Y}{\mathsf{Y}}
\newcommand{\Z}{\mathsf{Z}}
\newcommand{\Br}[1]{\mleft( #1 \mright)}
\newcommand{\Cb}[1]{\func{\operatorname{C_{b}}}{#1}}
\newcommand{\Cc}[1]{\func{\operatorname{C_{c}}}{#1}}
\newcommand{\CC}[1]{\func{\operatorname{C}}{#1}}
\newcommand{\cl}[1]{\overline{#1}}
\newcommand{\Cl}[2]{\overline{#1}^{#2}}
\newcommand{\Co}[1]{\func{\operatorname{C}_{0}}{#1}}
\newcommand{\df}{\stackrel{\textnormal{df}}{=}}
\newcommand{\ds}{\displaystyle}

\newcommand{\GL}[2]{\func{\operatorname{GL}_{#1}}{#2}}
\newcommand{\Id}{\operatorname{Id}}
\newcommand{\Ll}[2]{\func{L^{#1}}{#2}}
\newcommand{\lt}{\operatorname{lt}}
\newcommand{\MO}{\widehat{P}_{\hbar}}
\newcommand{\PO}{\widehat{X}}
\newcommand{\qq}{\widetilde{q}}
\newcommand{\RS}[1]{\func{\mathcal{RS}}{#1}}
\newcommand{\Abs}[1]{\mleft| #1 \mright|}
\newcommand{\Adj}[1]{\func{\mathbb{L}}{#1}}
\newcommand{\Aut}[1]{\func{\operatorname{Aut}}{#1}}
\newcommand{\Bdd}[1]{\func{\mathbb{B}}{#1}}

\newcommand{\CCI}[2]{\SqBr{#1,#2}}
\newcommand{\COI}[2]{\mleft[ #1,#2 \mright)}
\newcommand{\Dom}[1]{\func{\operatorname{Dom}}{#1}}
\newcommand{\FCP}[3]{\func{C^{\ast}}{#1,#2,#3}}
\newcommand{\Int}[4]{\int_{#1} #2 ~ \d{\func{#3}{#4}}}
\newcommand{\Inv}[1]{\func{\operatorname{Inv}}{#1}}
\newcommand{\Map}[4]{\mleft\{ \begin{matrix} #1 & \to & #2 \\ #3 & \mapsto & \displaystyle #4 \end{matrix} \mright\}}

\newcommand{\OOI}[2]{\mleft] #1,#2 \mright[}
\newcommand{\RCP}[3]{\func{C^{\ast}_{\operatorname{red}}}{#1,#2,#3}}
\newcommand{\Rep}[2]{\func{\operatorname{Rep}}{#1,#2}}
\newcommand{\Seq}[2]{\Br{#1}_{#2}}
\newcommand{\Set}[2]{\SSet{#1 ~ \middle| ~ #2}}
\newcommand{\sgn}[1]{\func{\operatorname{sgn}}{#1}}
\newcommand{\Comm}[2]{\SqBr{#1,#2}}

\newcommand{\Diff}[2]{\frac{\mathrm{d}}{\mathrm{d}{#1}} \SqBr{#2}}
\newcommand{\func}[2]{#1 \Br{#2}}
\newcommand{\Func}[2]{\func{\Br{#1}}{#2}}
\newcommand{\FUNC}[2]{\func{\SqBr{#1}}{#2}}
\newcommand{\Mult}[1]{\func{M}{#1}}
\newcommand{\Norm}[1]{\mleft\| #1 \mright\|}
\newcommand{\Pair}[2]{\Br{#1,#2}}
\newcommand{\RInt}[4]{\int_{#1}^{#2} #3 ~ \mathrm{d}{#4}}
\newcommand{\Quad}[4]{\Br{#1,#2,#3,#4}}
\newcommand{\Size}[1]{\func{\#}{#1}}
\newcommand{\Span}[1]{\func{\operatorname{Span}}{#1}}
\newcommand{\Spec}[1]{\func{\sigma}{#1}}
\newcommand{\SqBr}[1]{\mleft[ #1 \mright]}
\newcommand{\SSet}[1]{\mleft\{ #1 \mright\}}
\newcommand{\Supp}[1]{\func{\operatorname{Supp}}{#1}}
\newcommand{\Trip}[3]{\Br{#1,#2,#3}}
\newcommand{\URep}[2]{\func{\operatorname{U-Rep}}{#1,#2}}
\newcommand{\Cstar}[1]{\func{C^{\ast}}{#1}}
\newcommand{\Graph}[1]{\func{\operatorname{Graph}}{#1}}
\newcommand{\Inner}[2]{\mleft\langle #1 \middle| #2 \mright\rangle}
\newcommand{\LCont}[2]{\func{L_{\textnormal{cont}}^{#1}}{#2}}

\newcommand{\Range}[1]{\func{\operatorname{Range}}{#1}}
\newcommand{\alphArg}[2]{\func{\alph{#1}}{#2}}
\newcommand{\SqInner}[2]{\SqBr{#1 \middle| #2}}
\newcommand{\Unitary}[1]{\func{\mathbb{U}}{#1}}
\newcommand{\ThreeMatrix}[9]{\begin{bmatrix} #1 & #2 & #3 \\ #4 & #5 & #6 \\ #7 & #8 & #9 \end{bmatrix}}

\renewcommand{\d}{\mathrm{d}}
\renewcommand{\H}{\mathcal{H}}
\renewcommand{\L}[3]{\func{\mathsf{L}^{2}}{#1,#2,#3}}
\renewcommand{\Im}[2]{#1 \SqBr{#2}}
\renewcommand{\alph}[1]{\alpha_{#1}}

\begin{document}


\title{The Infinitesimal Covariant Stone-von Neumann Theorem}
\author{Leonard Huang}
\address{School of Physical \& Mathematical Sciences (SPMS), Nanyang Technological University, Singapore}
\email{leonard.huang@ntu.edu.sg}
\author{Lara Ismert}
\address{Department of Mathematics \& Statistics, University of New Mexico, Albuquerque, New Mexico, USA}
\email{ismert@unm.edu}


\begin{abstract}
In a previous paper, the authors established the Covariant Stone-von Neumann Theorem, which generalizes the Stone-von Neumann Theorem in quantum mechanics and yields a classification of all unitary Heisenberg representations of an abelian elementary $ C^{\ast} $-dynamical system up to unitary equivalence. In this sequel to that paper, we establish an infinitesimal version of the Covariant Stone-von Neumann Theorem in the setting of real elementary $ C^{\ast} $-dynamical systems. Firstly, we generalize Stone's Theorem to secure a one-to-one correspondence between the strongly continuous one-parameter unitary groups and the regular self-adjoint densely defined operators on a Hilbert $ C^{\ast} $-module. Secondly, we define the infinitesimal Heisenberg representations and the infinitesimal Schr\"{o}dinger representation of a real $ C^{\ast} $-dynamical system. Thirdly, we construct a one-to-one correspondence between the unitary Heisenberg representations and the infinitesimal Heisenberg representations of a real $ C^{\ast} $-dynamical system, providing a rigorous two-way passage --- which we call ``Heisenberg differentiability-integrability'' --- between the global unitary formulation of Heisenberg covariance and its infinitesimal formulation in terms of regular self-adjoint densely defined operators. This allows us to prove the Infinitesimal Covariant Stone-von Neumann Theorem: For a real elementary $ C^{\ast} $-dynamical system, every infinitesimal Heisenberg representation is a direct sum of copies of the infinitesimal Schr\"{o}dinger representation. Lastly, we obtain explicit descriptions of the position and momentum operators for a real $ C^{\ast} $-dynamical system and construct a natural common invariant core of analytic vectors for them. All requisite technical domain and integrability issues are fully accounted for.
\end{abstract}


\maketitle

\section{Introduction} 

The Stone-von Neumann Theorem is one of the most important mathematical results in the foundation of non-relativistic quantum mechanics, establishing the phenomenological equivalence between the matrix mechanics of Werner Heisenberg and the wave mechanics of Erwin Schr\"{o}dinger, which were the first two versions of quantum mechanics to be discovered. A precise statement of the theorem is as follows.

\begin{Thm}[the Stone-von Neumann Theorem \cite{VonNeumann}]
Let $ n \in \N $, and suppose that the ordered triple $ \Trip{\H}{R}{S} $ meets the following three conditions:
\begin{enumerate}
\item
$ \H $ is a Hilbert space.

\item
$ U $ and $ V $ are strongly continuous unitary representations of the topological group $ \R^{n} $ on $ \H $.

\item
For all $ \bm{x},\bm{y} \in \R^{n} $, we have $ V_{\bm{y}} U_{\bm{x}} = e^{- i \hbar \Inner{\bm{x}}{\bm{y}}} \cdot U_{\bm{x}} V_{\bm{y}} $ --- the Weyl Commutation Relation.
\end{enumerate}
Then $ \Trip{\H}{U}{V} $ is unitarily equivalent to a direct sum of copies of $ \Trip{\Ll{2}{\R^{n}}}{\U}{\V} $, where $ U $ and $ V $ are strongly continuous unitary representations of the locally compact Hausdorff abelian group $ \R^{n} $ on the Hilbert space $ \Ll{2}{\R^{n}} $, satisfying
$$
\forall \mathbf{x},\mathbf{y} \in \R^{n}, ~ \forall f \in \Ll{2}{\R^{n}}: \quad
\func{\U_{\bm{x}}}{f} \df \func{f}{\bullet + \hbar \bm{x}}
\quad \text{and} \quad
\func{\V_{\bm{y}}}{f} \df e^{i \Inner{\bm{y}}{\bullet}} \cdot f.
$$
\end{Thm}
Applying Stone's Theorem for strongly continuous one-parameter unitary groups in conjunction with well-known integrability results, the Stone-von Neumann Theorem may be given an infinitesimal reformulation, which involves the native materials of Heisenberg's matrix mechanics: the position operators $ X_{1},\ldots,X_{n} $ and the momentum operators $ P_{1},\ldots,P_{n} $. A precise statement of the infinitesimal version of the Stone-von Neumann Theorem for the case $ n = 1 $ is as follows:

\begin{Thm}[the Infinitesimal Stone-von Neumann Theorem \cite{VonNeumann}]
Suppose that the ordered triple $ \Trip{\H}{S}{T} $ meets the following three conditions:
\begin{enumerate}
\item
$ \H $ is a Hilbert space.

\item
$ S $ and $ T $ are self-adjoint densely defined linear operators on $ \H $.

\item
There exists an $ \SSet{S,T} $-analytic and $ \SSet{S,T} $-invariant dense linear subspace of $ \H $ such that
\begin{itemize}
\item
$ D $ is a core for both $ S $ and $ T $, and

\item
the Heisenberg Commutation Relation (HCR) holds for $ \Pair{S}{T} $ on $ D $, i.e., $ \Comm{S|_{D}}{T|_{D}} = - i \hbar \cdot \Id_{D} $.
\end{itemize}
\end{enumerate}
Then $ \Trip{\H}{S}{T} $ is unitarily equivalent to a direct sum of copies of $ \Trip{\Ll{2}{\R}}{P}{X} $.
\end{Thm}

The Stone-von Neumann Theorem was to become an integral part of harmonic analysis when George Mackey established it as a special case of a general result that classifies --- up to unitary equivalence --- all Heisenberg representations of an arbitrary locally compact Hausdorff abelian group, where the requisite harmonic-analytic generalization of the Weyl Commutation Relation is built into the very definition of a Heisenberg representation.

\begin{Thm}[the Stone-von Neumann Theorem-Mackey]
Let $ G $ be a locally compact Hausdorff abelian group, and suppose that the ordered triple $ \Trip{\H}{U}{V} $ is a Heisenberg representation of $ G $, i.e., satisfies the following three statements:
\begin{enumerate}
\item
$ \H $ is a Hilbert space.

\item
$ U $ and $ V $ are, respectively, strongly continuous unitary representations of $ G $ and $ \widehat{G} $ on $ \H $.

\item
The Weyl Commutation Relation: For all $ x \in G $ and $ \varphi \in \widehat{G} $, we have $ \func{V}{\varphi} \func{U}{x} = \func{\varphi}{x} \cdot \func{U}{x} \func{V}{\varphi} $.
\end{enumerate}
Then $ \Trip{H}{U}{V} $ is unitarily equivalent to a direct sum of copies of $ \Trip{\Ll{2}{\R}}{\U}{\V} $, where
\begin{itemize}
\item
$ U $ is the strongly continuous unitary representation of $ G $ on $ \Ll{2}{G} $ by left translations, i.e.,
$$
\forall x \in G, ~ \forall f \in \Ll{2}{G}: \quad
\FUNC{\func{\U}{x}}{f} \df \func{f}{x^{- 1} \bullet};
$$

\item
$ V $ is the strongly continuous unitary representation of $ \widehat{G} $ on $ \Ll{2}{G} $ by phase modulations, i.e.,
$$
\forall \varphi \in \widehat{G}, ~ \forall f \in \Ll{2}{G}: \quad
\FUNC{\func{\V}{\varphi}}{f} = \varphi \cdot f.
$$
\end{itemize}
\end{Thm}

Mackey's proof of the Stone-von Neumann-Mackey Theorem relied heavily on measure theory and required that $ G $ be second-countable. Indeed, Mackey's subsequent work on induced representations of locally compact Hausdorff groups --- including his proof of the Mackey Imprimitivity Theorem, which implies the Stone-von Neumann-Mackey Theorem --- was heavily measure-theoretic in nature and could not dispense with the second-countability hypothesis. Robert Blattner showed that this hypothesis was unnecessary; Marc Rieffel reformulated the Mackey Imprimitivity Theorem so that much of the measure-theoretic machinery was replaced by soft analysis, and then used his reformulation to reduce the Stone-von Neumann-Mackey Theorem to an essentially algebraic assertion, namely, the Morita equivalence of the $ C^{\ast} $-algebras $ \C $ and $ \Cstar{G,\Co{G},\lt} $.

Rieffel's algebraic approach to the Stone-von Neumann-Mackey Theorem provided the main inspiration for the paper \cite{Huang-Ismert} by the authors, which proves a far-reaching generalization of the theorem that classifies --- up to unitary equivalence --- all Heisenberg representations of an abelian $ C^{\ast} $-dynamical system on an elementary $ C^{\ast} $-algebra.

\begin{Thm}[the Covariant Stone-von Neumann Theorem \cite{Huang-Ismert}] \label{thm:CSVNT}
Let $ \Trip{G}{A}{\alpha} $ be a $ C^{\ast} $-dynamical system, with $ G $ a locally compact Hausdorff abelian group and $ A $ an elementary $ C^{\ast} $-algebra, and suppose that the ordered quadruple $ \Quad{X}{\rho}{U}{V} $ is a Heisenberg representation of $ \Trip{G}{A}{\alpha} $, i.e., satisfies the following six statements:
\begin{enumerate}
\item
$ \X $ is a full Hilbert $ A $-module.

\item
$ \rho $ is a non-degenerate $ \ast $-representation of $ A $ on $ \X $ by adjointable operators.

\item
$ R $ and $ S $ are, respectively, strongly continuous unitary representations of $ G $ and $ \widehat{G} $ on $ \X $.

\item
For all $ a \in A $ and $ x \in G $, we have $ \func{U}{x} \func{\rho}{a} = \func{\rho}{\alphArg{x}{a}} \func{U}{x} $.

\item
For all $ a \in A $ and $ \varphi \in \widehat{G} $, we have $ \func{V}{\varphi} \func{\rho}{a} = \func{\rho}{a} \func{V}{\varphi} $.

\item
The Weyl Commutation Relation: For all $ x \in G $ and $ \varphi \in \widehat{G} $, we have $ \func{V}{\varphi} \func{U}{x} = \func{\varphi}{x} \cdot \func{U}{x} \func{V}{\varphi} $.
\end{enumerate}
Then $ \Quad{\X}{\rho}{U}{V} $ is unitarily equivalent to a direct sum of copies of $ \Quad{\L{G}{A}{\alpha}}{\M}{\U}{\V} $, where
\begin{itemize}
\item
$ \L{G}{A}{\alpha} $ is the Hilbert $ A $-module completion of the pre-Hilbert $ A $-module $ \Cc{G,A} $ with respect to the $ A $-valued inner product $ \SqInner{\bullet}{\bullet} $ defined by
$$
\forall f,g \in \Cc{G,A}: \quad
\SqInner{f}{g} \df \Int{G}{\alphArg{x^{- 1}}{\func{f}{x}^{\ast} \func{g}{x}}}{\mu}{x},
$$
with the canonical dense linear embedding of $ \Cc{G,A} $ into $ \L{G}{A}{\alpha} $ to be denoted by $ q $;

\item
$ \M $ is a non-degenerate $ \ast $-representation of $ A $ on $ \L{G}{A}{\alpha} $, satisfying
$$
\forall a \in A, ~ \forall f \in \Cc{G,A}: \quad
\FUNC{\func{\M}{a}}{\func{q}{f}} = \func{q}{a f}.
$$

\item
$ \U $ is a strongly continuous unitary representation of $ G $ on $ \L{G}{A}{\alpha} $, satisfying
$$
\forall x \in G, ~ \forall f \in \Cc{G,A}: \quad
\FUNC{\func{\U}{x}}{\func{q}{f}} = \func{q}{\alphArg{x}{\func{f}{x^{- 1} \bullet}}};
$$

\item
$ \V $ is a strongly continuous unitary representation of $ \widehat{G} $ on $ \L{G}{A}{\alpha} $, satisfying
$$
\forall \varphi \in \widehat{G}, ~ \forall f \in \Cc{G,A}: \quad
\FUNC{\func{\V}{\varphi}}{\func{q}{f}} = \func{q}{\varphi \cdot f}.
$$
\end{itemize}
\end{Thm}

In this paper, we establish \autoref{The Infinitesimal Covariant Stone-von Neumann Theorem}, the infinitesimal version of \autoref{thm:CSVNT}, which returns the whole enterprise to the unbounded operator-theoretic roots of non-relativistic quantum mechanics. A primary technical challenge in developing this theory in the Hilbert \(C^*\)-module setting is that, unlike in the Hilbert space setting, the appropriate infinitesimal generators must be regular in addition to being self-adjoint. Two of the main results required to prove \autoref{The Infinitesimal Covariant Stone-von Neumann Theorem} are \autoref{Heisenberg Differentiability} and \autoref{Heisenberg Integrability}, which provide a two-way correspondence between unitary Heisenberg representations and their regular, self-adjoint infinitesimal generators. 

This paper is organized as follows:
\begin{itemize}
\item
In Section 2 (Preliminaries), we recall basic definitions and results concerning $ C^{\ast} $-dynamical systems, adjointable maps between Hilbert $ C^{\ast} $-modules, direct sums of Hilbert $ C^{\ast} $-modules, unbounded operators between Hilbert $ C^{\ast} $-modules, and symmetric unbounded operators on Hilbert $ C^{\ast} $-modules. Smooth vectors and analytic vectors shall be introduced here, which are integral to subsequent sections.

\item
In Section 3, we shall establish a bijection from the set of regular self-adjoint unbounded operators on an arbitrary Hilbert $ C^{\ast} $-module $ \X $ to the set of strongly continuous one-parameter unitary groups on $ \X $, by way of an exponentiation process made possible by S. L. Woronowicz's unbounded functional calculus for a Hilbert $ C^{\ast} $-module. Although this result was first announced in \cite{Amyari-Chakoshi}, its proof given there seems to have shortcomings. A correct proof may be found in \cite[Section 2.2]{Schlieter-Werner} that follows the proof in \cite[Section 5.3]{Pedersen} of Stone's Theorem for Hilbert spaces, but ours follows the proof in \cite[Section 5]{Conway} of the same theorem.

\item
In Section 4, infinitesimal Heisenberg representations and the infinitesimal Schr\"{o}dinger representation of an arbitrary real $ C^{\ast} $-dynamical system $ \Trip{\R}{A}{\alpha} $ are defined, which are all ordered quadruples of the form $ \Quad{\X}{\rho}{R}{S} $, where $ \X $ is a Hilbert $ A $-module, $ \rho $ is a non-degenerate $ \ast $-representation of $ A $ on $ \X $, and $ R $ and $ S $ are regular self-adjoint unbounded operators on $ \X $ --- the momentum and position operators within our framework, respectively --- that satisfy not only the Heisenberg Commutation Relation but also two additional commutation relations involving $ \rho $. These infinitesimal representations are the main ingredients of the infinitesimal version of \autoref{thm:CSVNT} --- in the context of $ G = \R $ --- that we are aiming for.

\item
In Section 5, the position and momentum operators are given explicit expressions as a multiplication operator and a differential operator on a dense $ A $-submodule of $ \L{\R}{A}{\alpha} $, respectively. Originally given as abstract infinitesimal generators of strongly continuous one-parameter unitary groups on $ \L{\R}{A}{\alpha} $, they lacked an explicit form, which obscured their connection to the traditional position and momentum operators of quantum mechanics. This section aims to rectify that.
\end{itemize}

\section{Preliminaries} \label{Preliminaries} 

We assume the reader is acquainted with the basic theory of $ C^{\ast} $-algebras and Hilbert $ C^{\ast} $-modules, though we will recall some definitions in this section to establish notation. Throughout, $ A $ is an arbitrary $ C^{\ast} $-algebra.

\subsection{Adjointable Maps Between Hilbert \texorpdfstring{$ C^{\ast} $}{C*}-Modules} 


Let $ \X $ and $ \Y $ be Hilbert $ A $-modules.
\begin{itemize}
\item
A map $ T: \X \to \Y $ is \emph{adjointable} if and only if there exists a map $ S: \Y \to \X $ such that
$$
\forall \zeta \in \X, ~ \forall \eta \in \Y: \quad
\Inner{\func{T}{\zeta}}{\eta} = \Inner{\zeta}{\func{S}{\eta}},
$$
in which case $ S $ is necessarily unique, so we denote it by $ T^{\ast} $ and call it the \emph{adjoint} of $ T $.

An adjointable map from $ \X $ to $ \Y $ is automatically linear, $ A $-linear, and bounded (see \cite{Lance} for details), and the set of adjointable maps from $ \X $ to $ \Y $ shall be denoted by $ \Adj{\X,\Y} $.

\item
$ \Adj{\X} \df \Adj{\X,\X} $ is a $ C^{\ast} $-algebra with respect to the obvious point-wise operations and the operator norm.

\item
A \emph{$ \ast $-representation} of a $ C^{\ast} $-algebra $ B $ on $ \X $ is a $ \ast $-homomorphism $ \rho: B \to \Adj{\X} $, which is then said to be \emph{non-degenerate} if and only if
$$
\Cl{\Span{\Set{\FUNC{\func{\rho}{b}}{\zeta} \in \X}{b \in B ~ \text{and} ~ \zeta \in \X}}}{\Norm{\cdot}_{\X}} = \X.
$$

The set of $ \ast $-representations of a $ C^{\ast} $-algebra $ B $ on $ \X $ shall be denoted by $ \Rep{B}{\X} $.

\item
A \emph{unitary map} from $ \X $ to $ \Y $ is an element $ U $ of $ \Adj{\X,\Y} $ such that $ U^{\ast} U = \Id_{\X} $ and $ U U^{\ast} = \Id_{\Y} $.

The set of unitary operators from $ \X $ to $ \Y $ shall be denoted by $ \Unitary{\X,\Y} $.

\item
$ \Unitary{\X} \df \Unitary{\X,\X} $ is a group with respect to composition.

\item
For $ G $ a locally compact Hausdorff group, a \emph{unitary representation} of $ G $ on a Hilbert $ A $-module $ \X $ is a group homomorphism $ U: G \to \Unitary{\X} $, which is then called \emph{strongly continuous} if and only if
$$
\forall g \in G, ~ \forall \zeta \in \X: \quad
\lim_{h \to g} \Norm{\FUNC{\func{U}{h}}{\zeta} - \FUNC{\func{U}{g}}{\zeta}}_{\X} = 0;
$$
an $ A $-submodule $ D $ of $ \X $ is said to be $ U $-\emph{invariant} if and only if, for all $ g \in G $, we have $ \Im{\SqBr{\func{U}{g}}}{D} \subseteq D $.

The set of strongly continuous unitary representations of $ G $ on $ \X $ shall be denoted by $ \URep{G}{\X} $.

\item
A \emph{one-parameter unitary group} on $ \X $ is a unitary representation of the additive group $ \R $ on $ \X $.
\end{itemize}

\subsection{Direct Sums of Hilbert \texorpdfstring{$ C^{\ast} $}{C*}-Modules} 

Let $ \Lambda $ be a set and $ \bm{\X} $ a $ \Lambda $-indexed family of Hilbert $ A $-modules. Recall from \cite{Lance} that
$$
    \bigoplus \bm{\X}
\df \Set{\bm{x} \in \prod \bm{\X}}{\sum_{\lambda \in \Lambda} \Inner{\bm{x}_{\lambda}}{\bm{x}_{\lambda}}_{\bm{\X}_{\lambda}} \text{ converges in } A}
$$
is a right $ A $-module with respect to the obvious entry-wise operations and can be equipped with an $ A $-valued inner product $ \Inner{\cdot}{\cdot}_{\bigoplus \bm{\X}} $, defined by
$$
\forall \bm{x},\bm{y} \in \bigoplus \bm{\X}: \quad
\Inner{\bm{x}}{\bm{y}}_{\bigoplus \bm{\X}} \df \sum_{\lambda \in \Lambda} \Inner{\bm{x}_{\lambda}}{\bm{y}_{\lambda}}_{\bm{\X}_{\lambda}}.
$$
It is routine to check that $ \ds \bigoplus \bm{\X} $ is already complete --- and therefore a Hilbert $ A $-module --- with respect to the norm induced by this $ A $-valued inner product.


\begin{Lem} \label{Corollary of the Pythaogrean Inequality for Orthogonal Elements of a Hilbert C*-Module}
Let $ \Lambda $ be a set and $ \bm{\X} $ a $ \Lambda $-indexed family of Hilbert $ A $-modules. Let $ \Pair{P}{Q} $ be a $ 2 $-partition of $ \Lambda $, i.e., $ P \cup Q = \Lambda $ and $ P \cap Q = \varnothing $. Then
$$
\forall \bm{x} \in \prod \bm{\X}: \quad
\bm{x} \in \bigoplus \bm{\X} \iff \bm{x}|_{P} \in \bigoplus \Br{\bm{\X}|_{P}} \text{ and } \bm{x}|_{Q} \in \bigoplus \Br{\bm{\X}|_{Q}}
$$
and
$$
\forall \bm{x} \in \bigoplus \bm{\X}: \quad
     \Norm{\bm{x}}_{\bigoplus \bm{\X}}
\leq \sqrt{\Norm{\Br{\bm{x}|_{P}}}_{\bigoplus \Br{\bm{\X}|_{P}}}^{2} + \Norm{\Br{\bm{x}|_{Q}}}_{\bigoplus \Br{\bm{\X}|_{Q}}}^{2}}.
$$
\end{Lem}

\begin{proof}
This follows from the Pythagorean Inequality for orthogonal elements of a Hilbert $ C^{\ast} $-module.
\end{proof}


\begin{Def} \label{The Amalgamation Map}
Let $ \Lambda $ be a set and $ \bm{\X} $ a $ \Lambda $-indexed family of Hilbert $ A $-modules. We shall denote by $ \ds \bigoplus \Adj{\bm{\X}} $ the $ C^{\ast} $-algebraic direct sum $ \ds \bigoplus_{\lambda \in \Lambda} \Adj{\bm{\X}_{\lambda}} $, and we shall define a map $ \ds \Phi_{\bm{\X}}: \bigoplus \Adj{\bm{\X}} \to \Br{\prod \bm{\X}}^{\bigoplus \bm{\X}} $ by
$$
\forall \bm{T} \in \bigoplus \Adj{\bm{\X}}, ~ \forall \bm{x} \in \bigoplus \bm{\X}: \quad
\FUNC{\func{\Phi_{\bm{\X}}}{\bm{T}}}{\bm{x}} \df \Seq{\func{\bm{T}_{\lambda}}{\bm{x}_{\lambda}}}{\lambda \in \Lambda}.
$$
For each $ \ds \bm{T} \in \bigoplus \Adj{\bm{\X}} $, it is customary to denote $ \func{\Phi_{\bm{\X}}}{\bm{T}} $ more intuitively by $ \ds \bigoplus \bm{T} $.
\end{Def}


\begin{Lem} \label{An Injectivity Result}
Let $ \Lambda $ be a set and $ \bm{\X} $ a $ \Lambda $-indexed family of Hilbert $ A $-modules. Then $ \Phi_{\bm{\X}} $ is a $ C^{\ast} $-monomorphism from $ \ds \bigoplus \Adj{\bm{\X}} $ to $ \ds \Adj{\bigoplus \bm{\X}} $.
\end{Lem}

\begin{proof}
When $ \Lambda $ is finite, the lemma is a fundamental result in the theory of Hilbert $ C^{\ast} $-modules of which proofs may easily be found in the literature. Consequently, when $ \Lambda $ is finite, we have
$$
\forall \bm{T} \in \bigoplus \Adj{\bm{\X}}: \quad
\func{\Phi_{\bm{\X}}}{\bm{T}} \in \Adj{\bigoplus \bm{\X}}
\quad \text{and} \quad
  \Norm{\func{\Phi_{\bm{\X}}}{\bm{T}}}_{\Adj{\bigoplus \bm{\X}}}
= \sup_{\lambda \in \Lambda} \Norm{\bm{T}_{\lambda}}_{\Adj{\bm{\X}_{\lambda}}},
$$
which we shall appeal to later on in this argument.

Let $ \ds \bm{T} \in \bigoplus \Adj{\bm{\X}} $ and $ \ds M \df \sup_{\lambda \in \Lambda} \Norm{\bm{T}_{\lambda}}_{\Adj{\bm{\X}_{\lambda}}} $, which is finite by definition. Let $ \ds \bm{x} \in \bigoplus \bm{\X} $. If $ \epsilon \in \R_{> 0} $, then there exists a finite subset $ \widetilde{F} $ of $ \Lambda $ such that
$$
\forall F \subseteq_{\text{finite}} \Lambda \setminus \widetilde{F}: \quad
  \Norm{\Br{\bm{x}|_{F}}}_{\bigoplus \Br{\bm{\X}|_{F}}}^{2}
= \Norm{\Inner{\Br{\bm{x}|_{F}}}{\Br{\bm{x}|_{F}}}_{\bigoplus \Br{\bm{\X}|_{F}}}}_{A}
= \Norm{\sum_{\lambda \in F} \Inner{\bm{x}_{\lambda}}{\bm{x}_{\lambda}}_{\bm{\X}_{\lambda}}}_{A}
< \frac{\epsilon}{1 + M^{2}};
$$
from the finite case of this lemma, it thus follows that
\begin{align*}
\forall F \subseteq_{\text{finite}} \Lambda \setminus \widetilde{F}: \quad
       \Norm{\sum_{\lambda \in F} \Inner{\func{\bm{T}_{\lambda}}{\bm{x}_{\lambda}}}{\func{\bm{T}_{\lambda}}{\bm{x}_{\lambda}}}_{\bm{\X}_{\lambda}}}_{A}
& =    \Norm{
            \Inner{\FUNC{\func{\Phi_{\bm{\X}|_{F}}}{\bm{T}|_{F}}}{\bm{x}|_{F}}}
                  {\FUNC{\func{\Phi_{\bm{\X}|_{F}}}{\bm{T}|_{F}}}{\bm{x}|_{F}}}_{\bigoplus \Br{\bm{\X}|_{F}}}
            }_{A} \\
& =    \Norm{\FUNC{\func{\Phi_{\bm{\X}|_{F}}}{\bm{T}|_{F}}}{\bm{x}|_{F}}}_{\bigoplus \Br{\bm{\X}|_{F}}}^{2} \\
& \leq \Norm{\func{\Phi_{\bm{\X}|_{F}}}{\bm{T}|_{F}}}_{\Adj{\bigoplus \Br{\bm{\X}|_{F}}}}^{2} \cdot \Norm{\Br{\bm{x}|_{F}}}_{\bigoplus \Br{\bm{\X}|_{F}}}^{2} \\
& =    \Br{\sup_{\lambda \in F} \Norm{\bm{T}_{\lambda}}_{\Adj{\bm{\X}_{\lambda}}}}^{2} \cdot \Norm{\Br{\bm{x}|_{F}}}_{\bigoplus \Br{\bm{\X}|_{F}}}^{2} \\
& \leq M^{2} \cdot \frac{\epsilon}{1 + M^{2}} \\
& <    \epsilon.
\end{align*}
As a result, $ \ds \Seq{\func{T_{\lambda}}{\bm{x}_{\lambda}}}{\lambda \in \Lambda} \in \bigoplus \bm{\X} $, so $ \ds \func{\Phi_{\bm{\X}}}{\bm{T}}: \bigoplus \bm{\X} \to \bigoplus \bm{\X} $. By the same token, we also have
$$
\func{\Phi_{\bm{\X}}}{\bm{T}^{\ast}} = \func{\Phi_{\bm{\X}}}{\Seq{\bm{T}_{\lambda}^{\ast}}{\lambda \in \Lambda}}: \bigoplus \bm{\X} \to \bigoplus \bm{\X}.
$$

Next, observe that
\begin{align*}
\forall \bm{x},\bm{y} \in \bigoplus \bm{\X}: \quad
    \Inner{\FUNC{\func{\Phi_{\bm{\X}}}{\bm{T}}}{\bm{x}}}{\bm{y}}_{\bigoplus \bm{\X}}
& = \Inner{\Seq{\func{\bm{T}_{\lambda}}{\bm{x}_{\lambda}}}{\lambda \in \Lambda}}{\Seq{\bm{y}_{\lambda}}{\lambda \in \Lambda}}_{\bigoplus \bm{\X}} \\
& = \sum_{\lambda \in \Lambda} \Inner{\func{\bm{T}_{\lambda}}{\bm{x}_{\lambda}}}{\bm{y}_{\lambda}}_{\bm{\X}_{\lambda}} \\
& = \sum_{\lambda \in \Lambda} \Inner{\bm{x}_{\lambda}}{\func{\bm{T}_{\lambda}^{\ast}}{\bm{y}_{\lambda}}}_{\bm{\X}_{\lambda}} \\
& = \Inner{\Seq{\bm{x}_{\lambda}}{\lambda \in \Lambda}}{\Seq{\func{\bm{T}_{\lambda}^{\ast}}{\bm{y}_{\lambda}}}{\lambda \in \Lambda}}_{\bigoplus \bm{\X}} \\
& = \Inner{\bm{x}}{\FUNC{\func{\Phi_{\bm{\X}}}{\bm{T}^{\ast}}}{\bm{y}}}_{\bigoplus \bm{\X}}.
\end{align*}
Consequently, $ \ds \func{\Phi_{\bm{\X}}}{\bm{T}} \in \Adj{\bigoplus \bm{\X}} $ and $ \func{\Phi_{\bm{\X}}}{\bm{T}^{\ast}} = \func{\Phi_{\bm{\X}}}{\bm{T}}^{\ast} $, and, as $ \ds \bm{T} \in \bigoplus \Adj{\bm{\X}} $ is arbitrary, we find that $ \Phi_{\bm{\X}} $ is $ \ast $-preserving. It is a rather routine verification that $ \Phi_{\bm{\X}} $ is linear and multiplicative, which means that $ \Phi_{\bm{\X}} $ is truly a $ C^{\ast} $-homomorphism from $ \ds \bigoplus \Adj{\bm{\X}} $ to $ \ds \Adj{\bigoplus \bm{\X}} $.

To prove the injectivity of $ \Phi_{\bm{\X}} $, suppose that $ \ds \bm{T} \in \bigoplus \Adj{\bm{\X}} $ satisfies $ \func{\Phi_{\bm{\X}}}{\bm{T}} = \bm{0}_{\Adj{\bigoplus \bm{\X}}} $. For arbitrary $ \lambda \in \Lambda $ and $ x \in \bm{\X}_{\lambda} $, if $ \bm{x} $ denotes the element of $ \ds \bigoplus \bm{\X} $ whose $ \lambda $-th entry is $ x $ and whose other entries are null vectors, then $ \Func{\func{\Phi_{\bm{\X}}}{\bm{T}}}{\bm{x}} = \bm{0}_{\bigoplus \bm{\X}} $, so $ \func{\bm{T}_{\lambda}}{x} = \bm{0}_{\bm{\X}_{\lambda}} $. We therefore have, for all $ \lambda \in \Lambda $, that $ \bm{T}_{\lambda} = \bm{0}_{\Adj{\bm{\X}_{\lambda}}} $, which leads us to conclude that $ \bm{T} = \bm{0}_{\bigoplus \Adj{\bm{\X}}} $ and that $ \Phi_{\bm{\X}} $ is a $ C^{\ast} $-monomorphism from $ \ds \bigoplus \Adj{\bm{\X}} $ to $ \ds \Adj{\bigoplus \bm{\X}} $.
\end{proof}


The following lemma verifies that the direct sum of strongly continuous unitary representations of a locally compact Hausdorff group $ G $ yields a strongly continuous unitary representation of $ G $.


\begin{Lem} \label{A Direct Sum of Strongly Continuous Unitary Representations Is a Strongly Continuous Unitary Representation}
Let $ G $ be a locally compact Hausdorff group, $ \Lambda $ a set, $ \bm{\X} $ a $ \Lambda $-indexed family of Hilbert $ A $-modules, and $ \ds \bm{U} \in \prod_{\lambda \in \Lambda} \URep{G}{\bm{\X}_{\lambda}} $. Define $ \ds \dot{\bm{U}}: G \to \prod_{\lambda \in \Lambda} \Unitary{\bm{\X}_{\lambda}} $ by
$$
\forall g \in G: \quad
\func{\dot{\bm{U}}}{g} \df \Seq{\func{\bm{U}_{\lambda}}{g}}{\lambda \in \Lambda}.
$$
Then the following two statements are true:
\begin{enumerate}
\item
For each $ g \in G $, we have $ \ds \func{\dot{\bm{U}}}{g} \in \bigoplus \Adj{\bm{\X}} $.

\item
The map $ \ds \Phi_{\bm{\X}} \circ \dot{\bm{U}}: G \to \Adj{\bigoplus \bm{\X}} $ belongs to $ \ds \URep{G}{\bigoplus \bm{\X}} $.
\end{enumerate}
\end{Lem}

\begin{proof}
The first statement follows from the fact that, for all $ \lambda \in \Lambda $ and $ g \in G $, we have $ \Norm{\func{\bm{U}_{\lambda}}{g}}_{\Adj{\bm{\X}_{\lambda}}} = 1 $.

It is a routine verification that $ \Phi_{\bm{\X}} \circ \dot{\bm{U}} $ is a unitary representation of $ G $ on $ \ds \bigoplus \bm{\X} $, so let us instead focus our efforts on establishing its strong continuity.

Let $ \ds \bm{x} \in \bigoplus \bm{\X} $ and $ \epsilon \in \R_{> 0} $ be arbitrary. Then there exists a finite subset $ F $ of $ \Lambda $ such that
\begin{equation}
  \Norm{\Br{\bm{x}|_{\Lambda \setminus F}}}_{\bigoplus \Br{\bm{\X}|_{\Lambda \setminus F}}}^{2}
= \Norm{\sum_{\lambda \in \Lambda \setminus F} \Inner{\bm{x}_{\lambda}}{\bm{x}_{\lambda}}_{\bm{\X}_{\lambda}}}_{A}
< \frac{\epsilon^{2}}{8}, \label{Inequality 1}
\end{equation}
with the finiteness of $ F $ guaranteeing a neighborhood $ \mathcal{O} $ of $ e_{G} $ such that
\begin{equation}
\forall \lambda \in F, ~ \forall g \in \mathcal{O}: \quad
  \Norm{\FUNC{\func{\bm{U}_{\lambda}}{g} - \Id_{\bm{\X}_{\lambda}}}{\bm{x}_{\lambda}}}_{\bm{\X}_{\lambda}}
< \frac{\Br{\frac{\epsilon}{\sqrt{2}}}}{1 + \Size{F}}, \label{Inequality 2}
\end{equation}
whereupon an application of \autoref{Corollary of the Pythaogrean Inequality for Orthogonal Elements of a Hilbert C*-Module} yields
\begin{align*}
     & ~ \Norm{\FUNC{\func{\Phi_{\bm{\X}}}{\func{\dot{\bm{U}}}{g}} - \Id_{\bigoplus \bm{\X}}}{\bm{x}}}_{\bigoplus \bm{\X}} \\
\leq & ~ \sqrt{
              \Norm{\Br{\FUNC{\func{\Phi_{\bm{\X}}}{\func{\dot{\bm{U}}}{g}} - \Id_{\bigoplus \bm{\X}}}{\bm{x}} \Big|_{F}}}_{\bigoplus \Br{\bm{\X}|_{F}}}^{2} +
              \Norm{
                   \Br{\FUNC{\func{\Phi_{\bm{\X}}}{\func{\dot{\bm{U}}}{g}} - \Id_{\bigoplus \bm{\X}}}{\bm{x}} \Big|_{\Lambda \setminus F}}
                   }_{\bigoplus \Br{\bm{\X}|_{\Lambda \setminus F}}}^{2}
              } \\
=    & ~ \sqrt{
              \Norm{
                   \FUNC{\func{\Phi_{\bm{\X}|_{F}}}{\func{\dot{\bm{U}}}{g}|_{F}} - \Id_{\bigoplus \Br{\bm{\X}|_{F}}}}{\bm{x}|_{F}}
                   }_{\bigoplus \Br{\bm{\X}|_{F}}}^{2}
              +
              \Norm{
                   \FUNC{
                        \func{\Phi_{\bm{\X}|_{\Lambda \setminus F}}}{\func{\dot{\bm{U}}}{g}|_{\Lambda \setminus F}} -
                        \Id_{\bigoplus \Br{\bm{\X}|_{\Lambda \setminus F}}}
                        }
                        {\bm{x}|_{\Lambda \setminus F}}
                   }_{\bigoplus \Br{\bm{\X}|_{\Lambda \setminus F}}}^{2}
              } \\
\leq & ~ \sqrt{
              \Br{\sum_{\lambda \in F} \Norm{\FUNC{\func{\bm{U}_{\lambda}}{g} - \Id_{\bm{\X}_{\lambda}}}{\bm{x}_{\lambda}}}_{\bm{\X}_{\lambda}}}^{2} +
              \Norm{
                   \func{\Phi_{\bm{\X}|_{\Lambda \setminus F}}}{\func{\dot{\bm{U}}}{g}|_{\Lambda \setminus F}} -
                   \Id_{\bigoplus \Br{\bm{\X}|_{\Lambda \setminus F}}}
                   }_{\Adj{\bigoplus \Br{\bm{\X}|_{\Lambda \setminus F}}}}^{2}
              \Norm{\Br{\bm{x}|_{\Lambda \setminus F}}}_{\bigoplus \Br{\bm{\X}|_{\Lambda \setminus F}}}^{2}
              } \\
\leq & ~ \sqrt{
              \SqBr{\sum_{\lambda \in F} \frac{\Br{\frac{\epsilon}{\sqrt{2}}}}{1 + \Size{F}}}^{2} +
              \Norm{
                   \func{\Phi_{\bm{\X}|_{\Lambda \setminus F}}}{\func{\dot{\bm{U}}}{g}|_{\Lambda \setminus F}} -
                   \Id_{\bigoplus \Br{\bm{\X}|_{\Lambda \setminus F}}}
                   }_{\Adj{\bigoplus \Br{\bm{\X}|_{\Lambda \setminus F}}}}^{2}
              \cdot
              \frac{\epsilon^{2}}{8}
              } \quad \Br{\text{Inequalities (\ref{Inequality 1}) and (\ref{Inequality 2})}} \\
\leq & ~ \sqrt{
              \SqBr{\sum_{\lambda \in F} \frac{\Br{\frac{\epsilon}{\sqrt{2}}}}{1 + \Size{F}}}^{2} +
              \SqBr{
                   \Norm{
                        \func{\Phi_{\bm{\X}|_{\Lambda \setminus F}}}{\func{\dot{\bm{U}}}{g}|_{\Lambda \setminus F}}
                        }_{\Adj{\bigoplus \Br{\bm{\X}|_{\Lambda \setminus F}}}}
                   +
                   \Norm{\Id_{\bigoplus \Br{\bm{\X}|_{\Lambda \setminus F}}}}_{\Adj{\bigoplus \Br{\bm{\X}|_{\Lambda \setminus F}}}}
                   }^{2}
              \cdot
              \frac{\epsilon^{2}}{8}
              } \\
=    & ~ \sqrt{\SqBr{\sum_{\lambda \in F} \frac{\Br{\frac{\epsilon}{\sqrt{2}}}}{1 + \Size{F}}}^{2} + \Br{1 + 1}^{2} \cdot \frac{\epsilon^{2}}{8}} \\
<    & ~ \sqrt{\frac{\epsilon^{2}}{2} + \frac{\epsilon^{2}}{2}} \\
=    & ~ \epsilon.
\end{align*}
Hence, $ \Phi_{\bm{\X}} \circ \dot{\bm{U}} $ is strongly continuous at $ e_{G} $ and therefore strongly continuous throughout $ G $.
\end{proof}

\subsection{Unbounded Operators Between Hilbert \texorpdfstring{$ C^{\ast} $}{C*}-Modules} 

Let us now review some basic definitions and results that are related to operators between Banach spaces and between Hilbert $ C^{\ast} $-modules that may only be partially defined.

Let $ E $, $ F $, and $ G $ be Banach spaces, and let $\Bdd{E,F}$ denote the set of all bounded linear maps from $E$ to $F$.
\begin{itemize}
\item
An operator from $ E $ to $ F $ is a $ \C $-linear map $ T $ from some $ \C $-linear subspace of $ E $ to $ F $, which is then said to be \emph{densely defined} if and only if $ \Dom{T} $ is dense in $ E $.

An operator on $ E $ is then an operator from $ E $ to $ E $.

\item
If $ T $ is an operator on $ E $, and $ D $ is a $ \C $-linear subspace of $ E $, then $ D $ is said to be $ T $-\emph{invariant} if and only if $ D \subseteq \Dom{T} $ and $ \Im{T}{D} \subseteq D $.

\item
An \emph{operator} $ T $ from $ E $ to $ F $ is said to be \emph{closed} if and only if its graph, denoted by $ \Graph{T} $, is a closed subset (in fact, a closed $ \C $-linear subspace) of $ E \times F $.

\item
An operator $ T $ from $ E $ to $ F $ is said to be \emph{closable} if and only if the closure of $ \Graph{T} $ in $ E \times F $ is the graph of an operator from $ E $ to $ F $, which is necessarily closed and shall be denoted by $ \cl{T} $.

\item
If $ S $ is a closable operator from $ E $ to $ F $ and $ T \in \Bdd{E,F} $, then $ S + T $ is closable and $ \cl{S + T} = \cl{S} + T $.

\item
If $ S $ is a closable operator from $ F $ to $ G $ and $ T \in \Bdd{E,F} $, then $ S T $ is closable and $ \cl{S T} = \cl{S} T $.

\item
If $ S \in \Bdd{F,G} $ is invertible and $ T $ is a closable operator from $ E $ to $ F $, then $ S T $ is closable and $ \cl{S T} = S \cl{T} $.

\item
A \emph{core} for an operator $ T $ from $ E $ to $ F $ is a $ \C $-linear subspace $ D $ of $ E $ contained within $ \Dom{T} $ for which $ T|_{D} $ is closable and $ \cl{T|_{D}} = T $ (hence, $ T $ is necessarily closed).

\item
Given a collection $ \mathcal{C} $ of operators on $ E $, we say that $ v \in E $ is \emph{$ \mathcal{C} $-analytic} if and only if, for all $ T \in \mathcal{C} $ and $ n \in \N_{0} $, we have $ v \in \Dom{T^{n}} $ and, furthermore, the real power-series map
$$
\Map{\R}{\R}{t}{\sum_{n \in \N_{0}} \frac{\Norm{\func{T^{n}}{v}}_{E}}{n!} \cdot t^{n}}
$$
has a positive radius of convergence. A $ \C $-linear subspace $ D $ of $ E $ is $ \mathcal{C} $-analytic if and only if all of its elements are $ \mathcal{C} $-analytic.
\end{itemize}

Let $ \X $, $ \Y $, and $ \Z $ be Hilbert $ A $-modules, which are already Banach spaces.
\begin{itemize}
\item
A \emph{modular operator from $ \X $ to $ \Y $} is a $ \C $-linear and $ A $-linear map $ T $ from some $ \C $-linear $ A $-submodule $\Dom{T}$ of $ \X $ to $ \Y $, which is then said to be \emph{densely defined} if and only if $ \Dom{T} $ is dense in $ \X $.

A {\em modular operator on $ \X $} is a modular operator from $ \X $ to $ \X $.

\item
The \emph{graph} of a modular operator $ T $ from $ \X $ to $ \Y $, denoted by $ \Graph{T} $, is the following subset of $ \X \oplus \Y $:
$$
\Set{\Pair{\zeta}{\eta}\in \X \oplus \Y}{\zeta \in \Dom{T} \text{ and } \func{T}{\zeta} = \eta}.
$$

Clearly, the graph of a modular operator from $ \X $ to $ \Y $ is a $ \C $-linear $ A $-submodule of $ \X \oplus \Y $.

\item
For $ T $ a densely defined modular operator from $ \X $ to $ \Y $, it may be shown that
$$
\Set{\Pair{\eta}{\xi} \in \Y \oplus \X}{\forall \zeta \in \Dom{T}: ~ \Inner{\func{T}{\zeta}}{\eta}_{\Y} = \Inner{\zeta}{\xi}_{\X}}
$$
is the graph of a modular operator from $ \Y $ to $ \X $, which shall be denoted by $ T^{\ast} $ and called the \emph{adjoint} of $ T $; if $ J: \X \oplus \Y \to \Y \oplus \X $ is defined by
$$
\forall \zeta \in \X, ~ \forall \eta \in \Y: \quad
\func{J}{\zeta,\eta} \df \Pair{- \eta}{\xi},
$$
then $ \Graph{T^{\ast}} = \Br{\Im{J}{\Graph{T}}}^{\perp_{\Y \oplus \X}} $, which is yet another basic result.

\item
As orthogonal complements in a Hilbert $ C^{\ast} $-module are closed $ \C $-linear $ A $-submodules, it follows from the previous bullet point that the adjoint of a densely defined modular operator from $ \X $ to $ \Y $ is closed.

\item
If $ T $ is a closable densely defined modular operator from $ \X $ to $ \Y $, then $ T^{\ast} = \cl{T}^{\ast} $.

\item
If $ S $ and $ T $ are densely defined modular operators from $ \Y $ to $ \Z $ and from $ \X $ to $ \Y $, respectively, and $ S T $ is densely defined, then $ T^{\ast} S^{\ast} \subseteq \Br{S T}^{\ast} $, with equality when $ T \in \Adj{\X,\Y} $ (cf. \cite[Theorem 13.2]{Rudin}).

\item
A \emph{core} for a modular operator $ T $ from $ \X $ to $ \Y $ is a $ \C $-linear $ A $-submodule $ D $ of $ \X $ contained within $ \Dom{T} $ for which $ T|_{D} $ is closable and $ \cl{T|_{D}} = T $ (hence, $ T $ is necessarily closed).

\item
A \emph{regular operator} from $ \X $ to $ \Y $ is a densely defined modular operator $ T $ from $ \X $ to $ \Y $ such that
\begin{enumerate}
\item
$ T $ is closed,

\item
$ T^{\ast} $ is densely defined, and

\item
$ \Cl{\Range{\Id_{\X} + T^{\ast} T}}{\X} = \X $,
\end{enumerate}
in which case the following three statements hold (\cite[Lemma 9.1, Lemma 9.2, and Theorem 10.4]{Lance}):
\begin{itemize}
\item
$ T^{*} T $ is densely defined, and $ \Dom{T^{*} T} $ is a core for $ T $.

\item
$ \Range{\Id_{\X} + T^{\ast} T} = \X $ and $ \Br{\Id_{\X} + T^{\ast} T}^{- 1} \in \Adj{\X}_{\geq 0} $.

\item
$ T \Br{\Id_{\X} + T^{\ast} T}^{- \frac{1}{2}} \in \Adj{\X,\Y} $.
\end{itemize}

\item
An \emph{essentially regular} operator from $ \X $ to $ \Y $ is a closable densely defined modular operator from $ \X $ to $ \Y $ whose closure is regular.

\item
A \emph{symmetric operator} on $ \X $ is a densely defined modular operator $ T $ on $ \X $ such that
$$
\Dom{T} \subseteq \Dom{T^{\ast}}
\quad \text{and} \quad
\forall \zeta \in \Dom{T}: \quad \func{T}{\zeta} = \func{T^{\ast}}{\zeta},
$$
which yields $ T \subseteq T^{\ast} $ and implies that $ T $ is closable. 

\item
A \emph{self-adjoint operator} on $ \X $ is a densely defined modular operator $ T $ on $ \X $ such that $ T = T^{\ast} $.

\item
An \emph{essentially self-adjoint operator} on $ \X $ is a symmetric operator on $ \X $ whose closure is self-adjoint.

\item
The set of regular self-adjoint operators on $ \X $ shall be denoted by $ \RS{\X} $.

\item
Given a collection $ \mathcal{C} $ of modular operators on $ \X $ and an $ A $-submodule $ D $ of $ \X $, we say that $ D $ is \emph{$ \mathcal{C} $-invariant} if and only if, for all $ T \in \mathcal{C} $, we have $ D \subseteq \Dom{T} $ and $ \Im{T}{D} \subseteq D $.

\item
Given a collection $ \mathcal{C} $ of modular operators on $ \X $, we say that $ \zeta \in \X $ is \emph{$ \mathcal{C} $-analytic} if and only if, for all $ T \in \mathcal{C} $ and $ n \in \N_{0} $, we have $ \zeta \in \Dom{T^{n}} $ and, furthermore, the real power-series map
$$
\Map{\R}{\R}{t}{\sum_{n \in \N_{0}} \frac{\Norm{\func{T^{n}}{\zeta}}_{\X}}{n!} \cdot t^{n}}
$$
has a positive radius of convergence; we then say that an $ A $-submodule $ D $ of $ \X $ is $ \mathcal{C} $-analytic if and only if all of its elements are $ \mathcal{C} $-analytic.
\end{itemize}

\subsection{Group Actions, Smooth and Analytic Vectors, and Infinitesimal Generators} 

We are ready to discuss group actions on Banach spaces, $ C^{\ast} $-algebras, and Hilbert $ C^{\ast} $-modules. Much of the material in this subsection has been adapted from \cite{Bratteli-Robinson,Taylor}; for a high-level summary of analytic vectors of Lie Banach dynamical systems, see \cite[Remark 3.1]{Gimperlein-Krotz-Schlichtkrull}.

We begin with Banach dynamical systems and $ C_{0} $-groups.


\begin{Def}
A \emph{Banach dynamical system} is an ordered triple $ \Trip{G}{E}{\pi} $ with the following two properties:
\begin{enumerate}
\item
$ G $ is a locally compact Hausdorff group and $ E $ is a Banach space.

\item
$ \pi $ is a strongly continuous action of $ G $ on $ E $, i.e., $ \pi $ is a group homomorphism from $ G $ to $ \Inv{E} $ --- the group of invertible elements of $ \Bdd{E} $ --- and, for each $ v \in E $, the map
$$
\Map{G}{E}{g}{\FUNC{\func{\pi}{g}}{v}}
$$
is continuous; a $ \C $-linear subspace $ D $ of $ E $ is said to be $ \pi $-\emph{invariant} if and only if, for all $ g \in G $, we have $ \Im{\SqBr{\func{\pi}{g}}}{D} \subseteq D $.
\end{enumerate}
A \emph{Lie} Banach dynamical system is a Banach dynamical system $ \Trip{G}{E}{\pi} $ where $ G $ is a real Lie group, and a \emph{$ C_{0} $-group} is a Banach dynamical system of the form $ \Trip{\R}{E}{\pi} $.

For $ \Trip{G}{E}{\pi} $ a Lie Banach dynamical system, its \emph{infinitesimal representation} is the map $ \d \pi $ from the Lie algebra $ \mathfrak{g} $ of $ G $ to the set of operators on $ E $ with the following two properties:
\begin{itemize}
\item
$ \ds \forall x \in \mathfrak{g}: \quad \Dom{\func{\d \pi}{x}} = \Set{v \in E}{\exists w \in E: ~ \lim_{t \to 0} \Norm{\frac{1}{t} \cdot \Br{\FUNC{\func{\pi}{\func{\exp_{G}}{t \cdot x}}}{v} - v} - w}_{E} = 0} $.

\item
$ \ds \forall x \in \mathfrak{g}, ~ \forall v \in \Dom{\func{\d \pi}{x}}: \quad \FUNC{\func{\d \pi}{x}}{v} \df \Diff{t}{\FUNC{\func{\pi}{\func{\exp_{G}}{t \cdot x}}}{v}} \bigg|_{t = 0} = \lim_{t \to 0} \frac{1}{t} \cdot \Br{\FUNC{\func{\pi}{\func{\exp_{G}}{t \cdot x}}}{v} - v} $.
\end{itemize}

For $ \Trip{\R}{E}{\pi} $ a $ C_{0} $-group, its \emph{infinitesimal generator} is the operator $ T $ on $ E $ with the following two properties:
\begin{itemize}
\item
$ \ds \Dom{T} = \Set{v \in E}{\exists w \in E: ~ \lim_{t \to 0} \Norm{\frac{1}{t} \cdot \Br{\FUNC{\func{\pi}{t}}{v} - v} - w}_{E} = 0} $.

\item
$ \ds \forall v \in \Dom{T}: \quad \func{T}{v} \df \Diff{t}{\FUNC{\func{\pi}{t}}{v}} \bigg|_{t = 0} = \lim_{t \to 0} \frac{1}{t} \cdot \Br{\FUNC{\func{\pi}{t}}{v} - v} $.
\end{itemize}
\end{Def}


\begin{Def}
Let $ \Trip{G}{E}{\pi} $ be a Lie Banach dynamical system.
\begin{itemize}
\item
The $ \Trip{G}{E}{\pi} $-\emph{smooth vectors} are those $ v \in E $ for which the map $ \Map{G}{E}{g}{\FUNC{\func{\pi}{g}}{v}} $ is smooth. \\
\textbf{Remark:} A map $ f: G \to E $ is said to be smooth if and only if, for all charts $ \Pair{O}{\phi} $ in the unique maximal smooth atlas defining $ G $, the map $ f \circ \phi^{- 1}: \Im{\phi}{O} \subseteq \R^{\func{\dim}{G}} \to E $ has partial derivatives of all orders.

The set of $ \Trip{G}{E}{\pi} $-smooth vectors shall be denoted by $ \Trip{G}{E}{\pi}^{\infty} $.

\item
The $ \Trip{G}{E}{\pi} $-\emph{analytic vectors} are those $ v \in E $ for which the map $ \Map{G}{E}{g}{\FUNC{\func{\pi}{g}}{v}} $ is real-analytic. \\
\textbf{Remark:} A map $ f: G \to E $ is said to be real-analytic if and only if, for all charts $ \Pair{O}{\phi} $ in the unique maximal analytic atlas defining $ G $, the map $ f \circ \phi^{- 1}: \Im{\phi}{O} \subseteq \R^{\func{\dim}{G}} \to E $ has a local real power-series expansion centered at each point in $ \Im{\phi}{O} $.

The set of $ \Trip{G}{E}{\pi} $-analytic vectors shall be denoted by $ \Trip{G}{E}{\pi}^{\omega} $.
\end{itemize}
Both $ \Trip{G}{E}{\pi}^{\infty} $ and $ \Trip{G}{E}{\pi}^{\omega} $ are $ \pi $-invariant dense $ \C $-linear subspaces of $ E $ (\cite[Theorem 4]{Nelson}); furthermore, $ \Trip{G}{E}{\pi}^{\omega} \subseteq \Trip{G}{E}{\pi}^{\infty} $.
\end{Def}


\begin{Lem}
For $ \Trip{G}{E}{\pi} $ a Lie Banach dynamical system and $ \mathfrak{g} $ the Lie algebra of $ G $, the following two statements hold:
\begin{enumerate}
\item
We have, for each $ x \in \mathfrak{g} $, that $ \func{\d \pi}{x} $ is a closed and densely defined operator on $ E $.

\item
We have, for each $ x \in \mathfrak{g} $, that $ \Trip{G}{E}{\pi}^{\infty} $ and $ \Trip{G}{E}{\pi}^{\omega} $ are $ \func{\d \pi}{x} $-invariant $ \C $-linear subspaces of $ E $.
\end{enumerate}
\end{Lem}


The following lemma contains all the foundational results about $ C_{0} $-groups that we shall need later.


\begin{Lem} \label{Infinitesimal Generators of C0-Groups}
Let $ \Trip{\R}{E}{\pi} $ and $ \Trip{\R}{E}{\rho} $ be $ C_{0} $-groups, and let $ S $ and $ T $ denote their infinitesimal generators, respectively. Then the following three statements hold:
\begin{enumerate}
\item
$ S $ and $ T $ are closed and densely defined operators on $ E $.

\item
We have $ \pi = \rho \iff S = T $.

\item
If $ S \subseteq T $, then, actually, $ S = T $ and $ \pi = \rho $; hence, infinitesimal generators of $ C_{0} $-groups are maximal.
\end{enumerate}
\end{Lem}


Let us now recall the definition of $ C^{\ast} $-dynamical systems and related concepts.


\begin{Def}
A \emph{$ C^{\ast} $-dynamical system} is an ordered triple $ \Trip{G}{A}{\alpha} $ with the following two properties:
\begin{enumerate}
\item
$ G $ is a locally compact Hausdorff group (and $ A $ is a $ C^{\ast} $-algebra).

\item
$ \alpha $ is a strongly continuous action of $ G $ on $ A $, i.e., $ \alpha $ is a group homomorphism from $ G $ to $ \Aut{A} $ --- the group of $ C^{\ast} $-automorphisms of $ A $ --- and, for each $ a \in A $, the map
$$
\Map{G}{A}{g}{\alphArg{g}{a}}
$$
is continuous; a $ \C $-linear subspace $ D $ of $ A $ is said to be $ \alpha $-\emph{invariant} if and only if, for all $ g \in G $, we have $ \Im{\alpha_{g}}{D} \subseteq D $.
\end{enumerate}
A $ C^{\ast} $-dynamical system $ \Trip{G}{A}{\alpha} $ is said to be:
\begin{itemize}
\item
\emph{Lie} if and only if $ G $ is a real Lie group;

\item
\emph{real} if and only if $ G = \R $;

\item
\emph{elementary} if and only if $ A $ is an elementary $ C^{\ast} $-algebra.
\end{itemize}

For $ \Trip{\R}{A}{\alpha} $ a real $ C^{\ast} $-dynamical system, its \emph{infinitesimal generator} is the operator $ \delta $ on $ A $ with the following two properties:
\begin{itemize}
\item
$ \ds \Dom{\delta} = \Set{a \in A}{\exists b \in A: ~ \lim_{t \to 0} \Norm{\frac{1}{t} \cdot \Br{\alphArg{t}{a} - a} - b}_{A} = 0} $.

\item
$ \ds \forall a \in \Dom{\delta}: \quad \func{\delta}{a} \df \Diff{t}{\alphArg{t}{a}} \bigg|_{t = 0} = \lim_{t \to 0} \frac{1}{t} \cdot \Br{\alphArg{t}{a} - a} $.
\end{itemize}
\end{Def}


To every $ C^{\ast} $-dynamical system $ \Trip{G}{A}{\alpha} $, there are two canonically associated $ C^{\ast} $-algebras:
\begin{itemize}
\item
The full crossed product of $ \Trip{G}{A}{\alpha} $, denoted by $ \FCP{G}{A}{\alpha} $.

\item
The reduced crossed product of $ \Trip{G}{A}{\alpha} $, denoted by $ \RCP{G}{A}{\alpha} $.
\end{itemize}
As all the necessary work involving crossed-product $ C^{\ast} $-algebras has already been completed in \autoref{thm:CSVNT}, this paper does not demand their use, so we shall not discuss them further.


\begin{Def}
Let $ \Trip{G}{A}{\alpha} $ be a Lie $ C^{\ast} $-dynamical system.
\begin{itemize}
\item
The $ \Trip{G}{A}{\alpha} $-\emph{smooth vectors} are those $ a \in A $ for which the map $ \Map{G}{A}{g}{\alphArg{g}{a}} $ is smooth.

The set of $ \Trip{G}{A}{\alpha} $-smooth vector shall be denoted by $ \Trip{G}{A}{\alpha}^{\infty} $.

\item
The $ \Trip{G}{A}{\alpha} $-\emph{analytic vectors} are those $ a \in A $ for which the map $ \Map{G}{A}{g}{\alphArg{g}{a}} $ is real-analytic.

The set of $ \Trip{G}{A}{\alpha} $-analytic vectors shall be denoted by $ \Trip{G}{A}{\alpha}^{\omega} $.
\end{itemize}
Both $ \Trip{G}{A}{\alpha}^{\infty} $ and $ \Trip{G}{A}{\alpha}^{\omega} $ are $ \alpha $-invariant dense $ \C $-linear subspaces of $ A $ (\cite[Theorem 4]{Nelson}); furthermore, $ \Trip{G}{A}{\alpha}^{\omega} \subseteq \Trip{G}{A}{\alpha}^{\infty} $.
\end{Def}


\begin{Lem} \label{Infinitesimal Generators of Real C*-Dynamical Systems}
For $ \Trip{\R}{A}{\alpha} $ a real $ C^{\ast} $-dynamical system and $ \delta $ its infinitesimal generator, the following three statements hold:
\begin{enumerate}
\item
$ \delta $ is a closed and densely defined operator on $ A $.

\item
$ \Trip{\R}{A}{\alpha}^{\infty} $ and $ \Trip{\R}{A}{\alpha}^{\omega} $ are $ \delta $-invariant $ \ast $-algebras of $ A $.

\item
$ \delta|_{\Trip{\R}{A}{\alpha}^{\infty}} $ and $ \delta|_{\Trip{\R}{A}{\alpha}^{\omega}} $ are $ \ast $-derivations on $ \Trip{\R}{A}{\alpha}^{\infty} $ and $ \Trip{\R}{A}{\alpha}^{\omega} $, respectively; furthermore,
\begin{align*}
  \forall t \in \R, ~ \forall a \in \Dom{\delta}:                                      \quad
& \alphArg{t}{a} \in \Dom{\delta}                  \quad \text{and} \quad \func{\delta}{\alphArg{t}{a}}     = \alphArg{t}{\func{\delta}{a}}; \\
  \forall t \in \R, ~ \forall a \in \Trip{\R}{A}{\alpha}^{\infty}, ~ \forall n \in \N: \quad
& \alphArg{t}{a} \in \Trip{\R}{A}{\alpha}^{\infty} \quad \text{and} \quad \func{\delta^{n}}{\alphArg{t}{a}} = \alphArg{t}{\func{\delta^{n}}{a}}; \\
  \forall t \in \R, ~ \forall a \in \Trip{\R}{A}{\alpha}^{\omega}, ~ \forall n \in \N: \quad
& \alphArg{t}{a} \in \Trip{\R}{A}{\alpha}^{\omega} \quad \text{and} \quad \func{\delta^{n}}{\alphArg{t}{a}} = \alphArg{t}{\func{\delta^{n}}{a}}.
\end{align*}
\end{enumerate}
\end{Lem}


Having defined strongly continuous unitary representations of locally compact Hausdorff groups in Section 2.1, we may similarly talk about smoothness and analyticity of vectors for such representations of Lie groups.


\begin{Def}
Let $ G $ be a real Lie group, $ \X $ a Hilbert $ A $-module, and $ U \in \URep{G}{\X} $.
\begin{itemize}
\item
The $ U $-\emph{smooth vectors} are those $ \zeta \in \X $ for which the map $ \Map{G}{\X}{g}{\FUNC{\func{U}{g}}{\zeta}} $ is smooth.

The set of $ U $-smooth vectors shall be denoted by $ \Trip{G}{\X}{U}^{\infty} $.

\item
The $ U $-\emph{analytic vectors} are those $ \zeta \in \X $ for which the map $ \Map{G}{\X}{g}{\FUNC{\func{U}{g}}{\zeta}} $ is real-analytic.

The set of $ U $-analytic vectors shall be denoted by $ \Trip{G}{\X}{U}^{\omega} $.
\end{itemize}

Both $ \Trip{G}{\X}{U}^{\infty} $ and $ \Trip{G}{\X}{U}^{\omega} $ are $ U $-invariant dense $ \C $-linear subspaces of $ \X $ (\cite[Theorem 4]{Nelson}); furthermore, $ \Trip{G}{\X}{U}^{\omega} \subseteq \Trip{G}{\X}{U}^{\infty} $.
\end{Def}


\begin{Def} \label{The Infinitesimal Generator of a Strongly Continuous One-Parameter Unitary Group}
For $ \X $ a Hilbert $ A $-module and $ U \in \URep{\R}{\X} $, the \emph{infinitesimal generator} of $ U $ is the unique operator $ T $ on $ \X $ which satisfies:
\begin{itemize}
\item
$ \ds \Dom{T} \df \Set{\zeta \in \X}{\exists \eta \in \X: ~ \lim_{t \to 0} \Norm{- \frac{i}{t} \cdot \Func{U_{t} - \Id_{\X}}{\zeta} - \eta}_{\X} = 0} $ and

\item
$ \ds \forall \zeta \in \Dom{T}: \quad \func{T}{\zeta} \df \lim_{t \to 0} \SqBr{- \frac{i}{t} \cdot \Func{U_{t} - \Id_{\X}}{\zeta}} $.
\end{itemize}
\end{Def}


\begin{Lem} \label{The Infinitesimal Generator and Analytic Vectors of a Strongly Continuous One-Parameter Unitary Group}
For $ \X $ a Hilbert $ A $-module, $ U \in \URep{\R}{\X} $, and $ T $ the infinitesimal generator of $ U $, the following three statements hold:
\begin{enumerate}
\item
$ T $ is a closed and densely defined modular operator on $ \X $.

\item
$ \Trip{G}{\X}{U}^{\infty} $ and $ \Trip{G}{\X}{U}^{\omega} $ are $ T $-invariant $ A $-submodules of $ \X $.

\item
We have
\begin{align*}
  \forall t \in \R, ~ \forall \zeta \in \Dom{T}:                                       \quad
& \func{U_{t}}{\zeta} \in \Dom{T}                   \quad \text{and} \quad \func{T}{\func{U_{t}}{\zeta}}     = \func{U_{t}}{\func{T}{\zeta}}; \\
  \forall t \in \R, ~ \forall \zeta \in \Trip{\R}{\X}{U}^{\infty}, ~ \forall n \in \N: \quad
& \func{U_{t}}{\zeta} \in \Trip{\R}{\X}{U}^{\infty} \quad \text{and} \quad \func{T^{n}}{\func{U_{t}}{\zeta}} = \func{U_{t}}{\func{T^{n}}{a}}; \\
  \forall t \in \R, ~ \forall \zeta \in \Trip{\R}{\X}{U}^{\omega}, ~ \forall n \in \N: \quad
& \func{U_{t}}{\zeta} \in \Trip{\R}{\X}{U}^{\omega} \quad \text{and} \quad \func{T^{n}}{\func{U_{t}}{\zeta}} = \func{U_{t}}{\func{T^{n}}{a}}.
\end{align*}
\end{enumerate}
\end{Lem}


We are now going to quote two important theorems.

The first, Rusinek's Integrability Theorem, says that a representation of the Lie algebra of a real Lie group $ G $, by operators on a common dense invariant $ \C $-linear subspace $ D $ of a Banach space $ E $, is integrable to a $ G $-Banach dynamical system on $ E $, if $ D $ is analytic for the operators associated to a Lie generating subset; this result is precisely the tool for establishing Heisenberg integrability in Section 4.

The second permits us to use analyticity as a criterion for deciding whether a dense $ \C $-linear subspace of a Banach space $ E $ is a core for the infinitesimal generator of a contractive $ C_{0} $-group on $ E $.

These theorems are based on the fact that, under the right hypotheses, a finite-dimensional real Lie algebra of operators on a Banach space $ E $, possessing a common dense invariant analytic $ \C $-linear subspace, may --- via a power series argument --- be used to construct a local representation of the associated Lie group $ G $ on $ E $, which may then be extended to a strongly continuous global representation of $ G $ on $ E $.


\begin{Thm}[{\cite[Theorem 1]{Rusinek}}] \label{Rusinek's Integrability Theorem}
Let $ E $ be a Banach space, $ D $ a dense $ \C $-linear subspace of $ E $, and $ G $ a real Lie group. Let $ \mathfrak{g} $ denote the Lie algebra of $ G $, and let $ T $ be a Lie-algebra representation of $ \mathfrak{g} $ on $ D $, i.e., the following three conditions hold:
\begin{enumerate}
\item
T is a function from $ \mathfrak{g} $ to the set of operators on $ E $ with invariant domain $ D $.

\item
For all $ \lambda,\mu \in \R $ and $ x,y \in \mathfrak{g} $, we have $ \lambda \cdot \func{T}{x} + \mu \cdot \func{T}{y} = \func{T}{\lambda \cdot x + \mu \cdot y} $.

\item
For all $ x,y \in \mathfrak{g} $, we have $ \func{T}{x} \func{T}{y} - \func{T}{y} \func{T}{x} = \func{T}{\Comm{x}{y}_{\mathfrak{g}}} $.
\end{enumerate}
Suppose that $ S \subseteq \mathfrak{g} $ generates $ \mathfrak{g} $ as a Lie algebra and that the following two conditions hold for each $ x \in S $:
\begin{itemize}
\item
$ \func{T}{x} $ is closable, and $ \cl{\func{T}{x}} $ is the infinitesimal generator of some $ C_{0} $-group on $ E $.

\item
$ D $ is $ \func{T}{x} $-analytic.
\end{itemize}
Then there exists a $ \pi $ making $ \Trip{G}{E}{\pi} $ a Lie Banach dynamical system such that
$$
\forall x \in \mathfrak{g}: \quad
\func{T}{x} = \func{\d \pi}{x}|_{D}.
$$
\end{Thm}


\begin{Thm}[{\cite[Theorem 3.1.20]{Bratteli-Robinson}}] \label{Contractivity and Invariance with Respect to a Contractive C0-Group's Infinitesimal Generator}
Let $ \Trip{\R}{E}{\pi} $ be a $ C_{0} $-group with infinitesimal generator $ T $, and let $ D $ be a dense $ \C $-linear subspace of $ E $. Suppose that the following two conditions hold:
\begin{enumerate}
\item
$ \Trip{\R}{E}{\pi} $ is a contractive $ C_{0} $-group, i.e., for each $ t \in \R $, we have that $ \func{\pi}{t} $ is a contraction on $ E $.

\item
$ D $ is $ S $-invariant and $ S $-analytic.
\end{enumerate}
Then $ D $ is a core for $ T $.
\end{Thm}

\subsection{Properties of Symmetric Operators} 

Finally, we conclude this preliminary section by formulating and proving certain statements on properties of symmetric operators on Hilbert $ C^{\ast} $-modules.


\begin{Prop}[Criterion for Regularity and Self-Adjointness] \label{Criterion for Regularity and Self-Adjointness}
Let $ T $ be a symmetric operator on a Hilbert $ A $-module $ \X $. Then $ T \in \RS{\X} $ if and only if $ \Range{T \pm i \cdot \Id_{\X}} = \X $.
\end{Prop}

\begin{proof}
Suppose that $ T \in \RS{\X} $. Then Lemma 9.8 of \cite{Lance} says that $ \Range{T \pm i \cdot \Id_{\X}} = \X $.

Conversely, suppose that $ \Range{T \pm i \cdot \Id_{\X}} = \X $. By way of contradiction, assume that $ T $ is not self-adjoint, which means that $ T^{\ast} $ is a proper extension of $ T $. Then $ \Br{T \pm i \cdot \Id_{\X}}^{\ast} = T^{\ast} \mp i \cdot \Id_{\X} $ is a proper extension of $ T \mp i \cdot \Id_{\X} $, so $ \Br{T \pm i \cdot \Id_{\X}}^{\ast} $ cannot be injective. However,
$$
  \ker{\Br{T \pm i \cdot \Id_{\X}}^{\ast}}
= \Range{T \pm i \cdot \Id_{\X}}^{\perp}
= X^{\perp}
= \SSet{0_{\X}},
$$
so $ \Br{T \pm i \cdot \Id_{\X}}^{\ast} $ is injective --- a contradiction. Therefore, $ T $ is indeed self-adjoint, and Lemma 9.8 of \cite{Lance} says that $ T $ is regular as well.
\end{proof}


\begin{Prop}[Criterion for Essential Regularity and Essential Self-Adjointness] \label{Criterion for Essential Regularity and Essential Self-Adjointness}
Let $ T $ be a symmetric operator on a Hilbert $ A $-module $ \X $. Then $ T $ is essentially regular and essentially self-adjoint if and only if $ \Cl{\Range{T \pm i \cdot \Id_{\X}}}{\X} = \X $.
\end{Prop}

\begin{proof}
Suppose that $ T $ is essentially regular and essentially self-adjoint, so that $ \cl{T} \in \RS{\X} $; by \autoref{Criterion for Regularity and Self-Adjointness}, we then have $ \Range{\cl{T} \pm i \cdot \Id_{\X}} = \X $. Thus, given an arbitrary $ \eta \in \X $, there exists $ \zeta \in \Dom{\cl{T}} $ such that $ \Pair{\zeta}{\eta} \in \Graph{\cl{T} \pm i \cdot \Id_{\X}} $. As
$$
  \Cl{\Graph{T \pm i \cdot \Id_{\X}}}{\X \oplus \X}
= \Graph{\cl{T \pm i \cdot \Id_{\X}}}
= \Graph{\cl{T} \pm i \cdot \Id_{\X}},
$$
there exists a sequence in $ \Graph{T \pm i \cdot \Id_{\X}} $ that converges in $ \X \oplus \X $ to $ \Pair{\zeta}{\eta} $, and, therefore, a sequence in $ \Range{T \pm i \cdot \Id_{\X}} $ that converges in $ \X $ to $ \eta $. Consequently, $ \eta \in \Cl{\Range{T \pm i \cdot \Id_{\X}}}{\X} $, so we may conclude that $ \Cl{\Range{T \pm i \cdot \Id_{\X}}}{\X} = \X $.

Conversely, suppose that $ \Cl{\Range{T \pm i \cdot \Id_{\X}}}{\X} = \X $. We shall show that
$$
\Range{\cl{T} \pm i \cdot \Id_{\X}}=\Cl{\Range{T \pm i \cdot \Id_{\X}}}{\X}
$$
and then appeal to \autoref{Criterion for Regularity and Self-Adjointness} to conclude that $ \cl{T} $ is self-adjoint. As the closure of a symmetric operator is symmetric, it follows that $ \cl{T} $ is symmetric, and so
$$
\forall \zeta \in \Dom{\cl{T}}: \quad
         \Inner{\Func{\cl{T} \pm i \cdot \Id_{\X}}{\zeta}}{\Func{\cl{T} \pm i \cdot \Id_{\X}}{\zeta}}_{\X}
=        \Inner{\func{\cl{T}}{\zeta}}{\func{\cl{T}}{\zeta}}_{\X} + \Inner{\zeta}{\zeta}_{\X}
\geq_{A} \Inner{\zeta}{\zeta}_{\X},
$$
which yields
\begin{equation}
\forall \zeta \in \Dom{\cl{T}}: \quad
\Norm{\Func{\cl{T} \pm i \cdot \Id_{\X}}{\zeta}}_{\X} \geq \Norm{\zeta}_{\X}. \label{Inequality 3}
\end{equation}

\textbf{Claim:} $ \cl{T} \pm i \cdot \Id_{\X} $ has closed range.

\emph{Proof of claim}. Given an arbitrary $ \eta \in \Cl{\Range{\cl{T} \pm i \cdot \Id_{\X}}}{\X} $, there exists a sequence $ \Seq{\zeta_{n}}{n \in \N} $ in $ \Dom{\cl{T}} $ such that $ \Seq{\Func{\cl{T} \pm i \cdot \Id_{\X}}{\zeta_{n}}}{n \in \N} $ converges in $ \X $ to $ \eta $. In particular, $ \Seq{\Func{\cl{T} \pm i \cdot \Id_{\X}}{\zeta_{n}}}{n \in \N} $ is Cauchy in $ \X $, so, by Inequality (\ref{Inequality 3}), $ \Seq{\zeta_{n}}{n \in \N} $ is also Cauchy in $ \X $ and thus converges in $ \X $ to, say, $ \widetilde{\zeta} $. Hence,
$$
    \Pair{\widetilde{\zeta}}{\eta}
\in \Cl{\Graph{\cl{T} \pm i \cdot \Id_{\X}}}{\X \oplus \X}
=   \Cl{\Graph{\cl{T \pm i \cdot \Id_{\X}}}}{\X \oplus \X}
=   \Graph{\cl{T \pm i \cdot \Id_{\X}}}
=   \Graph{\cl{T} \pm i \cdot \Id_{\X}},
$$
which yields $ \eta \in \Range{\cl{T} \pm i \cdot \Id_{\X}} $. Therefore, $ \cl{T} \pm i \cdot \Id_{\X} $ has closed range.

As $ \Range{T \pm i \cdot \Id_{X}} \subseteq \Range{\cl{T} \pm i \cdot \Id_{\X}} $, we thus find that $ \Range{\cl{T} \pm i \cdot \Id_{\X}} = \X $, so we may conclude from \autoref{Criterion for Regularity and Self-Adjointness} that $ \cl{T} \in \RS{\X} $.
\end{proof}


\begin{Cor} \label{Defining the Direct Sum of Regular Self-Adjoint Operators}
Let $ \Lambda $ be a set, $ \bm{\X} $ a $ \Lambda $-indexed family of Hilbert $ A $-modules, and $ \ds \bm{T} \in \prod_{\lambda \in \Lambda} \RS{\bm{\X}_{\lambda}} $. Letting
$$
    D
\df \Set{\bm{x} \in \Br{\bigoplus \bm{\X}} \cap \SqBr{\prod_{\lambda \in \Lambda} \Dom{\bm{T}_{\lambda}}}}
        {\Seq{\func{\bm{T}_{\lambda}}{\bm{x}_{\lambda}}}{\lambda \in \Lambda} \in \bigoplus \bm{\X}},
$$
define $ \ds \bigoplus \bm{T}: D \to \bigoplus \bm{\X} $ by
$$
\forall \bm{x} \in D: \quad
\Func{\bigoplus \bm{T}}{\bm{x}} \df \Seq{\func{\bm{T}_{\lambda}}{\bm{x}_{\lambda}}}{\lambda \in \Lambda}.
$$
Then $ D $ is a dense $ A $-submodule of $ \ds \bigoplus \bm{\X} $, and $ \ds \bigoplus \bm{T} \in \RS{\bigoplus \bm{\X}} $.
\end{Cor}

\begin{proof}
It is straightforward to verify that $ D $ is, at least, an $ A $-submodule of $ \ds \bigoplus \bm{\X} $. With regard to density, recall, for all $ \lambda \in \Lambda $, that $ \Dom{\bm{T}_{\lambda}} $ is a dense $ A $-submodule of $ \bm{\X}_{\lambda} $, which readily implies that
$$
\widetilde{D} \df \Set{\bm{x} \in \prod_{\lambda \in \Lambda} \Dom{\bm{T}_{\lambda}}}{\bm{x} \text{ has finite support}}
$$
is a dense $ A $-submodule of $ \ds \bigoplus \bm{\X} $; as $ \widetilde{D} \subseteq D $, we find that $ D $ is a dense $ A $-submodule of $ \ds \bigoplus \bm{\X} $ as well.

Furthermore, we have
\begin{align*}
\forall \bm{x},\bm{y} \in D: \quad
    \Inner{\Func{\bigoplus \bm{T}}{\bm{x}}}{\bm{y}}_{\bigoplus \bm{\X}}
& = \Inner{\Seq{\func{\bm{T}_{\lambda}}{\bm{x}_{\lambda}}}{\lambda \in \Lambda}}{\Seq{\bm{y}_{\lambda}}{\lambda \in \Lambda}}_{\bigoplus \bm{\X}} \\
& = \sum_{\lambda \in \Lambda} \Inner{\func{\bm{T}_{\lambda}}{\bm{x}_{\lambda}}}{\bm{y}_{\lambda}}_{\bm{\X}_{\lambda}} \\
& = \sum_{\lambda \in \Lambda} \Inner{\bm{x}_{\lambda}}{\func{\bm{T}_{\lambda}}{\bm{y}_{\lambda}}}_{\bm{\X}_{\lambda}} \\
& = \Inner{\Seq{\bm{x}_{\lambda}}{\lambda \in \Lambda}}{\Seq{\func{\bm{T}_{\lambda}}{\bm{y}_{\lambda}}}{\lambda \in \Lambda}}_{\bigoplus \bm{\X}} \\
& = \Inner{\bm{x}}{\Func{\bigoplus \bm{T}}{\bm{y}}}_{\bigoplus \bm{\X}},
\end{align*}
so $ \ds \bigoplus \bm{T} $ is a symmetric operator on $ \ds \bigoplus \bm{\X} $. It thus remains to show that $ \ds \bigoplus \bm{T} \in \RS{\bigoplus \bm{\X}} $.

By \autoref{Criterion for Regularity and Self-Adjointness}, it suffices to show that
$$
\Range{\Br{\bigoplus \bm{T}} \pm i \cdot \Id_{\bigoplus \bm{\X}}} = \bigoplus \bm{\X}.
$$
Hence, let $ \ds \bm{y} \in \bigoplus \bm{\X} $ be arbitrary. For each $ \lambda \in \Lambda $, \autoref{Criterion for Regularity and Self-Adjointness} says that $ \Range{\bm{T}_{\lambda} \pm i \cdot \Id_{\bm{\X}_{\lambda}}} = \bm{\X}_{\lambda} $, so we may find an $ \ds \bm{x} \in \prod_{\lambda \in \Lambda} \Dom{\bm{T}_{\lambda}} $ such that
$$
\forall \lambda \in \Lambda: \quad
  \func{\bm{T}_{\lambda}}{\bm{x}_{\lambda}} \pm i \cdot \bm{x}_{\lambda}
= \Func{\bm{T}_{\lambda} \pm i \cdot \Id_{\bm{\X}_{\lambda}}}{\bm{x}_{\lambda}}
= \bm{y}_{\lambda}.
$$
As $ \ds \sum_{\lambda \in \Lambda} \Inner{\bm{y}_{\lambda}}{\bm{y}_{\lambda}}_{\bm{\X}_{\lambda}} $ converges in $ A $, for every $ \epsilon \in \R_{> 0} $, there exists a finite subset $ \widetilde{F} $ of $ \Lambda $ such that
$$
\forall F \subseteq_{\text{finite}} \Lambda \setminus \widetilde{F}: \quad
\Norm{\sum_{\lambda \in F} \Inner{\bm{y}_{\lambda}}{\bm{y}_{\lambda}}_{\bm{\X}_{\lambda}}}_{A} < \epsilon;
$$
by the symmetry of the $ \bm{T}_{\lambda} $'s, we have
$$
\forall \lambda \in \Lambda: \quad
  \Inner{\bm{y}_{\lambda}}{\bm{y}_{\lambda}}_{\bm{\X}_{\lambda}}
= \Inner{\func{\bm{T}_{\lambda}}{\bm{x}_{\lambda}} \pm i \cdot \bm{x}_{\lambda}}
        {\func{\bm{T}_{\lambda}}{\bm{x}_{\lambda}} \pm i \cdot \bm{x}_{\lambda}}_{\bm{\X}_{\lambda}} \\
= \Inner{\func{\bm{T}_{\lambda}}{\bm{x}_{\lambda}}}{\func{\bm{T}_{\lambda}}{\bm{x}_{\lambda}}} + \Inner{\bm{x}_{\lambda}}{\bm{x}_{\lambda}}_{\bm{\X}_{\lambda}},
$$
so
$$
\forall F \subseteq_{\text{finite}} \Lambda \setminus \widetilde{F}: \quad
         \sum_{\lambda \in F} \Inner{\bm{x}_{\lambda}}{\bm{x}_{\lambda}}_{\bm{\X}_{\lambda}}
\leq_{A} \sum_{\lambda \in F} \Inner{\bm{y}_{\lambda}}{\bm{y}_{\lambda}}_{\bm{\X}_{\lambda}}
\quad \text{and} \quad
         \sum_{\lambda \in F} \Inner{\func{\bm{T}_{\lambda}}{\bm{x}_{\lambda}}}{\func{\bm{T}_{\lambda}}{\bm{x}_{\lambda}}}_{\bm{\X}_{\lambda}}
\leq_{A} \sum_{\lambda \in F} \Inner{\bm{y}_{\lambda}}{\bm{y}_{\lambda}}_{\bm{\X}_{\lambda}},
$$
which leads to
$$
\forall F \subseteq_{\text{finite}} \Lambda \setminus \widetilde{F}: \quad
     \Norm{\sum_{\lambda \in F} \Inner{\bm{x}_{\lambda}}{\bm{x}_{\lambda}}_{\bm{\X}_{\lambda}}}_{A},
     \Norm{\sum_{\lambda \in F} \Inner{\func{\bm{T}_{\lambda}}{\bm{x}_{\lambda}}}{\func{\bm{T}_{\lambda}}{\bm{x}_{\lambda}}}_{\bm{\X}_{\lambda}}}_{A}
\leq \Norm{\sum_{\lambda \in F} \Inner{\bm{y}_{\lambda}}{\bm{y}_{\lambda}}_{\bm{\X}_{\lambda}}}_{A}
<    \epsilon.
$$
Consequently, both $ \ds \sum_{\lambda \in \Lambda} \Inner{\bm{x}_{\lambda}}{\bm{x}_{\lambda}}_{\bm{\X}_{\lambda}} $ and $ \ds \sum_{\lambda \in \Lambda} \Inner{\func{\bm{T}_{\lambda}}{\bm{x}_{\lambda}}}{\func{\bm{T}_{\lambda}}{\bm{x}_{\lambda}}}_{\bm{\X}_{\lambda}} $ converge in $ A $, which yields
$$
\bm{x} \in D \quad \text{and} \quad \FUNC{\Br{\bigoplus \bm{T}} \pm i \cdot \Id_{\bigoplus \bm{\X}}}{\bm{x}} = \bm{y}.
$$
Finally, as $ \ds \bm{y} \in \bigoplus \bm{\X} $ is arbitrary, we therefore conclude that $ \ds \Range{\Br{\bigoplus \bm{T}} \pm i \cdot \Id_{\bigoplus \bm{\X}}} = \bigoplus \bm{\X} $.
\end{proof}

\section{Stone's Theorem for Hilbert \texorpdfstring{$ C^{\ast} $}{C*}-Modules} 

In this section, we generalize Stone's Theorem for Hilbert spaces to the setting of Hilbert $ C^{\ast} $-modules to obtain a one-to-one correspondence between strongly continuous one-parameter unitary groups and regular self-adjoint operators.

Let $ \X $ be a Hilbert $ A $-module and $ T \in \RS{\X} $. Then $ f_{T} \df T \Br{\Id_{\X} + T^{2}}^{- \frac{1}{2}} $ is a self-adjoint element of $ \Adj{\X} $ and $ \Spec{f_{T}} \subseteq \CCI{- 1}{1} $ (\cite[Theorem 10.4]{Lance}). Define $ \theta: \R \to \OOI{- 1}{1} $ by
$$
\forall x \in \R: \quad \func{\theta}{x} \df x \Br{1 + x^{2}}^{- \frac{1}{2}}.
$$
Then $ \theta $ is a homeomorphism, and, for each $ t \in \OOI{- 1}{1} $, we have $ \func{\theta^{- 1}}{x} = x \Br{1 - x^{2}}^{- \frac{1}{2}} $. Equation 10.2 of \cite{Lance} says that $ T = f_{T} \Br{1 - f_{T}^{2}}^{- \frac{1}{2}} $, so a functional calculus may naturally be adopted for $ T $ as the nondegenerate $ C^{\ast} $-homomorphism $ \psi_{T} $ from $ \Co{\R} $ to $ \Adj{\X} $ defined by
$$
\forall g \in \Co{\R}: \quad
\func{\psi_{T}}{g} \df \Func{\widetilde{g \circ \theta^{- 1}}}{f_{T}},
$$
where
\begin{itemize}
\item
$ \widetilde{g \circ \theta^{- 1}} $ denotes the extension of $ g \circ \theta^{- 1} \in \Co{\OOI{- 1}{1}} $ to the domain $ \CCI{- 1}{1} $, and

\item
the usual $ C^{\ast} $-algebraic continuous functional calculus is being applied to $ f_{T} $ (\cite[Theorem 10.9]{Lance} and \cite{Woronowicz}).
\end{itemize}

By \cite[Proposition 2.3]{Lance}, for every locally compact Hausdorff space $ X $, the multiplier $ C^{\ast} $-algebra $ \Mult{\Co{X}} $ is $ C^{\ast} $-isomorphic to $ \Cb{X} $. By \cite[Proposition 2.5]{Lance}, the non-degenerate $ \ast $-homomorphism $ \psi_{T} $ extends to a unital $ \ast $-homomorphism $ \overline{\psi_{T}} $ from $ \Cb{\R} $ to $ \Adj{X} $ that is strictly continuous on the unit ball of $ \Cb{\R} $. Equipped with this technology, we may easily generate a strongly continuous one-parameter unitary group on a Hilbert $ C^{\ast} $-module using the functional calculus for a regular self-adjoint operator (see \autoref{A Regular Self-Adjoint Operator Generates a Strongly Continuous One-Parameter Unitary Group}).

A comment is in order with regard to the strict topology on $ \Cb{\R} $. It is easily seen to be strictly finer than the topology of uniform convergence on compact subsets of $ \R $ (aka the compact-open topology), but the two topologies coincide on uniformly bounded subsets of $ \Cb{\R} $ --- this is important enough to warrant a proof.


\begin{Lem} \label{On Uniformly Bounded Subsets of Cb(R), the Strict Topology Coincides with the Topology of Uniform Convergence on Compact Subsets of R}
On uniformly bounded subsets of $ \Cb{\R} $, the strict topology coincides with the topology of uniform convergence on compact subsets of $ \R $.
\end{Lem}

\begin{proof}
Let $ B $ be a uniformly bounded subset of $ \Cb{\R} $, so that there is an $ M \in \R_{> 0} $ such that, for all $ f \in B $, we have $ \Norm{f}_{\infty} < M $. Let $ \Seq{f_{i}}{i \in I} $ be a net in $ B $ that converges to $ F \in B $ uniformly on compact subsets of $ \R $, and let $ g \in \Co{\R} $ --- we claim that $ \ds \lim_{i} \Norm{\Br{f_{i} - F} g}_{\infty} = 0 $.

Indeed, let $ \epsilon \in \R_{> 0} $, and pick any $ h \in \Cc{\R} $ such that $ \Norm{g - h}_{\infty} < \dfrac{\epsilon}{4 M + 2} $. By assumption, there exists an $ i^{\ast} \in I $ such that
$$
\forall i \in I_{\geq i^{\ast}}: \quad
\Norm{f_{i}|_{\Supp{h}} - F|_{\Supp{h}}}_{\infty} < \frac{\epsilon}{2 \Br{1 + \Norm{h}_{\infty}}},
$$
which then yields
\begin{align*}
\forall i \in I_{\geq i^{\ast}}: \quad
       \Norm{\Br{f_{i} - F} g}_{\infty}
& \leq \Norm{\Br{f_{i} - F} \Br{g - h}}_{\infty} + \Norm{\Br{f_{i} - F} h}_{\infty} \\
& \leq \Norm{f_{i} - F}_{\infty} \Norm{g - h}_{\infty} + \Norm{\Br{f_{i} - F} h|_{\Supp{h}}}_{\infty} \\
& \leq \Br{\Norm{f_{i}}_{\infty} + \Norm{F}_{\infty}} \Norm{g - h}_{\infty} + \Norm{f_{i}|_{\Supp{h}} - F|_{\Supp{h}}}_{\infty} \Norm{h}_{\infty} \\
& <    2 M \cdot \frac{\epsilon}{4 M + 2} + \frac{\epsilon}{2 \Br{1 + \Norm{h}_{\infty}}} \cdot \Norm{h}_{\infty} \\
& =    \frac{2 M}{2 M + 1} \cdot \frac{\epsilon}{2} + \frac{\Norm{h}_{\infty}}{1 + \Norm{h}_{\infty}} \cdot \frac{\epsilon}{2} \\
& <    \frac{\epsilon}{2} + \frac{\epsilon}{2} \\
& =    \epsilon.
\end{align*}
Hence, as $ g \in \Co{\R} $ and $ \epsilon \in \R_{> 0} $ are arbitrary, we obtain $ f_{i} \xrightarrow[i \in I]{\text{strictly in $ \Cb{\R} $}} F $.

Hence, on uniformly bounded subsets of $ \Cb{\R} $, the topology of uniform convergence on compact subsets of $ \R $ is finer than the strict topology, but the latter topology is finer than the former topology on all of $ \Cb{\R} $, so the two topologies coincide on all uniformly bounded subsets of $ \Cb{\R} $.
\end{proof}


In order to state our generalization of Stone's Theorem, we shall require the following definition.


\begin{Def} \label{The Gamma Map}
Define a map $ \gamma: \R \to \Cb{\R} $ by
$$
\forall t,x \in \R: \quad
\func{\gamma_{t}}{x} \df e^{i t x}.
$$
For any Hilbert $ A $-module $ \X $, any $ t \in \R $, and any $ T \in \RS{\X} $, let $ e^{i t \cdot T} \df \func{\overline{\psi_{T}}}{\gamma_{t}} \in \Adj{\X} $.
\end{Def}


Although the following lemma may appear very basic, it plays a surprisingly important role throughout the remainder of this section.


\begin{Lem} \label{A Basic Inequality}
For all $ t,x \in \R $, we have $ \Abs{\func{\gamma_{t}}{x} - 1} \leq \Abs{t} \Abs{x} $.
\end{Lem}

\begin{proof}
By the Fundamental Theorem of Calculus, we have
$$
\forall t,x \in \R: \quad
     \Abs{\func{\gamma_{t}}{x} - 1}
=    \Abs{\RInt{0}{x}{i t e^{i t s}}{s}}
\leq \sgn{x} \RInt{0}{x}{\Abs{i t e^{i t s}}}{s}
=    \sgn{x} \RInt{0}{x}{\Abs{t}}{s}
=    \Abs{t} \Abs{x}.
$$
The proof is complete.
\end{proof}


With the preceding two lemmas, we may apply the functional calculus for a regular self-adjoint operator to generate a strongly continuous one-parameter unitary group by way of exponentiation.


\begin{Lem} \label{A Regular Self-Adjoint Operator Generates a Strongly Continuous One-Parameter Unitary Group}
Let $ \X $ be a Hilbert $ A $-module and $ T \in \RS{\X} $. Then $ \Seq{e^{i t \cdot T}}{t \in \R} \in \URep{\R}{\X} $.
\end{Lem}

\begin{proof}
Firstly, as $ \overline{\psi_{T}} $ is a unital $ \ast $-homomorphism, $ \Seq{e^{i t \cdot T}}{t \in \R} $ is a one-parameter unitary group on $ \X $.

Secondly, for any compact subset $ K $ of $ \R $, we have $ \ds \lim_{t \to 0} \Norm{\gamma_{t}|_{K} - \chi_{K}}_{\infty} = 0 $; to see this, pick an $ M \in \R_{> 0} $ so that $ K \subseteq \CCI{- M}{M} $, and apply \autoref{A Basic Inequality} to obtain
$$
\forall t \in \R, ~ \forall x \in K: \quad
     \Abs{\func{\gamma_{t}}{x} - 1}
\leq \Abs{t} \Abs{x}
\leq M \Abs{t},
$$
which then readily yields
$$
\forall \epsilon \in \R_{> 0}, ~ \forall t \in \R: \quad
\Abs{t} < \frac{\epsilon}{M}
\quad \Longrightarrow \quad
\Norm{\gamma_{t}|_{K} - \chi_{K}}_{\infty} < \epsilon.
$$

Thirdly, $ \gamma_{t} \xrightarrow[t \to 0]{\text{strictly in $ \Cb{\R} $}} \mathbf{1} $ by \autoref{On Uniformly Bounded Subsets of Cb(R), the Strict Topology Coincides with the Topology of Uniform Convergence on Compact Subsets of R}.

Fourthly, as $ \SSet{\gamma_{t}}_{t \in \R} $ is contained in the unit ball of $ \Cb{\R} $, we may invoke \cite[Proposition 2.5]{Lance} to obtain
$$
                                                       e^{i t \cdot T}
=                                                      \func{\overline{\psi_{T}}}{\gamma_{t}}
\xrightarrow[t \to 0]{\text{strictly in $ \Adj{\X} $}} \func{\overline{\psi_{T}}}{\mathbf{1}}
=                                                      \Id_{\X}.
$$
With the strict topology on $ \Adj{\X} $ (aka the strong-* topology) being finer than the strong topology on $ \Adj{\X} $, it follows that $ \Seq{e^{i t \cdot T}}{t \in \R} $ is strongly continuous at $ 0 $ and therefore strongly continuous throughout $ \R $.
\end{proof}


\autoref{A Regular Self-Adjoint Operator Generates a Strongly Continuous One-Parameter Unitary Group} says that regular self-adjoint operators are a source of strongly continuous one-parameter unitary groups; our next result says the converse, namely, strongly continuous one-parameter unitary groups are a source of regular self-adjoint operators.


\begin{Thm} \label{The Infinitesimal Generator of a Strongly Continuous One-Parameter Unitary Group is Self-Adjoint and Regular Operator}
Let $ \X $ be a Hilbert $ A $-module. For any  $ U \in \URep{\R}{\X} $, the infinitesimal generator $ T $ of $ U $ is regular and self-adjoint.
\end{Thm}

\begin{proof}
\autoref{The Infinitesimal Generator and Analytic Vectors of a Strongly Continuous One-Parameter Unitary Group} already says that $ T $ is a closed and densely defined modular operator on $ \X $. Observe that
\begin{align*}
\forall \zeta,\eta \in \Dom{T}: \quad
    \Inner{\func{T}{\zeta}}{\eta}_{\X}
& = \Inner{\lim_{t \to 0} \SqBr{- \frac{i}{t} \cdot \Func{U_{t} - \Id_{\X}}{\zeta}}}{\eta}_{\X} \\
& = \lim_{t \to 0} \frac{i}{t} \cdot \Inner{\Func{U_{t} - \Id_{\X}}{\zeta}}{\eta}_{\X} \\
& = \lim_{t \to 0} \frac{i}{t} \cdot \Inner{\zeta}{\func{\Br{U_{t} - \Id_{\X}}^{\ast}}{\eta}}_{\X} \\
& = \lim_{t \to 0} \frac{i}{t} \cdot \Inner{\zeta}{\Func{U_{t}^{\ast} - \Id_{\X}}{\eta}}_{\X} \\
& = \lim_{t \to 0} \frac{i}{t} \cdot \Inner{\zeta}{\Func{U_{- t} - \Id_{\X}}{\eta}}_{\X} \\
& = \Inner{\zeta}{\lim_{t \to 0} \SqBr{- \frac{i}{- t} \cdot \Func{U_{- t} - \Id_{\X}}{\eta}}}_{\X} \\
& = \Inner{\zeta}{\func{T}{\eta}},
\end{align*}
so $ T $ is symmetric.

To conclude that $ T $ is regular and self-adjoint, it suffices to show that $ \Range{T \pm i \cdot \Id_{\X}} = \X $, which would allow us to invoke \autoref{Criterion for Regularity and Self-Adjointness}.

Toward this end, let $ \mathcal{L} \df \Ll{1}{\R_{\geq 0}} \cap \CC{\R_{\geq 0}} $, and, for each $ f \in \mathcal{L} $, define $ \mathcal{T}^{\pm}(f): \X\to \X $ by
$$
\forall \zeta \in \X: \quad
\FUNC{\func{\mathcal{T}^{\pm}}{f}}{\zeta} \df \Int{\R_{\geq 0}}{\func{f}{t} \cdot \func{U_{\pm t}}{\zeta}}{\mu}{t},
$$
where Bochner's Integrability Criterion --- applied to the integrability of $ f $ --- ensures that the integral is a well-defined element of $ \X $. It then follows from the continuity of $ \Inner{\cdot}{\cdot}_{\X}: \X \times \X \to A $ that
\begin{align*}
\forall f \in \mathcal{L}, ~ \forall \zeta,\eta \in \X: \quad
    \Inner{\FUNC{\func{\mathcal{T}^{\pm}}{f}}{\zeta}}{\eta}_{\X}
& = \Inner{\Int{\R_{\geq 0}}{\func{f}{t} \cdot \func{U_{\pm t}}{\zeta}}{\mu}{t}}{\eta}_{\X} \\
& = \Int{\R_{\geq 0}}{\Inner{\func{f}{t} \cdot \func{U_{\pm t}}{\zeta}}{\eta}_{\X}}{\mu}{t} \\
& = \Int{\R_{\geq 0}}{\Inner{\zeta}{\overline{\func{f}{t}} \cdot \func{U_{\mp t}}{\eta}}_{\X}}{\mu}{t} \\
& = \Inner{\zeta}{\Int{\R_{\geq 0}}{\overline{\func{f}{t}} \cdot \func{U_{\mp t}}{\eta}}{\mu}{t}}_{\X} \\
& = \Inner{\zeta}{\FUNC{\func{\mathcal{T}^{\mp}}{\overline{f}}}{\eta}}_{\X},
\end{align*}
so $ T^{\pm} $ maps every $ f \in \mathcal{L} $ to an adjointable map $ \func{T^{\pm}}{f} $ on $ \X $.

Let $ \mathcal{L}^{\Br{1}} $ denote the set of all continuously differentiable elements $ f $ of $ \mathcal{L} $ such that $ f' $ is integrable on $ \R_{\geq 0} $, i.e., $ \mathcal{L}^{\Br{1}} = \Set{f \in \mathcal{L} \cap \func{C^{1}}{\R_{\geq 0}}}{f' \in \mathcal{L}} $. Our strategy for showing that $ \Range{T \pm i \cdot \Id_{\X}} = \X $ is to find a $ \phi \in \mathcal{L}^{\Br{1}} $ such that $ \Func{T \pm i \cdot \Id_{\X}}{\func{\mathcal{T}^{\pm}}{\phi}} = \phi $. To get there, two technical claims are necessary, the first being that, whenever $ f \in \mathcal{L}^{\Br{1}} $, we have
\begin{equation} \label{The composition of the infinitesimal generator of U with T±}
\forall \zeta \in \X: \quad
  \lim_{t \to 0} \frac{1}{t} \cdot \Func{U_{t} - \Id_{\X}}{\FUNC{\func{\mathcal{T}^{\pm}}{f}}{\zeta}}
= \FUNC{\func{\mp\mathcal{T}^{\pm}}{f'}}{\zeta} + \func{f}{0} \cdot \zeta.
\end{equation}

As a preparatory step, observe that
\begin{align*}
\forall t \in \R, ~ \forall f \in \mathcal{L}, ~ \forall \zeta \in \X: \quad
    \func{U_{t}}{\FUNC{\func{\mathcal{T}^{\pm}}{f}}{\zeta}}
& = \func{U_{t}}{\Int{\R_{\geq 0}}{\func{f}{s} \cdot \func{U_{\pm s}}{\zeta}}{\mu}{s}} \\
& = \Int{\R_{\geq 0}}{\func{f}{s} \cdot \func{U_{t \pm s}}{\zeta}}{\mu}{s} \\
& = \Int{\R_{\geq t}}{\func{f}{s \mp t} \cdot \func{U_{\pm s}}{\zeta}}{\mu}{s};
\end{align*}
consequently,
\begin{align*}
\forall t \in \mathbb{R} \setminus \SSet{0}, ~ \forall f \in \mathcal{L}, ~ \forall \zeta \in \X: \qquad
  & ~ \frac{1}{t}\cdot \Func{U_{t} - \Id_{\X}}{\FUNC{\func{\mathcal{T}^{\pm}}{f}}{\zeta}} \\
= & ~ \Int{\R_{\geq t}}{\frac{\func{f}{s \mp t} - \func{f}{s}}{t} \cdot \func{U_{\pm s}}{\zeta}}{\mu}{s} -
      \frac{1}{t} \cdot \SqBr{\Int{\COI{0}{t}}{\func{f}{s} \cdot \func{U_{\pm s}}{\zeta}}{\mu}{s}},
\end{align*}
which facilitates an analysis of the expression inside the limit on the left-hand side of \autoref{The composition of the infinitesimal generator of U with T±}.

\textbf{Claim 1:} For all $ f \in \mathcal{L}^{\Br{1}} $ and $ \zeta \in \X $, we have
\begin{equation} \label{Claim 1}
  \lim_{t \to 0} \Int{\R_{\geq t}}{\frac{\func{f}{s \mp t} - \func{f}{s}}{t} \cdot \func{U_{\pm s}}{\zeta}}{\mu}{s}
= \mp \FUNC{\func{\mathcal{T}^{\pm}}{f'}}{\zeta},
\end{equation}
where the limit is taken with respect to $ \Norm{\cdot}_{\X} $.

\emph{Proof of Claim 1}. Let $ f \in \mathcal{L}^{\Br{1}} $ and $ \zeta \in \X $ be arbitrary; then
\begin{align*}
\forall t \in \R \setminus \SSet{0}: \quad
     & ~ \Norm{
              \Int{\R_{\geq t}}{\frac{\func{f}{s \mp t} - \func{f}{s}}{\mp t} \cdot \func{U_{\pm s}}{\zeta}}{\mu}{s} -
              \FUNC{\func{\mathcal{T}^{\pm}}{f'}}{\zeta}
              }_{\X} \\
=    & ~ \Norm{
              \Int{\R_{\geq t}}{\frac{\func{f}{s \mp t} - \func{f}{s}}{\mp t} \cdot \func{U_{\pm s}}{\zeta}}{\mu}{s} -
              \Int{\R_{\geq 0}}{\func{f'}{s} \cdot \func{U_{\pm s}}{\zeta}}{\mu}{s}
              }_{\X} \\
=    & ~ \Norm{
              \Int{\R_{\geq t}}{\SqBr{\frac{\func{f}{s \mp t} - \func{f}{s}}{\mp t} - \func{f'}{s}} \cdot \func{U_{\pm s}}{\zeta}}{\mu}{s} -
              \Int{\COI{0}{t}}{\func{f'}{s} \cdot \func{U_{\pm s}}{\zeta}}{\mu}{s}
              }_{\X} \\
\leq & ~ \Norm{\Int{\R_{\geq t}}{\SqBr{\frac{\func{f}{s \mp t} - \func{f}{s}}{\mp t} - \func{f'}{s}} \cdot \func{U_{\pm s}}{\zeta}}{\mu}{s}}_{\X} +
         \Norm{\Int{\COI{0}{t}}{\func{f'}{s} \cdot \func{U_{\pm s}}{\zeta}}{\mu}{s}}_{\X} \\
\leq & ~ \Int{\R_{\geq t}}{\Norm{\SqBr{\frac{\func{f}{s \mp t} - \func{f}{s}}{\mp t} - \func{f'}{s}} \cdot \func{U_{\pm s}}{\zeta}}_{\X}}{\mu}{s} +
         \Int{\COI{0}{t}}{\Norm{\func{f'}{s} \cdot \func{U_{\pm s}}{\zeta}}_{\X}}{\mu}{s} \\
=    & ~ \Int{\R_{\geq t}}{\Abs{\frac{\func{f}{s \mp t} - \func{f}{s}}{\mp t} - \func{f'}{s}} \Norm{\zeta}_{\X}}{\mu}{s} +
         \Int{\COI{0}{t}}{\Abs{\func{f'}{s}} \Norm{\zeta}_{\X}}{\mu}{s} \\
=    & ~ \SqBr{\Int{\R_{\geq t}}{\Abs{\frac{\func{f}{s \mp t} - \func{f}{s}}{\mp t} - \func{f'}{s}}}{\mu}{s} + \Int{\COI{0}{t}}{\Abs{\func{f'}{s}}}{\mu}{s}}
         \Norm{\zeta}_{\X} \\
=    & ~ \SqBr{
              \Int{\R_{\geq t}}{\Abs{\Int{\CCI{0}{1}}{\SqBr{\func{f'}{s \mp t r} - \func{f'}{s}}}{\mu}{r}}}{\mu}{s} +
              \Int{\COI{0}{t}}{\Abs{\func{f'}{s}}}{\mu}{s}
              }
         \Norm{\zeta}_{\X} \\
     & ~ \Br{\text{by the First Fundamental Theorem of Calculus}} \\
\leq & ~ \SqBr{
              \Int{\CCI{0}{1}}{\SqBr{\Int{\R_{\geq t}}{\Abs{\func{f'}{s \mp t r} - \func{f'}{s}}}{\mu}{s}}}{\mu}{r} +
              \Int{\COI{0}{t}}{\Abs{\func{f'}{s}}}{\mu}{s}
              }
         \Norm{\zeta}_{\X} \\
     & ~ \Br{\text{by Minkowski's Integral Inequality}} \\
=    & ~ \SqBr{\Int{\CCI{0}{1}}{\Norm{\func{f'}{\bullet \mp t r} - \func{f'}{\bullet}}_{1,\R_{\geq t}}}{\mu}{r} + \Int{\COI{0}{t}}{\Abs{\func{f'}{s}}}{\mu}{s}}
         \Norm{\zeta}_{\X}.
\end{align*}
Letting $ \widetilde{f'} $ denote the extension of $ f' $ to all of $ \R $ by $ 0 $, we have
$$
\forall t \in \R_{> 0}, ~ \forall r \in \CCI{0}{1}: \quad
     \Norm{\func{f'}{\bullet \mp t r} - \func{f'}{\bullet}}_{1,\R_{\geq t}}
\leq \Norm{\func{\widetilde{f'}}{\bullet \mp t r} - \func{\widetilde{f'}}{\bullet}}_{1,\R},
$$
and, recalling the well-known measure-theoretic fact that the action of $ \R $ on $ \Ll{1}{\R} $ by left/right translations is strongly continuous, we may then apply the Squeeze Theorem to obtain
$$
\forall r \in \CCI{0}{1}: \quad
\lim_{t \to 0} \Norm{\func{f'}{\bullet \mp t r} - \func{f'}{\bullet}}_{1,\R_{\geq t}} = 0.
$$
Furthermore,
$$
\forall t \in \R_{> 0}, ~ \forall r \in \CCI{0}{1}: \quad
     \Norm{\func{f'}{\bullet \mp t r} - \func{f'}{\bullet}}_{1,\R_{\geq t}}
\leq \Norm{\func{f'}{\bullet \mp t r}}_{1,\R_{\geq t}} + \Norm{f'}_{1,\R_{\geq t}}
\leq 2 \Norm{f'}_{1,\R_{\geq 0}},
$$
so, by the Lebesgue Dominated Convergence Theorem (LDCT),
$$
\lim_{t \to 0} \Int{\CCI{0}{1}}{\Norm{\func{f'}{\bullet \mp t r} - \func{f'}{\bullet}}_{1,\R_{\geq t}}}{\mu}{r} = 0.
$$
Finally, another application of the LDCT yields $ \ds \lim_{t \to 0} \Int{\COI{0}{t}}{\Abs{\func{f'}{s}}}{\mu}{s} = 0 $, and Claim 1 follows. \hfill $ \triangle $

\textbf{Claim 2:} 
For all $ f \in \mathcal{L} $ and $ \zeta \in \X $, we have
\begin{equation} \label{Claim 2}
\lim_{t \to 0} \frac{1}{t} \cdot \Int{\COI{0}{t}}{\func{f}{s} \cdot \func{U_{\pm s}}{\zeta}}{\mu}{s} = \func{f}{0} \cdot \zeta.
\end{equation}

\emph{Proof of Claim 2}. This is just a quick application of the First Fundamental Theorem of Calculus. \hfill $ \triangle $

\autoref{Claim 1} and \autoref{Claim 2} now allow us to deduce the following:
$$
\forall f \in \mathcal{L}^{\Br{1}}, ~ \forall \zeta \in \X: \quad
\FUNC{\func{\mathcal{T}^{\pm}}{f}}{\zeta} \in \Dom{T}
\quad \text{and} \quad
\func{T}{\FUNC{\func{\mathcal{T}^{\pm}}{f}}{\zeta}} = \pm i \cdot \FUNC{\func{\mathcal{T}^{\pm}}{f'}}{\zeta} + i \func{f}{0} \cdot \zeta.
$$

Proceed to define $ \phi \in \mathcal{L}^{\Br{1}} $ by
$$
\forall t \in \R_{\geq 0}: \quad
\func{\phi}{t} \df - i e^{- t}.
$$
Noting that $ \phi' = - \phi $, we find that
$$
  \Br{T \pm i \cdot \Id_{\X}} \circ \func{\mathcal{T}^{\pm}}{\phi}
= T \circ \func{\mathcal{T}^{\pm}}{\phi} \pm i \cdot \func{\mathcal{T}^{\pm}}{\phi}
= \pm i \cdot \SqBr{\func{\mathcal{T}^{\pm}}{\phi'} + \func{\mathcal{T}^{\pm}}{\phi}} + i \func{\phi}{0} \cdot \Id_{\X}
= \Id_{\X}.
$$
Therefore, $ \Range{T \pm i \cdot \Id_{\X}} = \X $ as desired.
\end{proof}


We now arrive at the crux of this section, \autoref{Stone's Theorem for Hilbert C*-Modules}, which makes precise the correspondence between strongly continuous one-parameter unitary groups and regular self-adjoint operators.

 
\begin{Prop}[Stone's Theorem for Hilbert $ C^{\ast} $-Modules] \label{Stone's Theorem for Hilbert C*-Modules}
Let $ \X $ be a Hilbert $ A $-module. Then
$$
\forall U \in \URep{\R}{\X}, ~ \forall T \in \RS{\X}: \quad
U = \Seq{e^{i t \cdot T}}{t \in \R} \iff \text{The infinitesimal generator of } U \text{ is } T.
$$
Consequently, $ \Map{\RS{\X}}{\URep{\R}{\X}}{T}{\Seq{e^{i t \cdot T}}{t \in \R}} $ is a bijection.
\end{Prop}

\begin{proof}
Let $ U \in \URep{\R}{\X} $ and $ T \in \RS{\X} $.

Suppose that $ U = \Seq{e^{i t \cdot T}}{t \in \R} $; we wish to show that $ T $ is the infinitesimal generator of $ U $, i.e., the following two statements hold:
\begin{enumerate}
\item
$ \ds \Dom{T} = \Set{\zeta \in \X}{\exists \eta \in \X: ~ \lim_{t \to 0} \Norm{- \frac{i}{t} \cdot \Func{e^{i t \cdot T} - \Id_{\X}}{\zeta} - \eta}_{\X} = 0} $.

\item 
$ \ds \forall \zeta \in \Dom{T}: \quad \lim_{t \to 0} \Norm{- \frac{i}{t} \cdot \Func{e^{i t \cdot T} - \Id_{\X}}{\zeta} - \func{T}{\zeta}}_{\X} = 0 $.
\end{enumerate}
We shall employ the functional calculus for $ T $, as defined in \cite[Theorem 10.9]{Lance} and described at the beginning of this section. Accordingly, define $ q \in \Co{\R} $ by
$$ 
\forall x \in \R: \quad
\func{q}{x} \df \Br{1 + x^{2}}^{- \frac{1}{2}},
$$
and let
$$
    q_{T}
\df \func{\psi_{T}}{q}
=   \Func{\widetilde{q \circ \theta^{- 1}}}{f_{T}}
=   \Br{\Id_{\X} - f_{T}^{2}}^{\frac{1}{2}}
=   \Br{\Id_{\X} + T^{2}}^{- \frac{1}{2}},
$$
where the last identity may readily be deduced from \cite[Equations 9.4--9.6]{Lance}.

Define an $ \Br{\R \setminus \SSet{0}} $-indexed family $ \delta $ in $ \Cb{\R} $ by
$$
\forall t \in \R \setminus \SSet{0}, ~ \forall x \in \R: \quad
    \func{\delta_{t}}{x}
\df - \frac{i}{t}\SqBr{\func{\gamma_{t}}{x} - 1}
=   - \frac{i}{t} \Br{e^{i t x} - 1}.
 $$
Then the following four statements hold:
\begin{itemize}
\item
For each $ t \in \R \setminus \SSet{0} $, we have $ \func{\delta_{t}}{T} \df \func{\overline{\psi_{T}}}{\delta_{t}} = - \dfrac{i}{t} \cdot \Br{e^{i t \cdot T} - \Id_{\X}} $.

\item
$ \SSet{\delta_{t} q}_{t \in \R \setminus \SSet{0}} $ and $ \theta $ are contained in the unit ball of $ \Cb{\R} $; to see this, apply \autoref{A Basic Inequality} to obtain
$$
\forall t \in \R \setminus \SSet{0}, ~ \forall x \in \R: \quad
     \Abs{\Func{\delta_{t} q}{x}}
=    \frac{\Abs{\func{\gamma_{t}}{x} - 1}}{\Abs{t}} \cdot \frac{1}{\sqrt{1 + x^{2}}}
\leq \frac{\Abs{t} \Abs{x}}{\Abs{t}} \cdot \frac{1}{\sqrt{1 + x^{2}}}
=    \frac{\Abs{x}}{\sqrt{1 + x^{2}}}
=    \Abs{\func{\theta}{x}}
<    1.
$$

\item
For any compact subset $ K $ of $ \R $, we have $ \ds \lim_{t \to 0} \Norm{\Br{\delta_{t} q}|_{K} - \theta|_{K}}_{\infty} = 0 $; to see this, pick an $ M \in \R_{> 0} $ so that $ K \subseteq \CCI{- M}{M} $, and apply \autoref{A Basic Inequality} to obtain
\begin{align*}
\forall t \in \R \setminus \SSet{0}, ~ \forall x \in K: \quad
       \Abs{\Func{\delta_{t} q}{x} - \func{\theta}{x}}
& =    \frac{1}{\sqrt{1 + x^{2}}} \Abs{- \frac{i}{t} \SqBr{\func{\gamma_{t}}{x} - 1} - x} \\
& =    \frac{1}{\sqrt{1 + x^{2}}} \Abs{\RInt{0}{x}{\SqBr{\func{\gamma_{t}}{s} - 1}}{s}} \\
& \leq \frac{1}{\sqrt{1 + x^{2}}} \cdot \sgn{x} \RInt{0}{x}{\Abs{\func{\gamma_{t}}{s} - 1}}{s} \\
& \leq \frac{1}{\sqrt{1 + x^{2}}} \cdot \sgn{x} \RInt{0}{x}{\Abs{t} \Abs{s}}{s} \\
& \leq \frac{1}{\sqrt{1 + x^{2}}} \cdot \sgn{x} \RInt{0}{x}{\Abs{t} \Abs{x}}{s} \\
& =    \frac{\Abs{t} \Abs{x}^{2}}{\sqrt{1 + x^{2}}} \\
& \leq \Abs{t} \Abs{x} \\
& \leq M \Abs{t},
\end{align*}
which then readily yields
$$
\forall \epsilon \in \R_{> 0}, ~ \forall t \in \R: \quad
\Abs{t} < \frac{\epsilon}{M}
\quad \Longrightarrow \quad
\Norm{\Br{\delta_{t} q}|_{K} - \theta|_{K}}_{\infty} < \epsilon.
$$

\item
$ \delta_{t} q \xrightarrow[t \to 0]{\text{strictly in $ \Cb{\R} $}} \theta $, which follows from the preceding statement and \autoref{On Uniformly Bounded Subsets of Cb(R), the Strict Topology Coincides with the Topology of Uniform Convergence on Compact Subsets of R}.
\end{itemize}

It now follows from \cite[Proposition 2.5]{Lance} again that
$$
                                                    \func{\delta_{t}}{T} \circ q_{T}
=                                                   \func{\overline{\psi_{T}}}{\delta_{t}} \circ \func{\psi_{T}}{q}
=                                                   \func{\overline{\psi_{T}}}{\delta_{t} q}
\xrightarrow[t \to 0]{\text{strictly in } \Adj{\X}} \func{\overline{\psi_{T}}}{\theta}
=                                                   f_{T}
=                                                   T \circ q_{T},
$$
so
$$
\forall \zeta \in \X: \quad
\lim_{t \to 0} \Norm{\FUNC{\func{\delta_{t}}{T}}{\func{q_{T}}{\zeta}} - \func{T}{\func{q_{T}}{\zeta}}}_{\X} = 0.
$$
However, $ \Range{q_{T}} = \Dom{T} $ (\cite[page 98]{Lance}), which implies that
$$
\forall \zeta \in \Dom{T}: \quad
  \lim_{t \to 0} \Norm{- \frac{i}{t} \cdot \Func{e^{i t \cdot T} - \Id_{\X}}{\zeta} - \func{T}{\zeta}}_{\X}
= \lim_{t \to 0} \Norm{\FUNC{\func{\delta_{t}}{T}}{\zeta} - \func{T}{\zeta}}_{\X}
= 0.
$$
The following statements have thus been established:
\begin{itemize}
\item
$ \ds \Dom{T} \subseteq \Set{\zeta \in \X}{\exists \eta \in \X: ~ \lim_{t \to 0} \Norm{- \frac{i}{t} \cdot \Func{e^{i t \cdot T} - \Id_{\X}}{\zeta} - \eta}_{\X} = 0} $.

\item
$ \ds \forall \zeta \in \Dom{T}: \quad \lim_{t \to 0} \Norm{- \frac{i}{t} \cdot \Func{e^{i t \cdot  T} - \Id_{\X}}{\zeta} - \func{T}{\zeta}}_{\X} = 0 $.
\end{itemize}
Let $ S \in \RS{\X} $ denote the infinitesimal generator of $ U $. Clearly, $ T \subseteq S $, but then $ T \subseteq S = S^{\ast} \subseteq T^{\ast} = T $, so $ T = S $, thereby proving the forward direction.

Conversely, suppose that $ T $ is the infinitesimal generator of $ U $, and let $ V \df \Seq{e^{i t \cdot T}}{t \in \R} $; we wish to show that $ U = V $. Picking arbitrary $ t \in \R $ and $ \zeta \in \Dom{T} $, we have
\begin{align*}
\forall s \in \R \setminus \SSet{t}: \qquad
  & ~ \frac{1}{s - t} \cdot \SqBr{\Func{U_{- s} V_{s}}{\zeta} - \Func{U_{- t} V_{t}}{\zeta}} \\
= & ~ \frac{1}{s - t} \cdot \SqBr{\Func{U_{- s} V_{s}}{\zeta} - \Func{U_{- s} V_{t}}{\zeta} + \Func{U_{- s} V_{t}}{\zeta} - \Func{U_{- t} V_{t}}{\zeta}} \\
= & ~ \frac{1}{s - t} \cdot \SqBr{\Func{U_{- s} V_{s}}{\zeta} - \Func{U_{- s} V_{t}}{\zeta}} +
      \frac{1}{s - t} \cdot \SqBr{\Func{U_{- s} V_{t}}{\zeta} - \Func{U_{- t} V_{t}}{\zeta}} \\
= & ~ \underbrace{\func{U_{- s}}{\frac{1}{s - t} \cdot \SqBr{\func{V_{s}}{\zeta} - \func{V_{t}}{\zeta}}}}_{\Br{\clubsuit}} +
      \underbrace{\frac{1}{s - t} \cdot \SqBr{\func{U_{- s}}{\func{V_{t}}{\zeta}} - \func{U_{- t}}{\func{V_{t}}{\zeta}}}}_{\Br{\spadesuit}}.
\end{align*}
To show that $ \ds \lim_{s \to t} \Br{\clubsuit} $ exists, first observe that
\begin{align*}
\forall s \in \R \setminus \SSet{t}: \qquad
     & ~ \Norm{
              \func{U_{- s}}{\frac{1}{s - t} \cdot \SqBr{\func{V_{s}}{\zeta} - \func{V_{t}}{\zeta}}} - \func{U_{- t}}{i \cdot \func{V_{t}}{\func{T}{\zeta}}}
              }_{\X} \\
\leq & ~ \Norm{
              \func{U_{- s}}{\frac{1}{s - t} \cdot \SqBr{\func{V_{s}}{\zeta} - \func{V_{t}}{\zeta}}} - \func{U_{- s}}{i \cdot \func{V_{t}}{\func{T}{\zeta}}}
              }_{\X} ~ + \\
     & ~ \Norm{\func{U_{- s}}{i \cdot \func{V_{t}}{\func{T}{\zeta}}} - \func{U_{- t}}{i \cdot \func{V_{t}}{\func{T}{\zeta}}}}_{\X} \\
=    & ~ \Norm{\func{U_{- s}}{\frac{1}{s - t} \cdot \SqBr{\func{V_{s}}{\zeta} - \func{V_{t}}{\zeta}} - i \cdot \func{V_{t}}{\func{T}{\zeta}}}}_{\X} ~ + \\
     & ~ \Norm{\func{U_{- s}}{i \cdot \func{V_{t}}{\func{T}{\zeta}}} - \func{U_{- t}}{i \cdot \func{V_{t}}{\func{T}{\zeta}}}}_{\X} \\
=    & ~ \Norm{\frac{1}{s - t} \cdot \SqBr{\func{V_{s}}{\zeta} - \func{V_{t}}{\zeta}} - i \cdot \func{V_{t}}{\func{T}{\zeta}}}_{X} +
     \Norm{\func{U_{- s}}{i \cdot \func{V_{t}}{\func{T}{\zeta}}} - \func{U_{- t}}{i \cdot \func{V_{t}}{\func{T}{\zeta}}}}_{\X}.
\end{align*}
Recalling that $ U $ is strongly continuous and --- from \autoref{The Infinitesimal Generator and Analytic Vectors of a Strongly Continuous One-Parameter Unitary Group} --- that
$$
  \lim_{s \to t} \frac{1}{s - t} \cdot \SqBr{\func{V_{s}}{\zeta} - \func{V_{t}}{\zeta}}
= \Diff{s}{\func{V_{s}}{\zeta}} \bigg|_{s = t}
= i \cdot \func{V_{t}}{\func{T}{\zeta}},
$$
we may then apply the Squeeze Theorem to infer that
$$
  \lim_{s \to t}
  \Norm{\func{U_{- s}}{\frac{1}{s - t} \cdot \SqBr{\func{V_{s}}{\zeta} - \func{V_{t}}{\zeta}}} - \func{U_{- t}}{i \cdot \func{V_{t}}{\func{T}{\zeta}}}}_{\X}
= 0,
$$
which immediately yields
$$
\lim_{s \to t} \func{U_{- s}}{\frac{1}{s - t} \cdot \SqBr{\func{V_{s}}{\zeta} - \func{V_{t}}{\zeta}}} = i \cdot \func{U_{- t}}{\func{V_{t}}{\func{T}{\zeta}}}.
$$
To show that $ \ds \lim_{s \to t} \Br{\spadesuit} $ exists, apply \autoref{The Infinitesimal Generator and Analytic Vectors of a Strongly Continuous One-Parameter Unitary Group} another time to see that $ \func{V_{t}}{\zeta} $ is a $ U $-differentiable vector, and apply the Chain Rule also to obtain
$$
  \lim_{s \to t} \frac{1}{s - t} \cdot \SqBr{\func{U_{- s}}{\func{V_{t}}{\zeta}} - \func{U_{- t}}{\func{V_{t}}{\zeta}}}
= \Diff{s}{\func{U_{- s}}{\func{V_{t}}{\zeta}}} \bigg|_{s = t}
= - i \cdot \func{U_{- t}}{\func{T}{\func{V_{t}}{\zeta}}}
= - i \cdot \func{U_{- t}}{\func{V_{t}}{\func{T}{\zeta}}}.
$$
The foregoing allows us to see that the map $ \Map{\R}{\X}{s}{\Func{U_{- s} V_{s}}{\zeta}} $ is differentiable and that
$$
\forall t \in \R: \quad
  \Diff{s}{\Func{U_{- s} V_{s}}{\zeta}} \bigg|_{s = t}
= i \cdot \func{U_{- t}}{\func{V_{t}}{\func{T}{\zeta}}} - i \cdot \func{U_{- t}}{\func{V_{t}}{\func{T}{\zeta}}}
= \bm{0}_{\X},
$$
which implies that said map is constant. Furthermore,
$$
\forall t \in \R: \quad
  \Func{U_{- t} V_{t}}{\zeta}
= \FUNC{U_{- 0} V_{0}}{\zeta}
= \zeta
\quad \text{or, equivalently,} \quad
\func{V_{t}}{\zeta} = \func{U_{t}}{\zeta}.
$$
However, $ \Dom{T} $ is dense in $ \X $, so, for all $ t \in \R $, because $ U_{t} $ and $ V_{t} $ are continuous, we must have $ U_{t} = V_{t} $. Therefore, $ U = V $, and the proof is complete.
\end{proof}

Next, we show that the infinitesimal generator of a direct sum of strongly-continuous one-parameter unitary group representations is the direct sum of the infinitesimal generators for the constituent strongly-continuous one-parameter unitary group representations. This is an important component in the proof of the main result of \autoref{Infinitesimal Heisenberg and Schrodinger Representations}.


\begin{Cor} \label{Taking the Infinitesimal Generator Commutes with Direct Sums}
Let $ \Lambda $ be a set, $ \bm{\X} $ a $ \Lambda $-indexed family of Hilbert $ A $-modules, and $ \ds \bm{U} \in \prod_{\lambda \in \Lambda} \URep{\R}{\bm{\X}_{\lambda}} $. Let $ \ds \bm{A} \in \prod_{\lambda \in \Lambda} \RS{\bm{\X}_{\lambda}} $ be determined by taking the infinitesimal generators of the components of $ \bm{U} $: For each $ \lambda \in \Lambda $, the infinitesimal generator of $ \bm{U}_{\lambda} $ is $ \bm{A}_{\lambda} $. Then the infinitesimal generator of $ \Phi_{\bm{\X}} \circ \dot{\bm{U}} $ is $ \ds \bigoplus \bm{A} $.
\end{Cor}

\begin{proof}
\autoref{A Direct Sum of Strongly Continuous Unitary Representations Is a Strongly Continuous Unitary Representation} says that $ \ds \Phi_{\bm{\X}} \circ \dot{\bm{U}} \in \URep{\R}{\bigoplus \bm{\X}} $, so it possesses an infinitesimal generator, which we shall denote by $ T $.

Let $ \ds \bm{x} \in \Dom{\bigoplus \bm{A}} $. Then, for all finite subsets $ F $ of $ \Lambda $, we have
$$
\chi_{F} \cdot \bm{x} \in \Dom{T}
\quad \text{and} \quad
\func{T}{\chi_{F} \cdot \bm{x}} = \chi_{F} \cdot \Func{\bigoplus \bm{A}}{\bm{x}},
$$
where $ \chi_{F}: \Lambda \to \SSet{0,1} $ denotes the $ F $-indicator function on $ \Lambda $. As $ T $ is closed, and as
$$
\Seq{\chi_{F} \cdot \bm{x}}{F \subseteq_{\text{finite}} \Lambda}
\quad \text{and} \quad
\Seq{\chi_{F} \cdot \Func{\bigoplus \bm{A}}{\bm{x}}}{F \subseteq_{\text{finite}} \Lambda}
$$
are $ \subseteq $-directed nets in $ \ds \bigoplus \bm{\X} $ that converge, respectively, to $ \bm{x} $ and $ \ds \Func{\bigoplus \bm{A}}{\bm{x}} $, we obtain
$$
\Pair{\bm{x}}{\Func{\bigoplus \bm{A}}{\bm{x}}} \in \Graph{T}.
$$
Hence, $ \ds \bigoplus \bm{A} \subseteq T $. However, $ \ds \bigoplus \bm{A} \in \RS{\bigoplus \bm{\X}} $ by \autoref{Defining the Direct Sum of Regular Self-Adjoint Operators}, and self-adjoint operators are maximally symmetric, so, therefore, $ \ds \bigoplus \bm{A} = T $.
\end{proof}

The next result shows that conjugation by a unitary $W\in \Unitary{\X}{\Y}$ of a regular self-adjoint operator $T\in \RS{\X}$ preserves self-adjointness and regularity, and similarly, conjugation by $W\in \Unitary{\X,\Y}$ of $U\in \URep{\R}{\X}$ gives rise to a representation in $\URep{\R}{\Y}$.


\begin{Cor}
\label{Infinitesimal Generators and Unitary Equivalence}
Let $ \X $ and $ \Y $ be Hilbert $ A $-modules and $ W \in \Unitary{\X,\Y} $.
\begin{enumerate}
\item
For every $ T \in \RS{\X} $, we have $ W T W^{\ast} \in \RS{\Y} $.

\item
For every $ U \in \URep{\R}{\X} $, the following two statements hold:
\begin{enumerate}
\item[(a)]
$ V \df \Seq{W U_{t} W^{\ast}}{t \in \R} \in \URep{\R}{\Y} $.

\item[(b)]
If $ T $ denotes the infinitesimal generator of $ U $, then the infinitesimal generator of $ V $ is $ W T W^{\ast} $.
\end{enumerate}
\end{enumerate}
\end{Cor}

\begin{proof}
Let $ T \in \RS{\X} $. Then $ \Dom{W T W^{\ast}} = \Im{W}{\Dom{T}} $ is a dense $ \C $-linear $ A $-submodule of $ \Y $, and
\begin{align*}
\forall \zeta,\eta \in \Dom{T}: \quad
    \Inner{\Func{W T W^{\ast}}{\func{W}{\zeta}}}{\func{W}{\eta}}_{\Y}
& = \Inner{\Func{W T}{\zeta}}{\func{W}{\eta}}_{\Y} \\
& = \Inner{\func{W}{\func{T}{\zeta}}}{\func{W}{\eta}}_{\Y} \\
& = \Inner{\func{T}{\zeta}}{\eta}_{\X} \quad \Br{\text{as $ W $ is unitary}} \\
& = \Inner{\zeta}{\func{T}{\eta}}_{\X} \quad \Br{\text{as $ T $ is self-adjoint}} \\
& = \Inner{\func{W}{\zeta}}{\func{W}{\func{T}{\eta}}}_{\Y} \quad \Br{\text{as $ W $ is unitary}} \\
& = \Inner{\func{W}{\zeta}}{\Func{W T}{\eta}}_{\Y} \\
& = \Inner{\func{W}{\zeta}}{\Func{W T W^{\ast}}{\func{W}{\eta}}}_{\Y}.
\end{align*}
Hence, $ W T W^{\ast} $ is a symmetric operator on $ \Y $. By \autoref{Criterion for Regularity and Self-Adjointness}, $ \Range{T \pm i \cdot \Id_{\X}} = \X $, so
\begin{align*}
    \Range{W T W^{\ast} \pm i \cdot \Id_{\Y}}
& = \Im{\Br{W T W^{\ast} \pm i \cdot \Id_{\Y}}}{\Im{W}{\Dom{T}}} \\
& = \Im{W}{\Im{\Br{T \pm i \cdot \Id_{\X}}}{\Im{W^{\ast}}{\Im{W}{\Dom{T}}}}} \\
& = \Im{W}{\Im{\Br{T \pm i \cdot \Id_{\X}}}{\Dom{T}}} \\
& = \Im{W}{\X} \\
& = \Y.
\end{align*}
By \autoref{Criterion for Regularity and Self-Adjointness} again, $ W T W^{\ast} \in \RS{\Y} $.

Now, let $ U \in \URep{\R}{\X} $. Then, clearly, $ V \df \Seq{W U_{t} W^{\ast}}{t \in \R} \in \URep{\R}{\Y} $, and
\begin{align*}
\forall \zeta \in \Dom{T}: \qquad
  & ~ \lim_{t \to 0} \Norm{- \frac{i}{t} \cdot \SqBr{\Func{W V_{t} W^{\ast}}{\func{W}{\zeta}} - \func{W}{\zeta}} - \Func{W T W^{\ast}}{\func{W}{\zeta}}}_{\Y} \\
= & ~ \lim_{t \to 0} \Norm{- \frac{i}{t} \cdot \SqBr{\Func{W V_{t} W^{\ast}}{\func{W}{\zeta}} - \func{W}{\zeta}} - \func{W}{\func{T}{\zeta}}}_{\Y} \\
= & ~ \lim_{t \to 0} \Norm{- \frac{i}{t} \cdot \SqBr{\Func{W V_{t}}{\zeta} - \func{W}{\zeta}} - \func{W}{\func{T}{\zeta}}}_{\Y} \\
= & ~ \lim_{t \to 0} \Norm{\func{W}{- \frac{i}{t} \cdot \SqBr{\func{V_{t}}{\zeta} - \zeta} - \func{T}{\zeta}}}_{\Y} \\
= & ~ \lim_{t \to 0} \Norm{- \frac{i}{t} \cdot \SqBr{\func{V_{t}}{\zeta} - \zeta} - \func{T}{\zeta}}_{\X} \quad \quad \Br{\text{as $ W $ is an isometry}} \\
= & ~ 0.
\end{align*}
Hence, the infinitesimal generator of $ V $ extends $ W T W^{\ast} $, but since self-adjoint operators on a Hilbert $ C^{\ast} $-module are maximally symmetric, the infinitesimal generator of $ V $ is precisely $ W T W^{\ast} $.
\end{proof}

\section{Infinitesimal Heisenberg and Schr\"{o}dinger Representations}\label{Infinitesimal Heisenberg and Schrodinger Representations}

Having developed the necessary unbounded operator-theoretic machinery, we shall now define infinitesimal Heisenberg representations and the infinitesimal Schr\"{o}dinger representation of a real $ C^{\ast} $-dynamical system, which are the main objects of study in this paper.


\begin{Def} \label{Infinitesimal Heisenberg Representation}
An \emph{infinitesimal Heisenberg representation} of a real $ C^{\ast} $-dynamical system $ \Trip{\R}{A}{\alpha} $ is an ordered quadruple $ \Quad{\X}{\rho}{S}{T} $ with the following four properties:
\begin{enumerate}
\item
$ \X $ is a full Hilbert $ A $-module, $ \rho $ is a non-degenerate $ \ast $-representation of $ A $ on $ \X $, and $ S,T \in \RS{\X} $.

\item
For all $ a \in \Trip{\R}{A}{\alpha}^{\infty} $, we have $ \func{\rho}{a} \circ S \subseteq S \circ \func{\rho}{a} - i \hbar \cdot \func{\rho}{\func{\delta}{a}} $. \\
\textbf{Note:} It is to be recalled that $ \delta $ denotes the infinitesimal generator of $ \alpha $.

\item
For all $ a \in A $, we have $ \func{\rho}{a} \circ T \subseteq T \circ \func{\rho}{a} $.

\item
There exists an $ \SSet{S,T} $-analytic and $ \SSet{S,T} $-invariant dense $ A $-submodule $ D $ of $ \X $ such that
\begin{itemize}
\item
$ D $ is a core for both $ S $ and $ T $, and

\item
the HCR holds for $ \Pair{S}{T} $ on $ D $, i.e., $ \Comm{S|_{D}}{T|_{D}} = - i \hbar \cdot \Id_{D} $.
\end{itemize}
\end{enumerate}
\end{Def}


In addition to the HCR, there are thus two additional commutation relations to work with.

Our eventual goal, the Infinitesimal Covariant Stone-von Neumann Theorem, rests in the assurance that the class of infinitesimal Heisenberg representations is closed under direct sums, which we must now establish.


\begin{Lem} \label{A Direct Sum of Infinitesimal Heisenberg Representations}
The direct sum of infinitesimal Heisenberg representations of a real $ C^{\ast} $-dynamical system $ \Trip{\R}{A}{\alpha} $ is an infinitesimal Heisenberg representation of $ \Trip{\R}{A}{\alpha} $.
\end{Lem}

\begin{proof}
Let $ \Lambda $ be a set and $ \Quad{\bm{\X}}{\bm{\rho}}{\bm{S}}{\bm{T}} $ an ordered quadruple with the following two properties:
\begin{itemize}
\item
$ \bm{\X} $ is a $ \Lambda $-indexed family of Hilbert $ A $-modules, $ \ds \bm{\rho} \in \prod_{\lambda \in \Lambda} \Rep{A}{\bm{\X}_{\lambda}} $, and $ \ds \bm{S},\bm{T} \in \prod_{\lambda \in \Lambda} \RS{\bm{\X}_{\lambda}} $.

\item
For each $ \lambda \in \Lambda $, we have that $ \Quad{\bm{\X}_{\lambda}}{\bm{\rho}_{\lambda}}{\bm{S}_{\lambda}}{\bm{T}_{\lambda}} $ is an infinitesimal Heisenberg representation of $ \Trip{\R}{A}{\alpha} $.
\end{itemize}
Then let $ \ds \bigoplus \Quad{\bm{\X}}{\bm{\rho}}{\bm{S}}{\bm{T}} \df \Quad{\bigoplus \bm{\X}}{\bigoplus \bm{\rho}}{\bigoplus \bm{S}}{\bigoplus \bm{T}} $, where $ \ds \bm{\rho}: A \to \Adj{\bigoplus \bm{\X}} $ is defined by
$$
\forall a \in A: \quad
\Func{\bigoplus \bm{\rho}}{a} \df \bigoplus \func{\bm{\rho}_{\lambda}}{a}.
$$

For all $ a \in \Trip{\R}{A}{\alpha}^{\infty} $ and $ \ds \bm{\zeta} \in \Dom{\bm{S}} $, we have $ \ds \bm{\zeta} \in \Dom{\Func{\bigoplus \bm{\rho}}{a} \circ \Br{\bigoplus \bm{S}}} $, so
\begin{align*}
      \Seq{\FUNC{\bm{S}_{\lambda} \circ \func{\bm{\rho}_{\lambda}}{a}}{\bm{\zeta}_{\lambda}}}{\lambda \in \Lambda}
& =   \Seq{
          \FUNC{\func{\bm{\rho}_{\lambda}}{a} \circ \bm{S}_{\lambda}}{\bm{\zeta}_{\lambda}} +
          \FUNC{i \hbar \cdot \func{\bm{\rho}_{\lambda}}{\func{\delta}{a}}}{\bm{\zeta}_{\lambda}}
          }
          {\lambda \in \Lambda} \\
& =   \FUNC{\Func{\bigoplus \bm{\rho}}{a} \circ \Br{\bigoplus \bm{S}}}{\bm{\zeta}} +
      \FUNC{i \hbar \cdot \Func{\bigoplus \bm{\rho}}{\func{\delta}{a}}}{\bm{\zeta}} \\
& \in \bigoplus \bm{\X},
\end{align*}
which implies that $ \ds \bm{\zeta} \in \Dom{\Br{\bigoplus \bm{S}} \circ \Func{\bigoplus \bm{\rho}}{a}} $ and hence
$$
          \Func{\bigoplus \bm{\rho}}{a} \circ \Br{\bigoplus \bm{S}}
\subseteq \Br{\bigoplus \bm{S}} \circ \Func{\bigoplus \bm{\rho}}{a} - i \hbar \cdot \Func{\bigoplus \bm{\rho}}{\func{\delta}{a}}.
$$

For all $ a \in A $ and $ \ds \bm{\zeta} \in \Dom{\bigoplus \bm{T}} $, we have $ \ds \bm{\zeta} \in \Dom{\Func{\bigoplus \bm{\rho}}{a} \circ \Br{\bigoplus \bm{T}}} $, so
$$
    \Seq{\FUNC{\bm{T}_{\lambda} \circ \func{\bm{\rho}_{\lambda}}{a}}{\bm{\zeta}_{\lambda}}}{\lambda \in \Lambda}
=   \Seq{\FUNC{\func{\bm{\rho}_{\lambda}}{a} \circ \bm{T}_{\lambda}}{\bm{\zeta}_{\lambda}}}{\lambda \in \Lambda}
=   \FUNC{\Func{\bigoplus \bm{\rho}}{a} \circ \Br{\bigoplus \bm{T}}}{\bm{\zeta}}
\in \bigoplus \bm{\X},
$$
which implies that $ \ds \bm{\zeta} \in \Dom{\Br{\bigoplus \bm{T}} \circ \Func{\bigoplus \bm{\rho}}{a}} $ and hence
$$
\Func{\bigoplus \bm{\rho}}{a} \circ \Br{\bigoplus \bm{T}} \subseteq \Br{\bigoplus \bm{T}} \circ \Func{\bigoplus \bm{\rho}}{a}.
$$

Let $ \bm{D} $ be a $ \Lambda $-indexed set so that, for each $ \lambda \in \Lambda $, the following three statements hold:
\begin{itemize}
\item
$ \bm{D}_{\lambda} $ is an $ \SSet{\bm{S}_{\lambda},\bm{T}_{\lambda}} $-analytic and $ \SSet{\bm{S}_{\lambda},\bm{T}_{\lambda}} $-invariant dense $ A $-submodule of $ \bm{\X}_{\lambda} $.

\item
$ \bm{D}_{\lambda} $ is a core for both $ \bm{S}_{\lambda} $ and $ \bm{T}_{\lambda} $.

\item
The HCR holds for $ \Pair{\bm{S}_{\lambda}}{\bm{T}_{\lambda}} $ on $ \bm{D}_{\lambda} $, i.e., $ \Comm{\bm{S}_{\lambda}|_{\bm{D}_{\lambda}}}{\bm{T}_{\lambda}|_{\bm{D}_{\lambda}}} = - i \hbar \cdot \Id_{\bm{D}_{\lambda}} $.
\end{itemize}
Let $ \ds D \df \Set{\bm{\zeta} \in \Br{\bigoplus \bm{\X}} \cap \Br{\prod \bm{D}}}{\Supp{\bm{\zeta}} \text{ is finite}} $, which is obviously a $ \ds \SSet{\bigoplus \bm{S},\bigoplus \bm{T}} $-invariant dense $ A $-submodule of $ \ds \bigoplus \bm{\X} $.

To establish the $ \ds \SSet{\bigoplus \bm{S},\bigoplus \bm{T}} $-analyticity of $ D $, fix $ \bm{\zeta} \in D $, and then choose $ R \in \R_{> 0} $ so that
$$
\forall \lambda \in \Supp{\bm{\zeta}}, ~ \forall t \in \OOI{- R}{R}: \quad
\sum_{n \in \N_{0}} \frac{\Norm{\func{\bm{S}_{\lambda}^{n}}{\bm{\zeta}_{\lambda}}}_{\bm{\X}_{\lambda}}}{n!} \cdot t^{n}
\text{ and }
\sum_{n \in \N_{0}} \frac{\Norm{\func{\bm{T}_{\lambda}^{n}}{\bm{\zeta}_{\lambda}}}_{\bm{\X}_{\lambda}}}{n!} \cdot t^{n}
\text{ both converge}.
$$
Letting $ t \in \OOI{- R}{R} $ and $ \epsilon \in \R_{> 0} $ be arbitrary, we may find an $ N \in \N_{0} $ such that
$$
  \sum_{n \in \N_{\geq N}} \frac{\Norm{\func{\bm{S}_{\lambda}^{n}}{\bm{\zeta}_{\lambda}}}_{\bm{\X}_{\lambda}}}{n!} \cdot \Abs{t}^{n}
< \frac{\epsilon}{1 + \Size{\Supp{\bm{\zeta}}}}
\text{ and }
  \sum_{n \in \N_{\geq N}} \frac{\Norm{\func{\bm{T}_{\lambda}^{n}}{\bm{\zeta}_{\lambda}}}_{\bm{\X}_{\lambda}}}{n!} \cdot \Abs{t}^{n}
< \frac{\epsilon}{1 + \Size{\Supp{\bm{\zeta}}}},
$$
whereupon repeated applications of \autoref{Corollary of the Pythaogrean Inequality for Orthogonal Elements of a Hilbert C*-Module} guarantee that
\begin{align*}
       \sum_{n \in \N_{\geq N}} \frac{\ds \Norm{\func{\Br{\bigoplus \bm{S}}^{n}}{\bm{\zeta}}}_{\bigoplus \bm{\X}}}{n!} \cdot \Abs{t}^{n}
& \leq \sum_{n \in \N_{\geq N}}
       \frac{\ds \sqrt{\sum_{\lambda \in \Supp{\bm{\zeta}}} \Norm{\func{\bm{S}_{\lambda}^{n}}{\bm{\zeta}_{\lambda}}}_{\bm{\X}_{\lambda}}^{2}}}{n!} \cdot
       \Abs{t}^{n} \\
& \leq \sum_{n \in \N_{\geq N}}
       \frac{\ds \sum_{\lambda \in \Supp{\bm{\zeta}}} \Norm{\func{\bm{S}_{\lambda}^{n}}{\bm{\zeta}_{\lambda}}}_{\bm{\X}_{\lambda}}}{n!} \cdot \Abs{t}^{n} \\
& =    \sum_{n \in \N_{\geq N}}
       \SqBr{\sum_{\lambda \in \Supp{\bm{\zeta}}} \frac{\Norm{\func{\bm{S}_{\lambda}^{n}}{\bm{\zeta}_{\lambda}}}_{\bm{\X}_{\lambda}}}{n!} \cdot \Abs{t}^{n}} \\
& =    \sum_{\lambda \in \Supp{\bm{\zeta}}}
       \SqBr{\sum_{n \in \N_{\geq N}} \frac{\Norm{\func{\bm{S}_{\lambda}^{n}}{\bm{\zeta}_{\lambda}}}_{\bm{\X}_{\lambda}}}{n!} \cdot \Abs{t}^{n}} \\
& \leq \sum_{\lambda \in \Supp{\bm{\zeta}}} \frac{\epsilon}{1 + \Size{\Supp{\bm{\zeta}}}} \\
& =    \frac{\Size{\Supp{\bm{\zeta}}}}{1 + \Size{\Supp{\bm{\zeta}}}} \cdot \epsilon \\
& <    \epsilon;
\end{align*}
the same derivation for $ \bm{T} $ in place of $ \bm{S} $ yields $ \ds \sum_{n \in \N_{0}} \frac{\ds \Norm{\func{\Br{\bigoplus \bm{T}}^{n}}{\bm{\zeta}}}_{\bigoplus \bm{\X}}}{n!} \cdot \Abs{t}^{n} < \epsilon $ also. Consequently,
$$
\forall t \in \OOI{- R}{R}: \quad
\sum_{n \in \N_{0}} \frac{\ds \Norm{\func{\Br{\bigoplus \bm{S}}^{n}}{\bm{\zeta}}}_{\bigoplus \bm{\X}}}{n!} \cdot t^{n}
\text{ and }
\sum_{n \in \N_{0}} \frac{\ds \Norm{\func{\Br{\bigoplus \bm{T}}^{n}}{\bm{\zeta}}}_{\bigoplus \bm{\X}}}{n!} \cdot t^{n}
\text{ both converge},
$$
so $ \bm{\zeta} $ is a $ \ds \SSet{\bigoplus \bm{S},\bigoplus \bm{T}} $-analytic vector, and, as $ \bm{\zeta} \in D $ is arbitrary, $ D $ is thus $ \ds \SSet{\bigoplus \bm{S},\bigoplus \bm{T}} $-analytic.

Let $ \ds \Pair{\bm{\xi}}{\bm{\eta}} \in \Graph{\bigoplus \bm{S}} $ and $ \epsilon \in \R_{> 0} $ be arbitrary. Then there exists a finite $ F \subseteq \Lambda $ such that
$$
  \Norm{\Pair{\bm{\xi}}{\bm{\eta}} - \Pair{\chi_{F} \cdot \bm{\xi}}{\chi_{F} \cdot \bm{\eta}}}_{\Br{\bigoplus \bm{\X}} \oplus \Br{\bigoplus \bm{\X}}}
< \frac{\epsilon}{2},
$$
but we have, for each $ \lambda \in F $, that $ \bm{D}_{\lambda} $ is a core for $ \bm{S}_{\lambda} $ (i.e., $ \Cl{\Graph{\bm{S}_{\lambda}|_{\bm{D}_{\lambda}}}}{\bm{\X}_{\lambda} \oplus \bm{\X}_{\lambda}} = \Graph{\bm{S}_{\lambda}} $), so we may find a $ \bm{\zeta} \in D $ such that $ \Supp{\bm{\zeta}} \subseteq F $ and
$$
  \Norm{
       \Pair{\chi_{F} \cdot \bm{\xi}}{\chi_{F} \cdot \bm{\eta}} - \Pair{\bm{\zeta}}{\Func{\bigoplus \bm{S}}{\bm{\zeta}}}
       }_{\Br{\bigoplus \bm{\X}} \oplus \Br{\bigoplus \bm{\X}}}
< \frac{\epsilon}{2},
$$
which yields
$$
  \Norm{
       \Pair{\bm{\xi}}{\bm{\eta}} - \Pair{\chi_{F} \cdot \bm{\xi}}{\chi_{F} \cdot \bm{\eta}} - \Pair{\bm{\zeta}}{\Func{\bigoplus \bm{S}}{\bm{\zeta}}}
       }_{\Br{\bigoplus \bm{\X}} \oplus \Br{\bigoplus \bm{\X}}}
< \epsilon.
$$
Therefore,
$$
\Cl{\Graph{\Br{\bigoplus \bm{S}} \Big|_{D}}}{\Br{\bigoplus \bm{\X}} \oplus \Br{\bigoplus \bm{\X}}} = \Graph{\bigoplus \bm{S}},
$$
so $ D $ is a core for $ \ds \bigoplus \bm{S} $. Similarly, $ D $ is a core for $ \ds \bigoplus \bm{T} $, which completes the proof of the lemma.
\end{proof}


Before proceeding, we prove \autoref{A Special Case of Nelson's Analytic Vectors Theorem}, a special case of Nelson's Analytic Vectors Theorem. It says that a strongly continuous unitary representation of a real Lie group on a Hilbert $ C^{\ast} $-module automatically produces a dense invariant submodule of analytic vectors.


\begin{Thm} \label{A Special Case of Nelson's Analytic Vectors Theorem}
Let $ G $ be a real Lie group, $ \X $ a Hilbert $ A $-module, $ W \in \URep{G}{\X} $, and $ \mathcal{B} $ a collection of continuous group homomorphisms from $ \R $ to $ G $. If
$$
\mathcal{C} \df \Set{T \in \RS{\X}}{\exists \Phi \in \mathcal{B}: ~ T \text{ is the infinitesimal generator of } W \circ \Phi \in \URep{\R}{\X}},
$$
then $ \Trip{G}{\X}{W}^{\omega} $ is a $ \mathcal{C} $-analytic and $ \mathcal{C} $-invariant dense $ A $-submodule of $ \X $ that is a core for each $ T \in \mathcal{C} $.
\end{Thm}

\begin{proof}
Let $ G $ be a real Lie group, $ \X $ a Hilbert $ A $-module, $ W \in \URep{G}{\X} $, and $ \mathcal{B} $ a collection of continuous group homomorphisms from $ \R $ to $ G $. As stated in \autoref{Preliminaries}, $ \Trip{G}{\X}{W}^{\omega} $ is a $ W $-invariant dense linear subspace of $ \X $.

For all $ \zeta \in \Trip{G}{\X}{W}^{\omega} $ and all charts $ \Pair{O}{\phi} $ in the unique maximal analytic atlas defining $ G $, the map
$$
\Map{\Im{\phi}{O}}{\X}{\bm{x}}{\FUNC{\func{W}{\func{\phi^{- 1}}{\bm{x}}}}{\zeta}}
$$
has a local real power-series expansion centered at each point in $ \Im{\phi}{O} $, so, for all $ a \in A $, the map
$$
  \Map{\Im{\phi}{O}}{\X}{\bm{x}}{\FUNC{\func{W}{\func{\phi^{- 1}}{\bm{x}}}}{\zeta \bullet a}}
= \Map{\Im{\phi}{O}}{\X}{\bm{x}}{\FUNC{\func{W}{\func{\phi^{- 1}}{\bm{x}}}}{\zeta} \bullet a}
$$
does as well. Hence, $ \Trip{G}{\X}{W}^{\omega} $ is closed under the right $ A $-action $ \bullet $, making it an $ A $-submodule of $ \X $.

For all $ \Phi \in \mathcal{B} $ and $ \zeta \in \Trip{G}{\X}{W}^{\omega} $, because the map $ \Map{G}{\X}{g}{\FUNC{\func{W}{g}}{\zeta}} $ is real-analytic by definition and $ \Phi $ is real-analytic by a theorem of \'{E}lie Cartan, their composition
$$
\Map{G}{\X}{g}{\FUNC{\func{W}{g}}{\zeta}} \circ \Phi = \Map{\R}{\X}{t}{\FUNC{\Func{W \circ \Phi}{t}}{\zeta}}
$$
is also real-analytic, so it has a local real power-series expansion centered at $ 0 $, i.e., there exist an $ R \in \R_{> 0} $ and a sequence $ \eta: \N_{0} \to \X $ such that
$$
\forall t \in \OOI{- R}{R}: \quad
\FUNC{\Func{W \circ \Phi}{t}}{\zeta} = \sum_{n \in \N_{0}} \frac{\Br{i t}^{n}}{n!} \cdot \eta_{n};
$$
letting $ T $ denote the infinitesimal generator of $ W \circ \Phi \in \URep{\R}{\X} $, an induction argument reveals that
$$
\forall n \in \N_{0}: \quad
\zeta \in \Dom{T^{n}}
\quad \text{and} \quad
\eta_{n} = \func{T^{n}}{\zeta}.
$$
Hence, $ \Trip{G}{\X}{W}^{\omega} $ is $ \mathcal{C} $-analytic.

Consequently, for all $ \Phi \in \mathcal{B} $, $ \zeta \in \Trip{G}{\X}{W}^{\omega} $, and $ \Phi \in \mathcal{B} $, if $ T $ denotes the infinitesimal generator of $ W \circ \Phi $, then $ \zeta \in \Dom{T} $ by the $ \mathcal{C} $-analyticity of $ \Trip{G}{\X}{W}^{\omega} $, and, furthermore,
\begin{align*}
    \Map{G}{\X}{g}{\FUNC{\func{W}{g}}{\func{T}{\zeta}}}
& = \Map{G}{\X}{g}{\FUNC{\func{W}{g}}{- i \cdot \Diff{t}{\FUNC{\Func{W \circ \Phi}{t}}{\zeta}} \bigg|_{t = 0}}} \\
& = \Map{G}{\X}{g}{\FUNC{\func{W}{g}}{- i \cdot \Diff{t}{\FUNC{\func{W}{\func{\Phi}{t}}}{\zeta}} \bigg|_{t = 0}}} \\
& = \Map{G}{\X}{g}{\Diff{t}{- i \cdot \FUNC{\func{W}{g}}{\FUNC{\func{W}{\func{\Phi}{t}}}{\zeta}}} \bigg|_{t = 0}} \quad
    \Br{\text{as $ \func{W}{g} $ is continuous}} \\
& = \Map{G}{\X}{g}{\Diff{t}{- i \cdot \FUNC{\func{W}{g \func{\Phi}{t}}}{\zeta}} \bigg|_{t = 0}} \\
& = \Map{G \times \R}{\X}{\Pair{g}{t}}{\Diff{t}{- i \cdot \FUNC{\func{W}{g \func{\Phi}{t}}}{\zeta}}} \Bigg|_{t = 0} \\
& = \frac{\partial}{\partial t} \SqBr{\Map{G \times \R}{\X}{\Pair{g}{t}}{- i \cdot \FUNC{\func{W}{g \func{\Phi}{t}}}{\zeta}}} \bigg|_{t = 0};
\end{align*}
as the map $ \sigma \df \Map{G}{\X}{g}{\FUNC{\func{W}{g}}{\zeta}} $ is real-analytic by definition and the map $ \tau \df \Map{G \times \R}{G}{\Pair{g}{t}}{g \func{\Phi}{t}} $ is real-analytic by the real-analyticity of $ \Phi $ and the multiplication law of $ G $, so are
$$
  \frac{\partial}{\partial t} \SqBr{- i \cdot \Br{\sigma \circ \tau}}
= \frac{\partial}{\partial t} \SqBr{\Map{G \times \R}{\X}{\Pair{g}{t}}{- i \cdot \FUNC{\func{W}{g \func{\Phi}{t}}}{\zeta}}}
$$
and, therefore, $ \dfrac{\partial}{\partial t} \SqBr{- i \cdot \Br{\sigma \circ \tau}} \bigg|_{t = 0} = \Map{G}{\X}{g}{\FUNC{\func{W}{g}}{\func{T}{\zeta}}} $. Hence, $ \Trip{G}{\X}{W}^{\omega} $ is $ \mathcal{C} $-invariant.

Let $ \Phi \in \mathcal{B} $, let $ T $ denote the infinitesimal generator of $ W \circ \Phi $, and let $ \xi \in \Dom{T} $. We claim that
$$
\Pair{\xi}{\func{T}{\xi}} \in \Cl{\Graph{T|_{\Trip{G}{\X}{W}^{\omega}}}}{\X \oplus \X}.
$$
As $ \Trip{G}{\X}{W}^{\omega} $ is dense in $ \X $, there exists a sequence $ \Seq{\zeta_{n}}{n \in \N} $ in $ \Trip{G}{\X}{W}^{\omega} $ that converges in $ \X $ to $ \zeta $. Fixing $ t \in \R_{> 0} $, we may then appeal to the $ W $-invariance of $ \Trip{G}{\X}{W}^{\omega} $ and the fact that $ \Trip{G}{\X}{W}^{\omega} \subseteq \Dom{T} $ to define a sequence $ \Seq{F_{n}: \CCI{0}{t} \to \Graph{T|_{\Trip{G}{\X}{W}^{\omega}}}}{n \in \N} $ by
$$
\forall n \in \N, ~ \forall s \in \CCI{0}{t}: \quad
\func{F_{n}}{s} \df \Pair{\FUNC{\Func{W \circ \Phi}{s}}{\zeta_{n}}}{\func{T}{\FUNC{\Func{W \circ \Phi}{s}}{\zeta_{n}}}}.
$$
For each $ n \in \N $, the function $ F_{n} $ is continuous thus integrable, so
\begin{align*}
      \Cl{\Graph{T|_{\Trip{G}{\X}{W}^{\omega}}}}{\X \oplus \X}
& \ni \RInt{0}{t}{\func{F_{n}}{s}}{s} \quad
      \Br{\text{as $ \Cl{\Graph{T|_{\Trip{G}{\X}{W}^{\omega}}}}{\X \oplus \X} $ is a closed $ A $-submodule of $ \X \oplus \X $}} \\
& =   \RInt{0}{t}{\Pair{\FUNC{\Func{W \circ \Phi}{s}}{\zeta_{n}}}{\func{T}{\FUNC{\Func{W \circ \Phi}{s}}{\zeta_{n}}}}}{s} \\
& =   \RInt{0}{t}{\SqBr{\Pair{\FUNC{\Func{W \circ \Phi}{s}}{\zeta_{n}}}{\bm{0}_{\X}} + \Pair{\bm{0}_{\X}}{\func{T}{\FUNC{\Func{W \circ \Phi}{s}}{\zeta_{n}}}}}}
           {s} \\
& =   \RInt{0}{t}{\Pair{\FUNC{\Func{W \circ \Phi}{s}}{\zeta_{n}}}{\bm{0}_{\X}}}{s} +
      \RInt{0}{t}{\Pair{\bm{0}_{\X}}{\func{T}{\FUNC{\Func{W \circ \Phi}{s}}{\zeta_{n}}}}}{s} \\
& =   \Pair{\RInt{0}{t}{\FUNC{\Func{W \circ \Phi}{s}}{\zeta_{n}}}{s}}{\bm{0}_{\X}} +
      \Pair{\bm{0}_{\X}}{\RInt{0}{t}{\func{T}{\FUNC{\Func{W \circ \Phi}{s}}{\zeta_{n}}}}{s}} \\
& =   \Pair{\RInt{0}{t}{\FUNC{\Func{W \circ \Phi}{s}}{\zeta_{n}}}{s}}{\RInt{0}{t}{\func{T}{\FUNC{\Func{W \circ \Phi}{s}}{\zeta_{n}}}}{s}} \\
& =   \Pair{\RInt{0}{t}{\FUNC{\Func{W \circ \Phi}{s}}{\zeta_{n}}}{s}}{\RInt{0}{t}{\Diff{s}{\FUNC{\Func{W \circ \Phi}{s}}{\zeta_{n}}}}{s}} \quad
      \Br{\text{by \autoref{The Infinitesimal Generator and Analytic Vectors of a Strongly Continuous One-Parameter Unitary Group}}} \\
& =   \Pair{\RInt{0}{t}{\FUNC{\Func{W \circ \Phi}{s}}{\zeta_{n}}}{s}}{\FUNC{\Func{W \circ \Phi}{t}}{\zeta_{n}} - \zeta_{n}}.
\end{align*}
As $ \Seq{\Map{\R}{\X}{s}{\FUNC{\Func{W \circ \Phi}{s}}{\zeta_{n}}}}{n \in \N} $ converges uniformly in $ \CC{\CCI{0}{t},\X} $ to $ \Map{\R}{\X}{s}{\FUNC{\Func{W \circ \Phi}{s}}{\xi}} $, we have
\begin{align*}
      \Pair{\RInt{0}{t}{\FUNC{\Func{W \circ \Phi}{s}}{\xi}}{s}}{\FUNC{\Func{W \circ \Phi}{t}}{\xi} - \xi}
& =   \Pair{\lim_{n \to \infty} \RInt{0}{t}{\FUNC{\Func{W \circ \Phi}{s}}{\zeta_{n}}}{s}}
           {\lim_{n \to \infty} \FUNC{\Func{W \circ \Phi}{t}}{\zeta_{n}} - \zeta_{n}} \\
& =   \lim_{n \to \infty} \Pair{\RInt{0}{t}{\FUNC{\Func{W \circ \Phi}{s}}{\zeta_{n}}}{s}}{\FUNC{\Func{W \circ \Phi}{t}}{\zeta_{n}} - \zeta_{n}} \\
& \in \Cl{\Graph{T|_{\Trip{G}{\X}{W}^{\omega}}}}{\X \oplus \X},
\end{align*}
which yields
\begin{align*}
      \Pair{\xi}{\func{T}{\xi}}
& =   \Pair{\lim_{t \to 0^{+}} \frac{1}{t} \cdot \RInt{0}{t}{\FUNC{\Func{W \circ \Phi}{s}}{\xi}}{s}}
           {\Diff{t}{\FUNC{\Func{W \circ \Phi}{t}}{\xi}} \bigg|_{t = 0}} \\
& =   \Pair{\lim_{t \to 0^{+}} \frac{1}{t} \cdot \RInt{0}{t}{\FUNC{\Func{W \circ \Phi}{s}}{\xi}}{s}}
           {\lim_{t \to 0^{+}} \frac{1}{t} \cdot \SqBr{\FUNC{\Func{W \circ \Phi}{t}}{\xi} - \xi}} \\
& =   \lim_{t \to 0^{+}} \frac{1}{t} \cdot \Pair{\RInt{0}{t}{\FUNC{\Func{W \circ \Phi}{s}}{\xi}}{s}}{\FUNC{\Func{W \circ \Phi}{t}}{\xi} - \xi} \\
& \ni \Cl{\Graph{T|_{\Trip{G}{\X}{W}^{\omega}}}}{\X \oplus \X}.
\end{align*}
As $ \xi \in \Dom{T} $ is arbitrary, we therefore obtain $ \cl{T|_{\Trip{G}{\X}{W}^{\omega}}} = T $, so $ \Trip{G}{\X}{W}^{\omega} $ is, indeed, a core for $ T $.
\end{proof}


The following two results --- one on the differentiability of Heisenberg representations and the other on the integrability of infinitesimal Heisenberg representations --- make up the essence of this paper. To facilitate the statements of these results, we shall employ the map $ \gamma $ from \autoref{The Gamma Map}, which is a topological group isomorphism from the locally compact Hausdorff abelian group $ \R $ to its Pontryagin dual $ \widehat{\R} $.


\begin{Prop}[Heisenberg differentiability] \label{Heisenberg Differentiability}
Let $ \Quad{\X}{\rho}{U}{V} $ be a Heisenberg representation of a real $ C^{\ast} $-dynamical system $ \Trip{\R}{A}{\alpha} $, and let $ S $ and $ T $ denote, respectively, the infinitesimal generators of
$$
\Seq{U_{- \hbar x}}{x \in \R} \in \URep{\R}{\X}
\quad \text{and} \quad
\Seq{\func{V}{\gamma_{y}}}{y \in \R} \in \URep{\R}{\X}.
$$
Then $ \Quad{\X}{\rho}{S}{T} $ is an infinitesimal Heisenberg representation of $ \Trip{\R}{A}{\alpha} $.
\end{Prop}

\begin{proof}
We shall begin our proof with a series of five claims.

\textbf{Claim 1:} For each $ a \in \Trip{\R}{A}{\alpha}^{\infty} $ and $ \zeta \in \Dom{S} $, the map
$$
\Map{\R}{\X}{x}{\FUNC{\func{\rho}{\alphArg{- \hbar x}{a}} \circ U_{- \hbar x}}{\zeta}}
$$
is differentiable; furthermore,
$$
\forall t \in \R: \quad
  \Diff{x}{\FUNC{\func{\rho}{\alphArg{- \hbar x}{a}} \circ U_{- \hbar x}}{\zeta}} \bigg|_{x = t}
= i \cdot \FUNC{\func{\rho}{\alphArg{- \hbar t}{a}}}{\func{U_{- \hbar t}}{\func{S}{\zeta}}}
  - \hbar \cdot \FUNC{\func{\rho}{\alphArg{- \hbar t}{\func{\delta}{a}}}}{\func{U_{- \hbar t}}{\zeta}}.
$$

\emph{Proof of Claim 1}. Picking arbitrary $ t \in \R $, $ a \in \Trip{\R}{A}{\alpha}^{\infty} $, and $ \zeta \in \Dom{S} $, we have
\begin{align*}
      \forall x \in \R \setminus \SSet{t}: \qquad
  & ~ \frac{1}{x - t} \cdot
      \SqBr{\FUNC{\func{\rho}{\alphArg{- \hbar x}{a}} \circ U_{- \hbar x}}{\zeta} - \FUNC{\func{\rho}{\alphArg{- \hbar t}{a}} \circ U_{- \hbar t}}{\zeta}} \\
= & ~ \frac{1}{x - t} \cdot
      \SqBr{\FUNC{\func{\rho}{\alphArg{- \hbar x}{a}} \circ U_{- \hbar x}}{\zeta} - \FUNC{\func{\rho}{\alphArg{- \hbar x}{a}} \circ U_{- \hbar t}}{\zeta}} ~ + \\
  & ~ \frac{1}{x - t} \cdot
      \SqBr{\FUNC{\func{\rho}{\alphArg{- \hbar x}{a}} \circ U_{- \hbar t}}{\zeta} - \FUNC{\func{\rho}{\alphArg{- \hbar t}{a}} \circ U_{- \hbar t}}{\zeta}} \\
= & ~ \frac{1}{x - t} \cdot \FUNC{\func{\rho}{\alphArg{- \hbar x}{a}}}{\func{U_{- \hbar x}}{\zeta} - \func{U_{- \hbar t}}{\zeta}} +
      \frac{1}{x - t} \cdot \FUNC{\func{\rho}{\alphArg{- \hbar x}{a} - \func{\rho}{\alphArg{- \hbar t}{a}}}}{\func{U_{- \hbar t}}{\zeta}} \\
= & ~ \underbrace{
                 \FUNC{\func{\rho}{\alphArg{- \hbar x}{a}}}{\frac{1}{x - t} \cdot \SqBr{\func{U_{- \hbar x}}{\zeta} - \func{U_{- \hbar t}}{\zeta}}}
                 }_{\Br{\clubsuit}} + \\
  & ~ \underbrace{
                 \FUNC{\func{\rho}{\frac{1}{x - t} \cdot \SqBr{\alphArg{- \hbar x}{a} - \alphArg{- \hbar t}{a}}}}{\func{U_{- \hbar t}}{\zeta}}
                 }_{\Br{\spadesuit}}.
\end{align*}

To show that $ \ds \lim_{x \to t} \Br{\clubsuit} $ exists, first observe that
\begin{align*}
         \forall x \in \R \setminus \SSet{t}: \qquad
     & ~ \Norm{
              \FUNC{\func{\rho}{\alphArg{- \hbar x}{a}}}{\frac{1}{x - t} \cdot \SqBr{\func{U_{- \hbar x}}{\zeta} - \func{U_{- \hbar t}}{\zeta}}} -
              i \cdot \FUNC{\func{\rho}{\alphArg{- \hbar t}{a}}}{\func{U_{- \hbar t}}{\func{S}{\zeta}}}
              }_{\X} \\
\leq & ~ \Norm{
              \FUNC{\func{\rho}{\alphArg{- \hbar x}{a}}}{\frac{1}{x - t} \cdot \SqBr{\func{U_{- \hbar x}}{\zeta} - \func{U_{- \hbar t}}{\zeta}}} -
              i \cdot \FUNC{\func{\rho}{\alphArg{- \hbar x}{a}}}{\func{U_{- \hbar t}}{\func{S}{\zeta}}}
              }_{\X} ~ + \\
     & ~ \Norm{
              i \cdot \FUNC{\func{\rho}{\alphArg{- \hbar x}{a}}}{\func{U_{- \hbar t}}{\func{S}{\zeta}}} -
              i \cdot \FUNC{\func{\rho}{\alphArg{- \hbar t}{a}}}{\func{U_{- \hbar t}}{\func{S}{\zeta}}}
              }_{\X} \\
=    & ~ \Norm{
              \FUNC{\func{\rho}{\alphArg{- \hbar x}{a}}}
                   {\frac{1}{x - t} \cdot \SqBr{\func{U_{- \hbar x}}{\zeta} - \func{U_{- \hbar t}}{\zeta}} - i \cdot \func{U_{- \hbar t}}{\func{S}{\zeta}}}
              }_{\X} ~ + \\
     & ~ \Norm{i \cdot \FUNC{\func{\rho}{\alphArg{- \hbar x}{a}} - \func{\rho}{\alphArg{- \hbar t}{a}}}{\func{U_{- \hbar t}}{\func{S}{\zeta}}}}_{\X} \\
\leq & ~ \Norm{\func{\rho}{\alphArg{- \hbar x}{a}}}_{\Adj{\X}}
         \Norm{\frac{1}{x - t} \cdot \SqBr{\func{U_{- \hbar x}}{\zeta} - \func{U_{- \hbar t}}{\zeta}} - i \cdot \func{U_{- \hbar t}}{\func{S}{\zeta}}}_{\X} + \\
     & ~ \Norm{\func{\rho}{\alphArg{- \hbar x}{a}} - \func{\rho}{\alphArg{- \hbar t}{a}}}_{\Adj{\X}} \Norm{\func{U_{- \hbar t}}{\func{S}{\zeta}}}_{\X} \\
\leq & ~ \Norm{a}_{A}
         \Norm{\frac{1}{x - t} \cdot \SqBr{\func{U_{- \hbar x}}{\zeta} - \func{U_{- \hbar t}}{\zeta}} - i \cdot \func{U_{- \hbar t}}{\func{S}{\zeta}}}_{\X} + \\
     & ~ \Norm{\func{\rho}{\alphArg{- \hbar x}{a}} - \func{\rho}{\alphArg{- \hbar t}{a}}}_{\Adj{\X}} \Norm{\func{S}{\zeta}}_{\X}.
\end{align*}
Recalling that $ \alpha $ is strongly continuous and --- from \autoref{The Infinitesimal Generator and Analytic Vectors of a Strongly Continuous One-Parameter Unitary Group} --- that
$$
  \lim_{x \to t} \frac{1}{x - t} \cdot \SqBr{\func{U_{- \hbar x}}{\zeta} - \func{U_{- \hbar t}}{\zeta}}
= \Diff{x}{\func{U_{- \hbar x}}{\zeta}} \bigg|_{x = t}
= i \cdot \func{U_{- \hbar t}}{\func{S}{\zeta}},
$$
we may then apply the Squeeze Theorem to infer that
$$
  \lim_{x \to t}
  \Norm{
       \FUNC{\func{\rho}{\alphArg{- \hbar x}{a}}}{\frac{1}{x - t} \cdot \SqBr{\func{U_{- \hbar x}}{\zeta} - \func{U_{- \hbar t}}{\zeta}}} -
       i \cdot \FUNC{\func{\rho}{\alphArg{- \hbar t}{a}}}{\func{U_{- \hbar t}}{\func{S}{\zeta}}}
       }_{\X}
= 0,
$$
which immediately yields $ \ds \lim_{x \to t} \Br{\clubsuit} = i \cdot \FUNC{\func{\rho}{\alphArg{- \hbar t}{a}}}{\func{U_{- \hbar t}}{\func{S}{\zeta}}} $.

To show that $ \ds \lim_{x \to t} \Br{\spadesuit} $ exists, first observe that
\begin{align*}
         \forall x \in \R \setminus \SSet{t}: \qquad
     & ~ \Norm{
              \FUNC{\func{\rho}{\frac{1}{x - t} \cdot \SqBr{\alphArg{- \hbar x}{a} - \alphArg{- \hbar t}{a}}}}{\func{U_{- \hbar t}}{\zeta}} +
              \hbar \cdot \FUNC{\func{\rho}{\alphArg{- \hbar t}{\func{\delta}{a}}}}{\func{U_{- \hbar t}}{\zeta}}
              }_{\X} \\
=    & ~ \Norm{
              \FUNC{
                   \func{\rho}
                        {\frac{1}{x - t} \cdot \SqBr{\alphArg{- \hbar x}{a} - \alphArg{- \hbar t}{a}} + \hbar \cdot \alphArg{- \hbar t}{\func{\delta}{a}}}
                   }
                   {\func{U_{- \hbar t}}{\zeta}}
              }_{\X} \\
\leq & ~ \Norm{
              \func{\rho}{\frac{1}{x - t} \cdot \SqBr{\alphArg{- \hbar x}{a} - \alphArg{- \hbar t}{a}} + \hbar \cdot \alphArg{- \hbar t}{\func{\delta}{a}}}
              }_{\Adj{\X}}
         \Norm{\func{U_{- \hbar t}}{\zeta}}_{\X} \\
\leq & ~ \Norm{\frac{1}{x - t} \cdot \SqBr{\alphArg{- \hbar x}{a} - \alphArg{- \hbar t}{a}} + \hbar \cdot \alphArg{- \hbar t}{\func{\delta}{a}}}_{A}
         \Norm{\zeta}_{\X}.
\end{align*}
Harnessing \autoref{Infinitesimal Generators of Real C*-Dynamical Systems} and the Chain Rule to obtain
$$
\lim_{x \to t} \frac{1}{x - t} \cdot \SqBr{\alphArg{- \hbar x}{a} - \alphArg{- \hbar t}{a}} = - \hbar \cdot \alphArg{- \hbar t}{\func{\delta}{a}},
$$
we may then apply the Squeeze Theorem to infer that
$$
  \lim_{x \to t}
  \Norm{
       \FUNC{\func{\rho}{\frac{1}{x - t} \cdot \SqBr{\alphArg{- \hbar x}{a} - \alphArg{- \hbar t}{a}}}}{\func{U_{- \hbar t}}{\zeta}} +
       \hbar \cdot \FUNC{\func{\rho}{\alphArg{- \hbar t}{\func{\delta}{a}}}}{\func{U_{- \hbar t}}{\zeta}}
       }_{\X}
= 0,
$$
which immediately yields $ \ds \lim_{x \to t} \Br{\spadesuit} = - \hbar \cdot \FUNC{\func{\rho}{\alphArg{- \hbar t}{\func{\delta}{a}}}}{\func{U_{- \hbar t}}{\zeta}} $.

Hence, the map $ \Map{\R}{\X}{x}{\FUNC{\func{\rho}{\alphArg{- \hbar x}{a}} \circ U_{- \hbar x}}{\zeta}} $ is differentiable; furthermore,
$$
\forall t \in \R: \quad
  \Diff{x}{\FUNC{\func{\rho}{\alphArg{- \hbar x}{a}} \circ U_{- \hbar x}}{\zeta}} \bigg|_{x = t}
= i \cdot \FUNC{\func{\rho}{\alphArg{- \hbar t}{a}}}{\func{U_{- \hbar t}}{\func{S}{\zeta}}}
  - \hbar \cdot \FUNC{\func{\rho}{\alphArg{- \hbar t}{\func{\delta}{a}}}}{\func{U_{- \hbar t}}{\zeta}}.
$$
The proof of Claim 1 is thereby complete. \hfill $ \triangle $

\textbf{Claim 2:} For each $ a \in \Trip{\R}{A}{\alpha}^{\infty} $ and $ \zeta \in \Dom{S} $, the map
$$
\Map{\R}{\X}{x}{\FUNC{U_{- \hbar x} \circ \func{\rho}{a}}{\zeta}}
$$
is differentiable; furthermore, $ \FUNC{\func{\rho}{a}}{\zeta} \in \Dom{S} $ and
$$
\forall t \in \R: \quad
\Diff{x}{\FUNC{U_{- \hbar x} \circ \func{\rho}{a}}{\zeta}} \bigg|_{x = t} = i \cdot \func{U_{- \hbar t}}{\func{S}{\FUNC{\func{\rho}{a}}{\zeta}}}.
$$

\emph{Proof of Claim 2}. By the definition of a Heisenberg representation in \autoref{thm:CSVNT}, we have
$$
\forall x \in \R, ~ \forall a \in A: \quad
U_{- \hbar x} \circ \func{\rho}{a} = \func{\rho}{\alphArg{- \hbar x}{a}} \circ U_{- \hbar x},
$$
so, for each $ a \in \Trip{\R}{A}{\alpha}^{\infty} $ and $ \zeta \in \Dom{S} $, the maps
$$
\Map{\R}{\X}{x}{\FUNC{U_{- \hbar x} \circ \func{\rho}{a}}{\zeta}}
\quad \text{and} \quad
\Map{\R}{\X}{x}{\FUNC{\func{\rho}{\alphArg{- \hbar x}{a}} \circ U_{- \hbar x}}{\zeta}}
$$
are equal; as the latter map is differentiable by Claim 1, the former map is also differentiable, which means that $ \FUNC{\func{\rho}{a}}{\zeta} \in \Dom{S} $ and --- by \autoref{The Infinitesimal Generator and Analytic Vectors of a Strongly Continuous One-Parameter Unitary Group} --- that
$$
\forall t \in \R: \quad
\Diff{x}{\FUNC{U_{- \hbar x} \circ \func{\rho}{a}}{\zeta}} \bigg|_{x = t} = i \cdot \func{U_{- \hbar t}}{\func{S}{\FUNC{\func{\rho}{a}}{\zeta}}}.
$$
The proof of Claim 2 is thereby complete. \hfill $ \triangle $

\textbf{Claim 3:} For each $ a \in A $ and $ \zeta \in \Dom{T} $, the map
$$
\Map{\R}{\X}{y}{\FUNC{\func{\rho}{a} \circ \func{V}{\gamma_{y}}}{\zeta}}
$$
is differentiable; furthermore,
$$
\forall t \in \R: \quad
\Diff{y}{\FUNC{\func{\rho}{a} \circ \func{V}{\gamma_{y}}}{\zeta}} \bigg|_{y = t} = i \cdot \FUNC{\func{\rho}{a}}{\FUNC{\func{V}{\gamma_{t}}}{\func{T}{\zeta}}}.
$$

\emph{Proof of Claim 3}. This follows from \autoref{The Infinitesimal Generator and Analytic Vectors of a Strongly Continuous One-Parameter Unitary Group} and --- for each $ a \in A $ --- the continuity of $ \func{\rho}{a} $. \hfill $ \triangle $

\textbf{Claim 4:} For each $ a \in A $ and $ \zeta \in \Dom{T} $, the map
$$
\Map{\R}{\X}{y}{\FUNC{\func{V}{\gamma_{y}} \circ \func{\rho}{a}}{\zeta}}
$$
is differentiable; furthermore, $ \FUNC{\func{\rho}{a}}{\zeta} \in \Dom{T} $ and
$$
\forall t \in \R: \quad
\Diff{y}{\FUNC{\func{V}{\gamma_{y}} \circ \func{\rho}{a}}{\zeta}} \bigg|_{y = t} = i \cdot \FUNC{\func{V}{\gamma_{t}}}{\func{T}{\FUNC{\func{\rho}{a}}{\zeta}}}.
$$

\emph{Proof of Claim 4}. By the definition of a Heisenberg representation in \autoref{thm:CSVNT}, we have
$$
\forall y \in \R, ~ \forall a \in A: \quad
\func{V}{\gamma_{y}} \circ \func{\rho}{a} = \func{\rho}{a} \circ \func{V}{\gamma_{y}},
$$
so, for each $ a \in A $ and $ \zeta \in \Dom{T} $, the maps
$$
\Map{\R}{\X}{y}{\FUNC{\func{V}{\gamma_{y}} \circ \func{\rho}{a}}{\zeta}}
\quad \text{and} \quad
\Map{\R}{\X}{y}{\FUNC{\func{\rho}{a} \circ \func{V}{\gamma_{y}}}{\zeta}}
$$
are equal; as the latter map is differentiable by Claim 3, the former map is also differentiable, which means that $ \FUNC{\func{\rho}{a}}{\zeta} \in \Dom{T} $ and --- by \autoref{The Infinitesimal Generator and Analytic Vectors of a Strongly Continuous One-Parameter Unitary Group} --- that
$$
\forall t \in \R: \quad
\Diff{y}{\FUNC{\func{V}{\gamma_{y}} \circ \func{\rho}{a}}{\zeta}} \bigg|_{y = t} = i \cdot \FUNC{\func{V}{\gamma_{t}}}{\func{T}{\FUNC{\func{\rho}{a}}{\zeta}}}.
$$
The proof of Claim 4 is thereby complete. \hfill $ \triangle $

\textbf{Claim 5:} There exists an $ \SSet{S,T} $-analytic and $ \SSet{S,T} $-invariant core for both $ S $ and $ T $ such that
$$
\Comm{S|_{D}}{T|_{D}} = - i \hbar \cdot \Id_{D}.
$$

\emph{Proof of Claim 5}. Recall that the Heisenberg group $ H $ is a $ 3 $-dimensional real Lie group and is the subgroup of $ \GL{3}{\R} $ defined by
$$
H \df \Set{\ThreeMatrix{1}{x}{z}{0}{1}{y}{0}{0}{1} \in \GL{3}{\R}}{x,y,z \in \R}.
$$
By the Weyl Commutation Relation in \autoref{thm:CSVNT}, we have
$$
\forall x,y \in \R: \quad
  \func{V}{\gamma_{y}} \circ U_{- \hbar x}
= \func{\gamma_{y}}{- \hbar x} \cdot \SqBr{U_{- \hbar x} \circ \func{V}{\gamma_{y}}}
= e^{- i \hbar x y} \cdot \SqBr{U_{- \hbar x} \circ \func{V}{\gamma_{y}}},
$$
which readily implies that the map $ W: H \to \Unitary{\X} $, defined by
$$
\forall x,y,z \in \R, ~ \forall \zeta \in \X: \quad
\FUNC{\func{W}{\ThreeMatrix{1}{x}{z}{0}{1}{y}{0}{0}{1}}}{\zeta} \df e^{i \hbar z} \cdot \FUNC{\func{V}{\gamma_{y}} \circ U_{- \hbar x}}{\zeta},
$$
is a strongly continuous unitary representation of $ H $ on $ \X $.

Define Lie-group homomorphisms $ \phi,\psi: \R \to G $ by
$$
\forall x,y \in \R: \quad
\func{\phi}{x} \df \ThreeMatrix{1}{x}{0}{0}{1}{0}{0}{0}{1}
\quad \text{and} \quad
\func{\psi}{y} \df \ThreeMatrix{1}{0}{0}{0}{1}{y}{0}{0}{1}.
$$
Then $ U = W \circ \phi $ and $ V = W \circ \psi $, whereupon \autoref{A Special Case of Nelson's Analytic Vectors Theorem} asserts that $ D \df \Trip{H}{\X}{W}^{\omega} $ is an $ \SSet{S,T} $-analytic and $ \SSet{S,T} $-invariant core for both $ S $ and $ T $, leaving us to verify the HCR for $ \Pair{S}{T} $ on $ D $.

For each $ y \in \R $ and $ \zeta \in D $, we have $ \FUNC{\func{V}{\gamma_{y}}}{\zeta} = \FUNC{\func{W}{\ThreeMatrix{1}{0}{0}{0}{1}{y}{0}{0}{1}}}{\zeta} \in D \subseteq \Dom{S} $, so the map
$$
\Map{\R}{\X}{x}{e^{- i \hbar x y} \cdot \func{U_{- \hbar x}}{\FUNC{\func{V}{\gamma_{y}}}{\zeta}}}
$$
is differentiable; furthermore,
\begin{align*}
    i \cdot \FUNC{\func{V}{\gamma_{y}}}{\func{S}{\zeta}}
& = \FUNC{\func{V}{\gamma_{y}}}{i \cdot \func{S}{\zeta}} \\
& = \FUNC{\func{V}{\gamma_{y}}}{\Diff{x}{\func{U_{- \hbar x}}{\zeta}} \bigg|_{x = 0}} \\
& = \Diff{x}{\FUNC{\func{V}{\gamma_{y}}}{\func{U_{- \hbar x}}{\zeta}}} \bigg|_{x = 0} \quad \Br{\text{as $ \func{V}{\gamma_{y}} $ is continuous}} \\
& = \Diff{x}{e^{- i \hbar x y} \cdot \func{U_{- \hbar x}}{\FUNC{\func{V}{\gamma_{y}}}{\zeta}}} \bigg|_{x = 0} \quad
    \Br{\text{by the Weyl Commutation Relation}} \\
& = \Diff{x}{e^{- i \hbar x y}} \bigg|_{x = 0} \cdot \func{U_{- \hbar x}}{\FUNC{\func{V}{\gamma_{y}}}{\zeta}} \bigg|_{x = 0} +
    e^{- i \hbar x y} \Big|_{x = 0} \cdot \Diff{x}{\func{U_{- \hbar x}}{\FUNC{\func{V}{\gamma_{y}}}{\zeta}}} \bigg|_{x = 0} \\
& = - i \hbar y \cdot \FUNC{\func{V}{\gamma_{y}}}{\zeta} + i \cdot \func{S}{\FUNC{\func{V}{\gamma_{y}}}{\zeta}},
\end{align*}
which, after a brief rearrangement of terms, becomes
\begin{equation} \label{A Basic Identity}
i \cdot \func{S}{\FUNC{\func{V}{\gamma_{y}}}{\zeta}} = i \cdot \FUNC{\func{V}{\gamma_{y}}}{\func{S}{\zeta}} + i \hbar y \cdot \FUNC{\func{V}{\gamma_{y}}}{\zeta}.
\end{equation}

Next, for each $ \zeta \in D $, we have $ \func{S}{\zeta} \in D \subseteq T $, so the map
$$
\Map{\R}{\X}{y}{i \cdot \FUNC{\func{V}{\gamma_{y}}}{\func{S}{\zeta}} + i \hbar y \cdot \FUNC{\func{V}{\gamma_{y}}}{\zeta}}
$$
is differentiable; furthermore,
\begin{align*}
  & ~ \Diff{y}{i \cdot \FUNC{\func{V}{\gamma_{y}}}{\func{S}{\zeta}} + i \hbar y \cdot \FUNC{\func{V}{\gamma_{y}}}{\zeta}} \bigg|_{y = 0} \\
= & ~ \Diff{y}{i \cdot \FUNC{\func{V}{\gamma_{y}}}{\func{S}{\zeta}}} \bigg|_{y = 0} +
      \Diff{y}{i \hbar y} \bigg|_{y = 0} \cdot \FUNC{\func{V}{\gamma_{y}}}{\zeta} \Big|_{y = 0} +
      i \hbar y \Big|_{y = 0} \cdot \Diff{y}{\FUNC{\func{V}{\gamma_{y}}}{\zeta}} \bigg|_{y = 0} \\
= & ~ - \func{T}{\func{S}{\zeta}} + i \hbar \cdot \zeta,
\end{align*}
which, in light of \autoref{A Basic Identity}, means that the map $ \Map{\R}{\X}{y}{i \cdot \func{S}{\FUNC{\func{V}{\gamma_{y}}}{\zeta}}} $ is differentiable and that
\begin{align*}
    - \func{S}{\func{T}{\zeta}}
& = i \cdot \func{S}{i \cdot \func{T}{\zeta}} \\
& = i \cdot \func{S}{\Diff{y}{\FUNC{\func{V}{\gamma_{y}}}{\zeta}} \bigg|_{y = 0}} \\
& = i \cdot \func{S}{\lim_{y \to 0} \frac{1}{y} \cdot \Br{\FUNC{\func{V}{\gamma_{y}}}{\zeta} - \zeta}} \\
& = i \cdot \lim_{y \to 0} \func{S}{\frac{1}{y} \cdot \Br{\FUNC{\func{V}{\gamma_{y}}}{\zeta} - \zeta}} \quad \Br{\text{as $ S $ is closed}} \\
& = i \cdot \lim_{y \to 0} \frac{1}{y} \cdot \SqBr{\func{S}{\FUNC{\func{V}{\gamma_{y}}}{\zeta}} - \func{S}{\zeta}} \\
& = i \cdot \Diff{y}{\func{S}{\FUNC{\func{V}{\gamma_{y}}}{\zeta}}} \bigg|_{y = 0} \\
& = \Diff{y}{i \cdot \func{S}{\FUNC{\func{V}{\gamma_{y}}}{\zeta}}} \bigg|_{y = 0} \\
& = \Diff{y}{i \cdot \FUNC{\func{V}{\gamma_{y}}}{\func{S}{\zeta}} + i \hbar y \cdot \FUNC{\func{V}{\gamma_{y}}}{\zeta}} \bigg|_{y = 0} \\
& = - \func{T}{\func{S}{\zeta}} + i \hbar \cdot \zeta.
\end{align*}
Therefore, $ \Comm{S|_{D}}{T|_{D}} = - i \hbar \cdot \Id_{D} $, which completes the proof of Claim 5. \hfill $ \triangle $

According to Claims 1 and 2, we have, for all $ t \in \R $, $ a \in \Trip{\R}{A}{\alpha}^{\infty} $, and $ \zeta \in \Dom{S} $, that
\begin{align*}
    i \cdot \func{U_{- \hbar t}}{\func{S}{\FUNC{\func{\rho}{a}}{\zeta}}}
& = \Diff{x}{\FUNC{U_{- \hbar x} \circ \func{\rho}{a}}{\zeta}} \bigg|_{x = t} \\
& = \Diff{x}{\FUNC{\func{\rho}{\alphArg{- \hbar x}{a}} \circ U_{- \hbar x}}{\zeta}} \bigg|_{x = t} \\
& = i \cdot \FUNC{\func{\rho}{\alphArg{- \hbar t}{a}}}{\func{U_{- \hbar t}}{\func{S}{\zeta}}}
    - \hbar \cdot \FUNC{\func{\rho}{\alphArg{- \hbar t}{\func{\delta}{a}}}}{\func{U_{- \hbar t}}{\zeta}};
\end{align*}
by setting $ t = 0 $, multiplying both sides by $ i $, and then rearranging terms, we find that
$$
\FUNC{\func{\rho}{a}}{\func{S}{\zeta}} = \func{S}{\FUNC{\func{\rho}{a}}{\zeta}} - i \hbar \cdot \FUNC{\func{\rho}{\func{\delta}{a}}}{\zeta},
$$
which yields
$$
\func{\rho}{a} \circ S \subseteq S \circ \func{\rho}{a} - i \hbar \cdot \func{\rho}{\func{\delta}{a}}.
$$
According to Claims 3 and 4, we have, for all $ t \in \R $, $ a \in A $, and $ \zeta \in \Dom{T} $, that
$$
    i \cdot \FUNC{\func{\rho}{a}}{\FUNC{\func{V}{\gamma_{t}}}{\func{T}{\zeta}}}
= \Diff{y}{\FUNC{\func{\rho}{a} \circ \func{V}{\gamma_{y}}}{\zeta}} \bigg|_{y = t}
= \Diff{y}{\FUNC{\func{V}{\gamma_{y}} \circ \func{\rho}{a}}{\zeta}} \bigg|_{y = t}
= i \cdot \FUNC{\func{V}{\gamma_{t}}}{\func{T}{\FUNC{\func{\rho}{a}}{\zeta}}};
$$
by setting $ t = 0 $ and multiplying both sides by $ - i $, we find that $ \FUNC{\func{\rho}{a}}{\func{T}{\zeta}} = \func{T}{\FUNC{\func{\rho}{a}}{\zeta}} $, which yields
$$
\func{\rho}{a} \circ T \subseteq T \circ \func{\rho}{a}.
$$
Claim 5 then completes the proof.
\end{proof}


\begin{Prop}[Heisenberg integrability] \label{Heisenberg Integrability}
Let $ \Quad{\X}{\rho}{S}{T} $ be an infinitesimal Heisenberg representation of a real $ C^{\ast} $-dynamical system $ \Trip{\R}{A}{\alpha} $. Then
$$
\Quad{\X}{\rho}{\Seq{e^{- \frac{i x}{\hbar} \cdot S}}{x \in \R}}{\Seq{e^{i \gamma^{- 1}_{\varphi} \cdot T}}{\varphi \in \widehat{\R}}}
$$
is a Heisenberg representation of $ \Trip{\R}{A}{\alpha} $, where $ \gamma^{- 1} $ denotes the inverse of $ \gamma: \R \to \widehat{\R} $.
\end{Prop}

\begin{proof}
For ease of notation, let $ U \df \Seq{e^{- \frac{i x}{\hbar} \cdot S}}{x \in \R} \in \URep{\R}{\X} $ and $ V \df \Seq{e^{i \gamma^{- 1}_{\varphi} \cdot T}}{\varphi \in \widehat{\R}} \in \URep{\widehat{\R}}{\X} $.

\textbf{Claim 1:} For each $ t \in \R $ and $ a \in \Trip{\R}{A}{\alpha}^{\infty} $, we have $ \func{\rho}{\alphArg{t}{a}} \circ S \subseteq S \circ \func{\rho}{\alphArg{t}{a}} - i \hbar \cdot \func{\rho}{\func{\delta}{\alphArg{t}{a}}} $.

\emph{Proof of Claim 1}. This follows from Property (2) of \autoref{Infinitesimal Heisenberg Representation} and Statement (2) of \autoref{Infinitesimal Generators of Real C*-Dynamical Systems}. \hfill $ \triangle $

\textbf{Claim 2:} For each $ a \in \Trip{\R}{A}{\alpha}^{\infty} $ and $ \zeta \in \Dom{S} $, the map
$$
\Map{\R}{A}{x}{\FUNC{\func{\rho}{\alphArg{x}{a}} \circ U_{x}}{\zeta}}
$$
is differentiable; furthermore,
$$
\forall t \in \R: \quad
  \Diff{x}{\FUNC{\func{\rho}{\alphArg{x}{a}} \circ U_{x}}{\zeta}} \bigg|_{x = t}
= \FUNC{\func{\rho}{\func{\delta}{\alphArg{t}{a}}}}{\func{U_{t}}{\zeta}} - \frac{i}{\hbar} \cdot \FUNC{\func{\rho}{\alphArg{t}{a}} \circ S}{\func{U_{t}}{\zeta}}.
$$

\emph{Proof of Claim 2}. A very minor modification of the argument used to prove Claim 1 of \autoref{Heisenberg Differentiability} yields, for each $ a \in \Trip{\R}{A}{\alpha}^{\infty} $ and $ \zeta \in \Dom{S} $, the differentiability of the map
$$
\Map{\R}{A}{x}{\FUNC{\func{\rho}{\alphArg{x}{a}} \circ U_{x}}{\zeta}}
$$
as well as the identity
$$
\forall t \in \R: \quad
  \Diff{x}{\FUNC{\func{\rho}{\alphArg{x}{a}} \circ U_{x}}{\zeta}} \bigg|_{x = t}
= \FUNC{\func{\rho}{\func{\delta}{\alphArg{t}{a}}}}{\func{U_{t}}{\zeta}} - \frac{i}{\hbar} \cdot \FUNC{\func{\rho}{\alphArg{t}{a}} \circ U_{t}}{\func{S}{\zeta}}.
$$
However, by Statement (2) of \autoref{The Infinitesimal Generator and Analytic Vectors of a Strongly Continuous One-Parameter Unitary Group}, we have, for all $ t \in \R $, that $ U_{t} \circ S \subseteq S \circ U_{t} $, so we are done. \hfill $ \triangle $

\textbf{Claim 3:} For each $ t \in \R $, $ a \in \Trip{\R}{A}{\alpha}^{\infty} $, and $ \zeta \in \Dom{S} $, we have $ \FUNC{\func{\rho}{\alphArg{t}{a}} \circ U_{t}}{\zeta} \in \Dom{S} $.

\emph{Proof of Claim 3}. By Statement (2) of \autoref{The Infinitesimal Generator and Analytic Vectors of a Strongly Continuous One-Parameter Unitary Group} and Claim 1, we have
$$
\forall t \in \R, ~ a \in \Trip{\R}{A}{\alpha}^{\infty}, ~ \forall \zeta \in \Dom{S}: \quad
          \func{U_{t}}{\zeta}
\in       \Dom{S}
=         \Dom{\func{\rho}{\alphArg{t}{a}} \circ S}
\subseteq \Dom{S \circ \func{\rho}{\alphArg{t}{a}}}.
$$
Claim 3 readily follows. \hfill $ \triangle $

\textbf{Claim 4:} For each $ t \in \R $, $ a \in A $, and $ \zeta \in \X $, we have $ \FUNC{U_{- x} \circ \func{\rho}{\alphArg{x}{a}} \circ U_{x}}{\zeta} = \FUNC{\func{\rho}{a}}{\zeta} $.

\emph{Proof of Claim 4}. Picking arbitrary $ t \in \R $, $ a \in \Trip{\R}{A}{\alpha}^{\infty} $, and $ \zeta \in \Dom{S} $, we have
\begin{align*}
      \forall x \in \R \setminus \SSet{t}: \qquad
  & ~ \frac{1}{x - t} \cdot
      \SqBr{\FUNC{U_{- x} \circ \func{\rho}{\alphArg{x}{a}} \circ U_{x}}{\zeta} - \FUNC{U_{- t} \circ \func{\rho}{\alphArg{t}{a}} \circ U_{t}}{\zeta}} \\
= & ~ \frac{1}{x - t} \cdot
      \SqBr{\FUNC{U_{- x} \circ \func{\rho}{\alphArg{x}{a}} \circ U_{x}}{\zeta} - \FUNC{U_{- x} \circ \func{\rho}{\alphArg{t}{a}} \circ U_{t}}{\zeta}} ~ + \\
  & ~ \frac{1}{x - t} \cdot
      \SqBr{\FUNC{U_{- x} \circ \func{\rho}{\alphArg{t}{a}} \circ U_{t}}{\zeta} - \FUNC{U_{- t} \circ \func{\rho}{\alphArg{t}{a}} \circ U_{t}}{\zeta}} \\
= & ~ \underbrace{
                 \func{U_{- x}}
                      {
                      \frac{1}{x - t} \cdot \SqBr{\FUNC{\func{\rho}{\alphArg{x}{a}} \circ U_{x}}{\zeta} - \FUNC{\func{\rho}{\alphArg{t}{a}} \circ U_{t}}{\zeta}}
                      }
                 }_{\Br{\clubsuit}} ~ + \\
  & ~ \underbrace{
                 \frac{1}{x - t} \cdot
                 \SqBr{
                      \func{U_{- x}}{\FUNC{\func{\rho}{\alphArg{t}{a}} \circ U_{t}}{\zeta}} -
                      \func{U_{- t}}{\FUNC{\func{\rho}{\alphArg{t}{a}} \circ U_{t}}{\zeta}}
                      }
                 }_{\Br{\spadesuit}}.
\end{align*}

To show that $ \ds \lim_{x \to t} \Br{\clubsuit} $ exists, first observe, for all $ x \in \R \setminus \SSet{t} $, that
\begin{align*}
     & ~ \Norm{
              \func{U_{- x}}
                   {\frac{1}{x - t} \cdot \SqBr{\FUNC{\func{\rho}{\alphArg{x}{a}} \circ U_{x}}{\zeta} - \FUNC{\func{\rho}{\alphArg{t}{a}} \circ U_{t}}{\zeta}}} -
              \func{U_{- t}}{\Diff{x}{\FUNC{\func{\rho}{\alphArg{x}{a}} \circ U_{x}}{\zeta}} \bigg|_{x = t}}
              }_{\X} \\
\leq & ~ \Norm{
              \func{U_{- x}}
                   {\frac{1}{x - t} \cdot \SqBr{\FUNC{\func{\rho}{\alphArg{x}{a}} \circ U_{x}}{\zeta} - \FUNC{\func{\rho}{\alphArg{t}{a}} \circ U_{t}}{\zeta}}} -
              \func{U_{- x}}{\Diff{x}{\FUNC{\func{\rho}{\alphArg{x}{a}} \circ U_{x}}{\zeta}} \bigg|_{x = t}}
              }_{\X} ~ + \\
     & ~ \Norm{
              \func{U_{- x}}{\Diff{x}{\FUNC{\func{\rho}{\alphArg{x}{a}} \circ U_{x}}{\zeta}} \bigg|_{x = t}} -
              \func{U_{- t}}{\Diff{x}{\FUNC{\func{\rho}{\alphArg{x}{a}} \circ U_{x}}{\zeta}} \bigg|_{x = t}}
              }_{\X} \\
=    & ~ \Norm{
              \func{U_{- x}}
                   {
                   \frac{1}{x - t} \cdot \SqBr{\FUNC{\func{\rho}{\alphArg{x}{a}} \circ U_{x}}{\zeta} - \FUNC{\func{\rho}{\alphArg{t}{a}} \circ U_{t}}{\zeta}} -
                   \Diff{x}{\FUNC{\func{\rho}{\alphArg{x}{a}} \circ U_{x}}{\zeta}} \bigg|_{x = t}
                   }
              }_{\X} ~ + \\
     & ~ \Norm{
              \func{U_{- x}}{\Diff{x}{\FUNC{\func{\rho}{\alphArg{x}{a}} \circ U_{x}}{\zeta}} \bigg|_{x = t}} -
              \func{U_{- t}}{\Diff{x}{\FUNC{\func{\rho}{\alphArg{x}{a}} \circ U_{x}}{\zeta}} \bigg|_{x = t}}
              }_{\X} \\
=    & ~ \Norm{
              \frac{1}{x - t} \cdot \SqBr{\FUNC{\func{\rho}{\alphArg{x}{a}} \circ U_{x}}{\zeta} - \FUNC{\func{\rho}{\alphArg{t}{a}} \circ U_{t}}{\zeta}} -
              \Diff{x}{\FUNC{\func{\rho}{\alphArg{x}{a}} \circ U_{x}}{\zeta}} \bigg|_{x = t}
              }_{\X} ~ + \\
     & ~ \Norm{
              \func{U_{- x}}{\Diff{x}{\FUNC{\func{\rho}{\alphArg{x}{a}} \circ U_{x}}{\zeta}} \bigg|_{x = t}} -
              \func{U_{- t}}{\Diff{x}{\FUNC{\func{\rho}{\alphArg{x}{a}} \circ U_{x}}{\zeta}} \bigg|_{x = t}}
              }_{\X}. 
\end{align*}
Applying Claim 2 and the strong continuity of $ U $, we may then apply the Squeeze Theorem to infer that
$$
  \lim_{x \to t}
  \Norm{
       \func{U_{- x}}
            {\frac{1}{x - t} \cdot \SqBr{\FUNC{\func{\rho}{\alphArg{x}{a}} \circ U_{x}}{\zeta} - \FUNC{\func{\rho}{\alphArg{t}{a}} \circ U_{t}}{\zeta}}} -
       \func{U_{- t}}{\Diff{x}{\FUNC{\func{\rho}{\alphArg{x}{a}} \circ U_{x}}{\zeta}} \bigg|_{x = t}}
       }_{\X}
= 0,
$$
which immediately yields
$$
  \lim_{x \to t} \Br{\clubsuit}
= \func{U_{- t}}{\Diff{x}{\FUNC{\func{\rho}{\alphArg{x}{a}} \circ U_{x}}{\zeta}} \bigg|_{x = t}}
= \func{U_{- t}}
       {
       \FUNC{\func{\rho}{\func{\delta}{\alphArg{t}{a}}}}{\func{U_{t}}{\zeta}} -
       \frac{i}{\hbar} \cdot \FUNC{\func{\rho}{\alphArg{t}{a}} \circ S}{\func{U_{t}}{\zeta}}
       }.
$$

By Claim 3 and the Chain Rule, the map
$$
\Map{\R}{\X}{x}{\func{U_{- x}}{\FUNC{\func{\rho}{\alphArg{t}{a}} \circ U_{t}}{\zeta}}}
$$
is differentiable; \autoref{The Infinitesimal Generator and Analytic Vectors of a Strongly Continuous One-Parameter Unitary Group} tells us additionally that
\begin{align*}
    \lim_{x \to t} \Br{\spadesuit}
& = \Diff{x}{\func{U_{- x}}{\FUNC{\func{\rho}{\alphArg{t}{a}} \circ U_{t}}{\zeta}}} \bigg|_{x = t} \\
& = \frac{i}{\hbar} \cdot \func{U_{- t}}{\func{S}{\FUNC{\func{\rho}{\alphArg{t}{a}} \circ U_{t}}{\zeta}}} \\
& = \func{U_{- t}}{\frac{i}{\hbar} \cdot \FUNC{S \circ \func{\rho}{\alphArg{t}{a}}}{\func{U_{t}}{\zeta}}}.
\end{align*}

The foregoing allows us to see that the map $ \Map{\R}{\X}{x}{\FUNC{U_{- x} \circ \func{\rho}{\alphArg{x}{a}} \circ U_{x}}{\zeta}} $ is differentiable and that
\begin{align*}
\forall t \in \R: \qquad
  & ~ \Diff{x}{\FUNC{U_{- x} \circ \func{\rho}{\alphArg{x}{a}} \circ U_{x}}{\zeta}} \bigg|_{x = t} \\
= & ~ \lim_{x \to t} \Br{\clubsuit} + \lim_{x \to t} \Br{\spadesuit} \\
= & ~ \func{U_{- t}}
           {
           \FUNC{\func{\rho}{\func{\delta}{\alphArg{t}{a}}}}{\func{U_{t}}{\zeta}} -
           \frac{i}{\hbar} \cdot \FUNC{\func{\rho}{\alphArg{t}{a}} \circ S}{\func{U_{t}}{\zeta}}
           } +
      \func{U_{- t}}{\frac{i}{\hbar} \cdot \FUNC{S \circ \func{\rho}{\alphArg{t}{a}}}{\func{U_{t}}{\zeta}}} \\
= & ~ \frac{i}{\hbar} \cdot
      \func{U_{- t}}
           {
           \FUNC{- i \hbar \cdot \func{\rho}{\func{\delta}{\alphArg{t}{a}}} - \func{\rho}{\alphArg{t}{a}} \circ S + S \circ \func{\rho}{\alphArg{t}{a}}}
                {\func{U_{t}}{\zeta}}
           } \\
= & ~ \frac{i}{\hbar} \cdot \func{U_{- t}}{\bm{0}_{\X}} \quad \Br{\text{by Claim 1}} \\
= & ~ \bm{0}_{\X},
\end{align*}
which implies that said map is constant. Furthermore,
\begin{equation} \label{Covariance Identity 1}
\forall x \in \R: \quad
  \FUNC{U_{- x} \circ \func{\rho}{\alphArg{x}{a}} \circ U_{x}}{\zeta}
= \FUNC{U_{- 0} \circ \func{\rho}{\alphArg{0}{a}} \circ U_{0}}{\zeta}
= \FUNC{\Id_{\X} \circ \func{\rho}{a} \circ \Id_{\X}}{\zeta}
= \FUNC{\func{\rho}{a}}{\zeta};
\end{equation}
as $ \Dom{S} $ is dense in $ \X $, the continuity of $ U_{- x} \circ \func{\rho}{\alphArg{x}{a}} \circ U_{x} $ and $ \func{\rho}{a} $ ensures that \autoref{Covariance Identity 1} holds for all $ \zeta \in \X $, and as $ \Trip{\R}{A}{\alpha}^{\infty} $ is dense in $ A $, the continuity of $ \rho $ ensures that \autoref{Covariance Identity 1} holds for all $ a \in A $ also. The proof of Claim 4 is thereby complete. \hfill $ \triangle $

\textbf{Claim 5:} For each $ a \in A $ and $ \zeta \in \Dom{T} $, the map
$$
\Map{\R}{A}{y}{\FUNC{\func{\rho}{a} \circ e^{i y \cdot T}}{\zeta}}
$$
is differentiable; furthermore,
$$
\forall t \in \R: \quad
\Diff{y}{\FUNC{\func{\rho}{a} \circ e^{i y \cdot T}}{\zeta}} \bigg|_{y = t} = i \cdot \FUNC{\func{\rho}{a} \circ T}{\Func{e^{i t \cdot T}}{\zeta}}.
$$

\emph{Proof of Claim 5}. This follows from \autoref{The Infinitesimal Generator and Analytic Vectors of a Strongly Continuous One-Parameter Unitary Group} and --- for each $ a \in A $ --- the continuity of $ \func{\rho}{a} $. \hfill $ \triangle $

\textbf{Claim 6:} For each $ t \in \R $, $ a \in A $, and $ \zeta \in \Dom{T} $, we have $ \FUNC{\func{\rho}{a} \circ e^{i t \cdot T}}{\zeta} \in \Dom{T} $.

\emph{Proof of Claim 6}. By Statement (2) of \autoref{The Infinitesimal Generator and Analytic Vectors of a Strongly Continuous One-Parameter Unitary Group}, we have
$$
\forall t \in \R, ~ a \in A, ~ \forall \zeta \in \Dom{T}: \quad
          \Func{e^{i t \cdot T}}{\zeta}
\in       \Dom{T}
=         \Dom{\func{\rho}{a} \circ T}
\subseteq \Dom{T \circ \func{\rho}{a}}.
$$
Claim 6 readily follows. \hfill $ \triangle $

\textbf{Claim 7:} For each $ \varphi \in \widehat{\R} $, $ a \in A $, and $ \zeta \in \X $, we have $ \FUNC{\func{V}{\varphi^{- 1}} \circ \func{\rho}{a} \circ \func{V}{\varphi}}{\zeta} = \FUNC{\func{\rho}{a}}{\zeta} $.

\emph{Proof of Claim 7}. Picking arbitrary $ t \in \R $, $ a \in A $, and $ \zeta \in \Dom{T} $, we have
\begin{align*}
\forall y \in \R \setminus \SSet{t}: \qquad
  & ~ \frac{1}{y - t} \cdot
      \SqBr{
           \FUNC{\func{V}{\gamma_{y}^{- 1}} \circ \func{\rho}{a} \circ \func{V}{\gamma_{y}}}{\zeta} -
           \FUNC{\func{V}{\gamma_{t}^{- 1}} \circ \func{\rho}{a} \circ \func{V}{\gamma_{t}}}{\zeta}
           } \\
= & ~ \frac{1}{y - t} \cdot
      \SqBr{
           \FUNC{\func{V}{\gamma_{- y}} \circ \func{\rho}{a} \circ \func{V}{\gamma_{y}}}{\zeta} -
           \FUNC{\func{V}{\gamma_{- t}} \circ \func{\rho}{a} \circ \func{V}{\gamma_{t}}}{\zeta}
           } \\
= & ~ \frac{1}{y - t} \cdot
      \SqBr{
           \FUNC{\func{V}{\gamma_{- y}} \circ \func{\rho}{a} \circ \func{V}{\gamma_{y}}}{\zeta} -
           \FUNC{\func{V}{\gamma_{- y}} \circ \func{\rho}{a} \circ \func{V}{\gamma_{t}}}{\zeta}
           } ~ + \\
  & ~ \frac{1}{y - t} \cdot
      \SqBr{
           \FUNC{\func{V}{\gamma_{- y}} \circ \func{\rho}{a} \circ \func{V}{\gamma_{t}}}{\zeta} -
           \FUNC{\func{V}{\gamma_{- t}} \circ \func{\rho}{a} \circ \func{V}{\gamma_{t}}}{\zeta}
           } \\
= & ~ \FUNC{\func{V}{\gamma_{- y}}}
           {\frac{1}{y - t} \cdot \SqBr{\FUNC{\func{\rho}{a} \circ \func{V}{\gamma_{y}}}{\zeta} - \FUNC{\func{\rho}{a} \circ \func{V}{\gamma_{t}}}{\zeta}}} ~ +
           \\
  & ~ \frac{1}{y - t} \cdot
      \SqBr{
           \FUNC{\func{V}{\gamma_{- y}}}{\FUNC{\func{\rho}{a} \circ \func{V}{\gamma_{t}}}{\zeta}} -
           \FUNC{\func{V}{\gamma_{- t}}}{\FUNC{\func{\rho}{a} \circ \func{V}{\gamma_{t}}}{\zeta}}
           } \\
= & ~ \underbrace{
                 \Func{e^{- i y \cdot T}}
                      {\frac{1}{y - t} \cdot \SqBr{\FUNC{\func{\rho}{a} \circ e^{i y \cdot T}}{\zeta} - \FUNC{\func{\rho}{a} \circ e^{i t \cdot T}}{\zeta}}}
                 }_{\Br{\diamondsuit}} ~ + \\
  & ~ \underbrace{
                 \frac{1}{y - t} \cdot
                 \SqBr{
                      \Func{e^{- i y \cdot T}}{\FUNC{\func{\rho}{a} \circ e^{i t \cdot T}}{\zeta}} -
                      \Func{e^{- i t \cdot T}}{\FUNC{\func{\rho}{a} \circ e^{i t \cdot T}}{\zeta}}
                      }
                 }_{\Br{\heartsuit}}.
\end{align*}

To show that $ \ds \lim_{y \to t} \Br{\diamondsuit} $ exists, first observe, for all $ y \in \R \setminus \SSet{t} $, that
\begin{align*}
     & ~ \Norm{
              \Func{e^{- i y \cdot T}}
                   {\frac{1}{y - t} \cdot \SqBr{\FUNC{\func{\rho}{a} \circ e^{i y \cdot T}}{\zeta} - \FUNC{\func{\rho}{a} \circ e^{i t \cdot T}}{\zeta}}} -
              \Func{e^{- i t \cdot T}}{\Diff{y}{\FUNC{\func{\rho}{a} \circ e^{i y \cdot T}}{\zeta}} \bigg|_{y = t}}
              }_{\X} \\
\leq & ~ \Norm{
              \Func{e^{- i y \cdot T}}
                   {\frac{1}{y - t} \cdot \SqBr{\FUNC{\func{\rho}{a} \circ e^{i y \cdot T}}{\zeta} - \FUNC{\func{\rho}{a} \circ e^{i t \cdot T}}{\zeta}}} -
              \Func{e^{- i y \cdot T}}{\Diff{y}{\FUNC{\func{\rho}{a} \circ e^{i y \cdot T}}{\zeta}} \bigg|_{y = t}}
              }_{\X} ~ + \\
     & ~ \Norm{
              \Func{e^{- i y \cdot T}}{\Diff{y}{\FUNC{\func{\rho}{a} \circ e^{i y \cdot T}}{\zeta}} \bigg|_{y = t}} -
              \Func{e^{- i t \cdot T}}{\Diff{y}{\FUNC{\func{\rho}{a} \circ e^{i y \cdot T}}{\zeta}} \bigg|_{y = t}}
              }_{\X} \\
=    & ~ \Norm{
              \Func{e^{- i y \cdot T}}
                   {
                   \frac{1}{y - t} \cdot \SqBr{\FUNC{\func{\rho}{a} \circ e^{i y \cdot T}}{\zeta} - \FUNC{\func{\rho}{a} \circ e^{i t \cdot T}}{\zeta}} -
                   \Diff{y}{\FUNC{\func{\rho}{a} \circ e^{i y \cdot T}}{\zeta}} \bigg|_{y = t}
                   }
              }_{\X} ~ + \\
     & ~ \Norm{
              \Func{e^{- i y \cdot T}}{\Diff{y}{\FUNC{\func{\rho}{a} \circ e^{i y \cdot T}}{\zeta}} \bigg|_{y = t}} -
              \Func{e^{- i t \cdot T}}{\Diff{y}{\FUNC{\func{\rho}{a} \circ e^{i y \cdot T}}{\zeta}} \bigg|_{y = t}}
              }_{\X} \\
=    & ~ \Norm{
              \frac{1}{y - t} \cdot \SqBr{\FUNC{\func{\rho}{a} \circ e^{i y \cdot T}}{\zeta} - \FUNC{\func{\rho}{a} \circ e^{i t \cdot T}}{\zeta}} -
              \Diff{y}{\FUNC{\func{\rho}{a} \circ e^{i y \cdot T}}{\zeta}} \bigg|_{y = t}
              }_{\X} ~ + \\
     & ~ \Norm{
              \Func{e^{- i y \cdot T}}{\Diff{y}{\FUNC{\func{\rho}{a} \circ e^{i y \cdot T}}{\zeta}} \bigg|_{y = t}} -
              \Func{e^{- i t \cdot T}}{\Diff{y}{\FUNC{\func{\rho}{a} \circ e^{i y \cdot T}}{\zeta}} \bigg|_{y = t}}
              }_{\X}.
\end{align*}
Applying Claim 5 and the strong continuity of $ \Seq{e^{i y \cdot T}}{y \in \R} $, we may then apply the Squeeze Theorem to infer that
$$
  \lim_{y \to t}
  \Norm{
       \Func{e^{- i y \cdot T}}
            {\frac{1}{y - t} \cdot \SqBr{\FUNC{\func{\rho}{a} \circ e^{i y \cdot T}}{\zeta} - \FUNC{\func{\rho}{a} \circ e^{i t \cdot T}}{\zeta}}} -
       \Func{e^{- i t \cdot T}}{\Diff{y}{\FUNC{\func{\rho}{a} \circ e^{i y \cdot T}}{\zeta}} \bigg|_{y = t}}
       }_{\X}
= 0,
$$
which immediately yields
$$
  \lim_{y \to t} \Br{\diamondsuit}
= \Func{e^{- i t \cdot T}}{\Diff{y}{\FUNC{\func{\rho}{a} \circ e^{i y \cdot T}}{\zeta}} \bigg|_{y = t}}
= i \cdot \Func{e^{- i t \cdot T}}{\FUNC{\func{\rho}{a} \circ T}{\Func{e^{i y \cdot T}}{\zeta}}}.
$$

By Claim 6 and the Chain Rule, the map
$$
\Map{\R}{\X}{x}{\Func{e^{- i y \cdot T}}{\FUNC{\func{\rho}{a} \circ e^{i t \cdot T}}{\zeta}}}
$$
is differentiable; \autoref{The Infinitesimal Generator and Analytic Vectors of a Strongly Continuous One-Parameter Unitary Group} tells us additionally that
\begin{align*}
    \lim_{y \to t} \Br{\heartsuit}
& = \Diff{y}{\Func{e^{- i y \cdot T}}{\FUNC{\func{\rho}{a} \circ e^{i t \cdot T}}{\zeta}}} \bigg|_{y = t} \\
& = - i \cdot \func{e^{- i t \cdot T}}{\func{T}{\FUNC{\func{\rho}{a} \circ e^{i t \cdot T}}{\zeta}}} \\
& = - i \cdot \func{e^{- i t \cdot T}}{\FUNC{T \circ \func{\rho}{a}}{\Func{e^{i t \cdot T}}{\zeta}}}.
\end{align*}

The foregoing allows us to see that the map $ \Map{\R}{\X}{y}{\FUNC{\func{V}{\gamma_{y}^{- 1}} \circ \func{\rho}{a} \circ \func{V}{\gamma_{y}}}{\zeta}} $ is differentiable and that
\begin{align*}
\forall t \in \R: \qquad
  & ~ \Diff{y}{\FUNC{\func{V}{\gamma_{y}^{- 1}} \circ \func{\rho}{a} \circ \func{V}{\gamma_{y}}}{\zeta}} \bigg|_{y = t} \\
= & ~ \lim_{y \to t} \Br{\diamondsuit} + \lim_{y \to t} \Br{\heartsuit} \\
= & ~ i \cdot \Func{e^{- i t \cdot T}}{\FUNC{\func{\rho}{a} \circ T}{\Func{e^{i y \cdot T}}{\zeta}}} -
      i \cdot \Func{e^{- i t \cdot T}}{\FUNC{T \circ \func{\rho}{a}}{\Func{e^{i t \cdot T}}{\zeta}}} \\
= & ~ i \cdot \Func{e^{- i t \cdot T}}{\FUNC{\func{\rho}{a} \circ T - T \circ \func{\rho}{a}}{\zeta}} \\
= & ~ i \cdot \Func{e^{- i t \cdot T}}{\bm{0}_{\X}} \quad \Br{\text{by Property (3) of \autoref{Infinitesimal Heisenberg Representation}}} \\
= & ~ \bm{0}_{\X},
\end{align*}
which implies that said map is constant. Furthermore,
\begin{equation} \label{Covariance Identity 2}
\forall y \in \R: \quad
  \FUNC{\func{V}{\gamma_{y}^{- 1}} \circ \func{\rho}{a} \circ \func{V}{\gamma_{y}}}{\zeta}
= \FUNC{\func{V}{\gamma_{0}^{- 1}} \circ \func{\rho}{a} \circ \func{V}{\gamma_{0}}}{\zeta}
= \FUNC{\Id_{\X} \circ \func{\rho}{a} \circ \Id_{\X}}{\zeta}
= \FUNC{\func{\rho}{a}}{\zeta};
\end{equation}
as $ \Dom{T} $ is dense in $ \X $, the continuity of $ \func{V}{\gamma_{y}^{- 1}} \circ \func{\rho}{a} \circ \func{V}{\gamma_{y}} $ and $ \func{\rho}{a} $ ensures that \autoref{Covariance Identity 2} holds for all $ \zeta \in \X $. Finally, as $ \gamma: \R \to \widehat{\R} $ is a group isomorphism, we may conclude that
$$
\forall \varphi \in \widehat{\R}, ~ \forall a \in A, ~ \forall \zeta \in \X: \quad
\FUNC{\func{V}{\varphi^{- 1}} \circ \func{\rho}{a} \circ \func{V}{\varphi}}{\zeta} = \FUNC{\func{\rho}{a}}{\zeta}.
$$
The proof of Claim 7 is thereby complete. \hfill $ \triangle $

\textbf{Claim 8:} For all $ x \in \R $ and $ \varphi \in \widehat{\R} $, we have $ \func{V}{\varphi} \circ U_{x} = \func{\varphi}{x} \cdot \SqBr{U_{x} \circ \func{V}{\varphi}} $.

\emph{Proof of Claim 8}. Recall that the Heisenberg Lie algebra $ \mathfrak{h} $ may be Lie-generated by a three-element subset $ \mathfrak{G} \df \SSet{X,Y,Z} $, where
$$
\Comm{X}{Y}_{\mathfrak{h}} = Z,                     \quad
\Comm{X}{Z}_{\mathfrak{h}} = \bm{0}_{\mathfrak{h}}, \quad
\Comm{Y}{Z}_{\mathfrak{h}} = \bm{0}_{\mathfrak{h}}.
$$
By Property (4) of \autoref{Infinitesimal Heisenberg Representation}, there exists an $ \SSet{S,T} $-analytic and $ \SSet{S,T} $-invariant dense $ A $-submodule $ D $ of $ \X $ such that $ D $ is a core for both $ S $ and $ T $, and the HCR holds for $ \Pair{S}{T} $ on $ D $, which allows us to define a Lie-algebra representation $ \mathscr{L} $ of $ \mathfrak{h} $ as $ A $-linear operators on $ D $ by
$$
\forall x,y,z \in \R: \quad
    \func{\mathscr{L}}{x X + y Y + z Z}
\df x \cdot \Br{- \frac{i}{\hbar} \cdot S|_{D}} + y \cdot \Br{i \cdot T|_{D}} + z \cdot \Br{- i \cdot \Id_{D}}.
$$
\autoref{Rusinek's Integrability Theorem} then supplies a $ W $ making $ \Trip{G}{\X}{W} $ a Lie Banach dynamical system so that, for all $ x,y,z \in \R $, the map
$$
\Map{\R}{\X}{t}{\FUNC{\func{W}{\func{\exp_{H}}{t \cdot \Br{x X + y Y + z Z}}}}{\zeta}}
$$
is differentiable and
$$
\forall \zeta \in D: \quad
  \Diff{t}{\FUNC{\func{W}{\func{\exp_{H}}{t \cdot \Br{x X + y Y + z Z}}}}{\zeta}} \bigg|_{t = 0}
= \FUNC{x \cdot \Br{- \frac{i}{\hbar} \cdot S|_{D}} + y \cdot \Br{i \cdot T|_{D}} + z \cdot \Br{- i \cdot \Id_{D}}}{\zeta}.
$$
\autoref{Stone's Theorem for Hilbert C*-Modules}, \autoref{Infinitesimal Generators of C0-Groups}, and the fact that $ D $ is a core for $ S $, $ T $, and $ \Id_{\X} $ now combine to yield the following three statements:
\begin{itemize}
\item
The infinitesimal generator of $ \Seq{\func{W}{\func{\exp_{H}}{x X}}}{x \in \R} $ extends $ - \dfrac{i}{\hbar} \cdot S|_{D} $ and thus equals $ - \dfrac{i}{\hbar} \cdot S $; hence,
$$
\forall x \in \R: \quad
  U_{x}
= e^{- \frac{i x}{\hbar} \cdot S}
= \func{W}{\func{\exp_{H}}{x X}}.
$$

\item
The infinitesimal generator of $ \Seq{\func{W}{\func{\exp_{H}}{y Y}}}{y \in \R} $ extends $ i \cdot T|_{D} $ and thus equals $ i \cdot T $; hence,
$$
\forall y \in \R: \quad
  \func{V}{\gamma_{y}}
= e^{i y \cdot T}
= \func{W}{\func{\exp_{H}}{y Y}}.
$$

\item
The infinitesimal generator of $ \Seq{\func{W}{\func{\exp_{H}}{z Z}}}{z \in \R} $ extends $ - i \cdot \Id_{D} $ and thus equals $ - i \cdot \Id_{\X} $; hence,
$$
\forall z \in \R: \quad
e^{- i z} \cdot \Id_{\X} = \func{W}{\func{\exp_{H}}{z Z}}.
$$
\end{itemize}
Consequently,
\begin{align*}
\forall x,y \in \R: \qquad
  & ~ \func{V}{\gamma_{y}} \circ U_{x} \\
= & ~ \func{W}{\func{\exp_{H}}{y Y}} \circ \func{W}{\func{\exp_{H}}{x X}} \\
= & ~ \func{W}{\func{\exp_{H}}{y Y} \func{\exp_{H}}{x X}} \\
= & ~ \func{W}{\func{\exp_{H}}{y Y + x X + \frac{1}{2} \Comm{y Y}{x X}_{\mathfrak{h}}}} \quad \Br{\text{by the Baker-Campbell-Hausdorff Formula}} \\
= & ~ \func{W}{\func{\exp_{H}}{y Y + x X - \frac{1}{2} x y Z}} \quad \Br{\text{as $ \Comm{Y}{X}_{\mathfrak{h}} = - Z $}} \\
= & ~ \func{W}{\func{\exp_{H}}{- x y Z} \func{\exp_{H}}{y Y + x X + \frac{1}{2} x y Z}} \quad \Br{\text{as $ Z $ is a central element of $ \mathfrak{h} $}} \\
= & ~ \func{W}{\func{\exp_{H}}{- x y Z} \func{\exp_{H}}{x X + y Y + \frac{1}{2} \Comm{x X}{y Y}_{\mathfrak{h}}}} \quad
      \Br{\text{as $ \Comm{X}{Y}_{\mathfrak{h}} = Z $}} \\
= & ~ \func{W}{\func{\exp_{H}}{- x y Z} \func{\exp_{H}}{x X} \func{\exp_{H}}{y Y}} \quad \Br{\text{by the Baker-Campbell-Hausdorff Formula}} \\
= & ~ \func{W}{\func{\exp_{H}}{- x y Z}} \circ \func{W}{\func{\exp_{H}}{x X}} \circ \func{W}{\func{\exp_{H}}{y Y}} \\
= & ~ \func{W}{\func{\exp_{H}}{x y Z}}^{- 1} \circ \func{W}{\func{\exp_{H}}{x X}} \circ \func{W}{\func{\exp_{H}}{y Y}} \\
= & ~ \func{\gamma_{y}}{x} \cdot \SqBr{U_{x} \circ \func{V}{\gamma_{y}}}.
\end{align*}
As $ \gamma: \R \to \widehat{\R} $ is a group isomorphism, we may conclude that
$$
\forall x \in \R, ~ \forall \varphi \in \widehat{\R}: \quad
\func{V}{\varphi} \circ U_{x} = \func{\varphi}{x} \cdot \SqBr{U_{x} \circ \func{V}{\varphi}}.
$$
The proof of Claim 8 is thereby complete. \hfill $ \triangle $

Finally, Claims 4, 7, and 8 confirm that $ \Quad{\X}{\rho}{U}{V} $ is a Heisenberg representation of $ \Trip{\R}{A}{\alpha} $.
\end{proof}


From Heisenberg differentiability and Heisenberg integrability, we get a one-to-one correspondence between Heisenberg representations and infinitesimal Heisenberg representations, which allows us to switch --- at will --- between the global and infinitesimal formulations of Heisenberg covariance.


\begin{Def}[infinitesimal Schr\"{o}dinger representation] \label{def:The Infinitesimal Schrodinger Representation}
The \emph{infinitesimal Schr\"{o}dinger representation} of a real $ C^{\ast} $-dynamical system $ \Trip{\R}{A}{\alpha} $ is the infinitesimal Heisenberg representation of $ \Trip{\R}{A}{\alpha} $ obtained, via \autoref{Heisenberg Differentiability}, from the Schr\"{o}dinger representation $ \Quad{\L{\R}{A}{\alpha}}{\M}{\U}{\V} $ of $ \Trip{\R}{A}{\alpha} $ defined in \autoref{thm:CSVNT}.
\end{Def}


\begin{Prop}[the Infinitesimal Covariant Stone-von Neumann Theorem] \label{The Infinitesimal Covariant Stone-von Neumann Theorem}
Let $ A $ be an elementary $ C^{\ast} $-algebra and $ \Trip{\R}{A}{\alpha} $ a real $ C^{\ast} $-dynamical system. Then every infinitesimal Heisenberg representation of $ \Trip{\R}{A}{\alpha} $ is unitarily equivalent to a direct sum of copies of the infinitesimal Schr\"{o}dinger representation of $ \Trip{\R}{A}{\alpha} $.
\end{Prop}

\begin{proof}
Let $ \Quad{\X}{\rho}{S}{T} $ be an infinitesimal Heisenberg representation of $ \Trip{\R}{A}{\alpha} $. Then, by \autoref{Heisenberg Integrability} and \autoref{thm:CSVNT}, we have
$$
      \Quad{\X}{\rho}{\Seq{e^{- \frac{i x}{\hbar} \cdot S}}{x \in \R}}{\Seq{e^{i \gamma^{- 1}_{\varphi} \cdot T}}{\varphi \in \widehat{\R}}}
\cong \Quad{\bigoplus_{i \in I} \L{\R}{A}{\alpha}}{\bigoplus_{i \in I} \M}{\bigoplus_{i \in I} \U}{\bigoplus_{i \in I} \V},
$$
which means that there exist an index set $ I $ and a $ \ds W \in \Unitary{\X,\bigoplus_{i \in I} \L{\R}{A}{\alpha}} $ such that
$$
\forall x,y \in \R: \quad
e^{- \frac{i x}{\hbar} \cdot S} = W^{\ast} \Br{\bigoplus_{i \in I} \U_{x}} W
\quad \text{and} \quad
e^{i y \cdot T}                 = W^{\ast} \Br{\bigoplus_{i \in I} \func{\V}{\gamma_{y}}} W,
$$
so
\begin{align*}
    S
& = i \hbar \cdot \Diff{x}{e^{- \frac{i x}{\hbar} \cdot S}} \bigg|_{x = 0} \\
& = i \hbar \cdot \Diff{x}{W^{\ast} \Br{\bigoplus_{i \in I} \U_{x}} W} \Bigg|_{x = 0} \quad \Br{\text{by \autoref{thm:CSVNT}, as mentioned}} \\
& = i \hbar \cdot \SqBr{W^{\ast} \Br{\Diff{x}{\bigoplus_{i \in I} \U_{x}} \Bigg|_{x = 0}} W} \quad
    \Br{\text{by \autoref{Infinitesimal Generators and Unitary Equivalence}}} \\
& = i \hbar \cdot \SqBr{W^{\ast} \Br{\bigoplus_{i \in I} \Diff{x}{\U_{x}} \bigg|_{x = 0}} W} \quad
    \Br{\text{by \autoref{A Direct Sum of Infinitesimal Heisenberg Representations}}} \\
& = W^{\ast} \Br{\bigoplus_{i \in I} i \hbar \cdot \Diff{x}{\U_{x}} \bigg|_{x = 0}} W \\
& = W^{\ast} \Br{\bigoplus_{i \in I} - i \cdot \Diff{x}{\U_{- \hbar x}} \bigg|_{x = 0}} W \quad \Br{\text{by the Chain Rule}}
\end{align*}
and
\begin{align*}
    T
& = - i \cdot \Diff{y}{e^{i y \cdot T}} \bigg|_{y = 0} \\
& = - i \cdot \Diff{y}{W^{\ast} \Br{\bigoplus_{i \in I} \func{\V}{\gamma_{y}}} W} \Bigg|_{y = 0} \quad \Br{\text{by \autoref{thm:CSVNT}, as mentioned}} \\
& = - i \cdot \SqBr{W^{\ast} \Br{\Diff{y}{\bigoplus_{i \in I} \func{\V}{\gamma_{y}}} \Bigg|_{y = 0}} W} \quad
    \Br{\text{by \autoref{Infinitesimal Generators and Unitary Equivalence}}} \\
& = - i \cdot \SqBr{W^{\ast} \Br{\bigoplus_{i \in I} \Diff{y}{\func{\V}{\gamma_{y}}} \bigg|_{y = 0}} W} \quad
    \Br{\text{by \autoref{A Direct Sum of Infinitesimal Heisenberg Representations}}} \\
& = W^{\ast} \Br{\bigoplus_{i \in I} - i \cdot \Diff{y}{\func{\V}{\gamma_{y}}} \bigg|_{y = 0}} W.
\end{align*}
Therefore, by \autoref{def:The Infinitesimal Schrodinger Representation} (which relies on \autoref{Heisenberg Differentiability}), the proposition is proven.
\end{proof}

\section{Properties of the Momentum and Position Operators} 

In this final section, we fix a real $ C^{\ast} $-dynamical system $ \Trip{\R}{A}{\alpha} $ and provide explicit formulas for the position and momentum operators of $ \Trip{\R}{A}{\alpha} $, so as to relate them to the classical position and momentum operators of quantum mechanics. 


\begin{Def}
Recall the notation of \autoref{thm:CSVNT}; the \emph{momentum operator} $ \MO $ and the \emph{position operator} $ \PO $ of $ \Trip{\R}{A}{\alpha} $ are defined as the infinitesimal generators of the strongly continuous one-parameter unitary groups $ \Seq{\U_{- \hbar x}}{x\in \R} $ and $ \Seq{\func{\V}{\gamma_{y}}}{y\in \R} $ on $ \L{\R}{A}{\alpha} $; hence, $ \Quad{\L{\R}{A}{\alpha}}{\M}{\MO}{\PO} $ is the infinitesimal Schr\"{o}dinger representation of $ \Trip{\R}{A}{\alpha} $.
\end{Def}

\subsection{A Useful Norm Inequality} 

Recall that $ \L{\R}{A}{\alpha} $ is the norm-completion of $ \Cc{\R,A} $ with respect to the pre-Hilbert $ A $-module structure given in \autoref{thm:CSVNT}. We shall show below that the canonical dense embedding $ q: \Cc{\R,A} \to \L{\R}{A}{\alpha} $ is continuous with respect to the traditional $ L^{2} $-norm on $ \Cc{\R,A} $.


\begin{Def}
Let $ A $ be a $ C^{\ast} $-algebra. A function $ \phi \in \CC{\R,A} $ is said to be \emph{square-integrable} if and only if
$$
\Map{\R}{\R_{\geq 0}}{x}{\Norm{\func{\phi}{x}}_{A}} \in \func{L^{2}}{\R},
$$
i.e., $ \displaystyle \Int{\R}{\Norm{\func{\phi}{x}}_{A}^{2}}{\mu}{x} < \infty $; the set of all such functions forms a $ \C $-vector space, which shall be denoted by $ \LCont{2}{\R,A} $.
\end{Def}


Given $ f: \R \to \C $ and $ a \in A $, define the function $ f \odot a: \R \to A $ by
$$
\forall x \in \R: \quad
\Func{f \odot a}{x} \df \func{f}{x} \cdot a.
$$
Letting $ \LCont{2}{\R} \df \LCont{2}{\R,\C} $, we then have
\begin{align*}
\forall f \in \LCont{2}{\R}, ~ \forall a \in A: \quad
    \Int{\R}{\Norm{\Func{f \odot a}{x}}_A^2}{\mu}{x}
& = \Int{\R}{\Norm{\func{f}{x} \cdot a}_{A}^{2}}{\mu}{x} \\
& = \Norm{a}_{A}^{2} \Int{\R}{\Abs{\func{f}{x}}^{2}}{\mu}{x} \\
& = \Norm{a}_{A}^{2} \Norm{f}_{2,\R}^{2} \\
& < \infty,
\end{align*}
where $\Norm{f}_{2,\mathbb{R}}$ is the usual $L^2$-norm of $f\in \LCont{2}{\R}$. Hence, 
\[
\LCont{2}{\R} \odot A 
\df \Span{\Set{f \odot a}{f \in \LCont{2}{\R} \text{ and } a \in A}} 
\subseteq \LCont{2}{\R,A}.
\]


\begin{Lem} \label{The 2-Norm}
Define a norm $ \Norm{\cdot}_{2,\R,A} $ on $ \LCont{2}{\R,A} $ by
$$
\forall \phi \in \LCont{2}{\R,A}: \qquad
\Norm{\phi}_{2,\R,A} \df \sqrt{\Int{\R}{\Norm{\func{\phi}{x}}_{A}^{2}}{\mu}{x}}.
$$
For all $ \phi \in \Cc{\R,A} $, we have $ \Norm{\func{q}{\phi}}_{\L{\R}{A}{\alpha}} \leq \Norm{\phi}_{2,\R,A} $, so $ q $ may be extended by continuity to a dense linear embedding $ \qq: \LCont{2}{\R,A} \to \L{\R}{A}{\alpha} $, which satisfies
$$
\forall f \in \LCont{2}{\R}, ~ \forall a \in A: \quad
     \Norm{\func{\qq}{f \odot a}}_{\L{\R}{A}{\alpha}}
\leq \Norm{f \odot a}_{2,\R,A}
=    \Norm{f}_{2,\R} \Norm{a}_{A}.
$$
\end{Lem}

We shall leave the proving of this result to the reader.

\subsection{The Hermite Functions} 

For each $ n \in \N_{0} $, the \emph{$ n $-th Hermite polynomial} is the degree-$ n $ polynomial function $ H_{n}: \R \to \R $ defined by
$$
\forall x \in \R: \quad
\func{H_{n}}{x} \df \Br{- 1}^n e^{x^{2}} \frac{\mathrm{d}^{n}}{\mathrm{d} x^{n}} \SqBr{e^{- x^{2}}}.
$$
For each $ n \in \N_{0} $, the \emph{$ n $-th Hermite function} is then the function $ \psi_{n}: \R \to \R $ defined by
$$
\forall x \in \R: \quad
\func{\psi_{n}}{x} \df \Br{2^{n} n! \sqrt{\pi}}^{- 1/2} e^{- x^{2}/2} \func{H_{n}}{x}.
$$
The following three properties of the Hermite functions are well known:
\begin{itemize}
\item
$ \SSet{\psi_{n}}_{n \in \N_{0}} $ is an orthonormal basis for $ \Ll{2}{\R} $ consisting of smooth functions.

\item
For each $ n \in \N $, we have $ \psi_{n}' = \dfrac{1}{\sqrt{2}} \cdot \Br{\sqrt{n} \cdot \psi_{n - 1} - \sqrt{n + 1} \cdot \psi_{n + 1}} $.

\item
For each $ n \in \N $, we have $ \Id_{\R} \psi_{n} = \dfrac{1}{\sqrt{2}} \cdot \Br{\sqrt{n} \cdot \psi_{n - 1} + \sqrt{n + 1} \cdot \psi_{n + 1}} $.
\end{itemize}
In particular, $ \HH \df \Span{\SSet{\psi_{n}}_{n \in \N_{0}}} $ is a dense linear subspace of $ \Ll{2}{\R} $ that is closed under differentiation and point-wise multiplication by $ \Id_{\R} $.

Lastly, note that $ \HH $ naturally embeds as a dense linear subspace of $ \Ll{2}{\R} $ that is analytic for the following two densely defined linear operators on $ \Ll{2}{\R} $:
\begin{itemize}
\item
The differential operator $ D \df \dfrac{\mathrm{d}}{\mathrm{d} x} $: For each $ f \in \HH $, there is an $ r \in \R_{> 0} $ making $ \ds \sum_{n = 0}^{\infty} \frac{\Norm{f^{\Br{n}}}_{2,\R}}{n!} \cdot r^{n} < \infty $.

\item
Multiplication by $ \Id_{\R} $: For each $ f \in \HH $, there is an $ s \in \R_{> 0} $ making $ \ds \sum_{n = 0}^{\infty} \frac{\Norm{\Id_{\R}^{n} f}_{2,\R}}{n!} \cdot s^{n} < \infty $.
\end{itemize}

\subsection{The Momentum and Position Operators} 

The rest of this section is devoted to proving that
$$
\D \df \Span{\Set{\func{\qq}{f \odot a} \bullet b}{f \in \HH, ~ a \in \Trip{\R}{A}{\alpha}^{\omega} \text{ and } b \in A}}
$$
is a $ \SSet{\MO,\PO} $-analytic and $ \SSet{\MO,\PO} $-invariant core for both $ \MO $ and $ \PO $. By definition, it is clear that $ \D $ is a dense $ A $-submodule of $ \L{\R}{A}{\alpha} $.


\begin{Prop} \label{An Explicit Description of the Momentum Operator}
We have $ \D \subseteq \Dom{\MO} $ and
$$
\forall f \in \HH, ~ \forall a \in \Trip{\R}{A}{\alpha}^{\omega}, ~ \forall b \in A: \quad
\func{\MO}{\func{\qq}{f \odot a} \bullet b} = i \hbar \cdot \SqBr{\func{\qq}{f \odot \func{\delta}{a} - f' \odot a} \bullet b}.
$$
In particular, $ \D $ is $ \MO $-invariant, i.e., $ \Im{\MO}{\D} \subseteq \D $.
\end{Prop}

\begin{proof}
Let $ f \in \HH $, $ a \in \Trip{\R}{A}{\alpha}^{\omega} $, and $ x \in \R \setminus \SSet{0} $ be arbitrary. Observe that
\begin{align*}
    \frac{1}{x} \cdot \SqBr{\func{\U_{x}}{\func{\qq}{f \odot a}} - \func{\qq}{f \odot a}}
& = \frac{1}{x} \cdot \SqBr{\func{\qq}{\alphArg{x}{\func{f}{\bullet - x} \cdot a}} - \func{\qq}{\func{f}{\bullet} \cdot a}} \\
& = \frac{1}{x} \cdot \SqBr{\func{\qq}{\func{f}{\bullet - x} \cdot \alphArg{x}{a}} - \func{\qq}{\func{f}{\bullet} \cdot a}} \\
& = \func{\qq}{\frac{1}{x} \cdot \SqBr{\SqBr{\func{f}{\bullet - x} - \func{f}{\bullet}} \cdot \alphArg{x}{a}}} +
    \func{\qq}{\func{f}{\bullet} \cdot \Br{\frac{1}{x} \cdot \SqBr{\alphArg{x}{a} - a}}},
\end{align*}
so
\begin{align*}
  & ~ \frac{1}{x} \cdot \SqBr{\func{\U_{x}}{\func{\qq}{f \odot a}} - \func{\qq}{f \odot a}} - \func{\qq}{f \odot \func{\delta}{a} - f' \odot a} \\
= & ~ \underbrace{
                 \func{\qq}{\frac{1}{x} \cdot \SqBr{\SqBr{\func{f}{\bullet - x} - \func{f}{\bullet}} \cdot \alphArg{x}{a}} + \func{f'}{\bullet} \cdot a}
                 }_{\sun} +
      \underbrace{\func{\qq}{\func{f}{\bullet} \cdot \SqBr{\frac{1}{x} \cdot \SqBr{\alphArg{x}{a} - a} - \func{\delta}{a}}}}_{\leftmoon}.
\end{align*}

Apply \autoref{The 2-Norm} to $ \sun $ and then invoke the Triangle and Minkowski's Integral Inequalities to find that
\begin{align*}
  \Norm{\sun}_{\L{\R}{A}{\alpha}}
& \leq \Norm{\frac{1}{x} \cdot \SqBr{\SqBr{\func{f}{\bullet - x} - \func{f}{\bullet}} \cdot \alphArg{x}{a}} + \func{f'}{\bullet} \cdot a}_{2,\R,A} \\
& \leq \Norm{\frac{1}{x} \cdot \SqBr{\SqBr{\func{f}{\bullet - x} - \func{f}{\bullet}} \cdot \alphArg{x}{a}} + \func{f'}{\bullet} \cdot \alphArg{x}{a}}_{2,\R,A} +
       \Norm{\func{f'}{\bullet} \cdot a - \func{f'}{\bullet} \cdot \alphArg{x}{a}}_{2,\R,A} \\
& =    \Norm{\frac{\func{f}{\bullet - x} - \func{f}{\bullet}}{- x} - \func{f'}{\bullet}}_{2,\R} \Norm{a}_{A} + \Norm{f'}_{2,\R} \Norm{a - \alphArg{x}{a}}_{A} \\
& =    \Norm{\Int{\CCI{0}{1}}{\SqBr{\func{f'}{\bullet - x r} - \func{f'}{\bullet}}}{\mu}{r}}_{2,\R} + \Norm{f'}_{2,\R} \Norm{a - \alphArg{x}{a}}_{A} \\
& =    \sqrt{\Int{\R}{\Abs{\Int{\CCI{0}{1}}{\SqBr{\func{f'}{s - x r} - \func{f'}{s}}}{\mu}{r}}^{2}}{\mu}{s}} + \Norm{f'}_{2,\R} \Norm{a - \alphArg{x}{a}}_{A} \\
& \leq \Int{\CCI{0}{1}}{\sqrt{\Int{\R}{\Abs{\func{f'}{s - x r} - \func{f'}{s}}^{2}}{\mu}{s}}}{\mu}{r} + \Norm{f'}_{2,\R} \Norm{a - \alphArg{x}{a}}_{A} \\
& =    \Int{\CCI{0}{1}}{\underbrace{\Norm{\func{f'}{\bullet - x r} - \func{f'}{\bullet}}_{2,\R}}_{\leq 2 \Norm{f'}_{2,\R}}}{\mu}{r} +
       \Norm{f'}_{2,\R} \Norm{a - \alphArg{x}{a}}_{A}.
\end{align*}
Recalling the well-known measure-theoretic fact that the action of $ \R $ on $ \Ll{2}{\R} $ by left/right translations is strongly continuous, we have
$$
\forall r \in \CCI{0}{1}: \quad
\lim_{x \to 0} \Norm{\func{f'}{\bullet - x r} - \func{f'}{\bullet}}_{2,\R} = 0,
$$
so, by the LDCT,
$$
\lim_{x \to 0} \Int{\CCI{0}{1}}{\Norm{\func{f'}{\bullet - x r} - \func{f'}{\bullet}}_{2,\R}}{\mu}{r} = 0.
$$
As $ \alpha $ is strongly continuous, we obtain $ \ds \Norm{f'}_{2,\R} \lim_{x \to 0} \Norm{a - \alphArg{x}{a}}_{A} = 0 $, and the Squeeze Theorem then yields $ \ds \lim_{x \to 0} \Norm{\sun}_{\L{\R}{A}{\alpha}} = 0 $.

Similarly, apply \autoref{The 2-Norm} to $ \leftmoon $ to find that
$$
     \Norm{\leftmoon}_{\L{\R}{A}{\alpha}}
\leq \Norm{\func{f}{\bullet} \cdot \SqBr{\frac{1}{x} \cdot \SqBr{\alphArg{x}{a} - a} - \func{\delta}{a}}}_{2,\R,A}
=    \Norm{f}_{2,\R} \Norm{\frac{1}{x} \cdot \SqBr{\alphArg{x}{a} - a} - \func{\delta}{a}}_{A}.
$$
As $ \ds \lim_{x \to 0} \Norm{\frac{1}{x} \cdot \SqBr{\alphArg{x}{a} - a} - \func{\delta}{a}}_{A} = 0 $, the Squeeze Theorem yields $ \ds \lim_{x \to 0} \Norm{\leftmoon}_{\L{\R}{A}{\alpha}} = 0 $ as well.

The foregoing argument reveals that
$$
\forall f \in \HH, ~ \forall a \in \Trip{\R}{A}{\alpha}^{\omega}: \quad
  \lim_{x \to 0}
  \Norm{
       \frac{1}{x} \cdot \SqBr{\func{\U_{x}}{\func{\qq}{f \odot a}} - \func{\qq}{f \odot a}} - \func{\qq}{f \odot \func{\delta}{a} - f' \odot a}
       }_{\L{\R}{A}{\alpha}}
= 0.
$$
Consequently,
$$
\Span{\Set{\func{\qq}{f \odot a}}{f \in \HH \text{ and } a \in \Trip{\R}{A}{\alpha}^{\omega}}} \subseteq \Dom{\Diff{x}{\func{\U_{x}}{\bullet}} \bigg|_{x = 0}}
$$
and
$$
\forall f \in \HH, ~ \forall a \in \Trip{\R}{A}{\alpha}^{\omega}: \quad
\Diff{x}{\func{\U_{x}}{\func{\qq}{f \odot a}}} \bigg|_{x = 0} = \func{\qq}{f \odot \func{\delta}{a} - f' \odot a}.
$$
As $ \ds \MO = - i \cdot \Diff{x}{\func{\U_{- \hbar x}}{\bullet}} \bigg|_{x = 0} $ is a modular operator, we have by the Chain Rule that $ \D \subseteq \Dom{\MO} $ and
\begin{align*}
\forall f \in \HH, ~ \forall a \in \Trip{\R}{A}{\alpha}^{\omega}, ~ \forall b \in A: \quad
    \func{\MO}{\func{\qq}{f \odot a} \bullet b}
& = \func{\MO}{\func{\qq}{f \odot a}} \bullet b \\
& = \SqBr{- i \cdot \Diff{x}{\func{\U_{- \hbar x}}{\func{\qq}{f \odot a}}} \bigg|_{x = 0}} \bullet b \\
& = \SqBr{- i \Br{- \hbar} \cdot \Diff{x}{\func{\U_{x}}{\func{\qq}{f \odot a}}} \bigg|_{x = 0}} \bullet b \\
& = i \hbar \cdot \SqBr{\func{\qq}{f \odot \func{\delta}{a} - f' \odot a} \bullet b}.
\end{align*}
Finally, as $ \HH $ is closed under differentiation and $ \Trip{\R}{A}{\alpha}^{\omega} $ is $ \delta $-invariant, we conclude that $ \Im{\MO}{\D} \subseteq \D $.
\end{proof}


\begin{Prop} \label{An Explicit Description of the Position Operator}
We have $ \D \subseteq \Dom{\PO} $ and
$$
\forall f \in \HH, ~ \forall a \in \Trip{\R}{A}{\alpha}^{\omega}, ~ \forall b \in A: \quad
\func{\PO}{\func{\qq}{f \odot a} \bullet b} = \func{\qq}{\Id_{\R} f \odot a} \bullet b.
$$
In particular, $ \D $ is $ \PO $-invariant, i.e., $ \Im{\PO}{\D} \subseteq \D $.
\end{Prop}

\begin{proof}
Let $ f \in \HH $, $ a \in \Trip{\R}{A}{\alpha}^{\omega} $, and $ y \in \R \setminus \SSet{0} $ be arbitrary. Observe that
\begin{align*}
    \frac{1}{y} \cdot \SqBr{\FUNC{\func{V}{\gamma_{y}}}{\func{\qq}{f \odot a}} - \func{\qq}{f \odot a}}
& = \frac{1}{y} \cdot \SqBr{\func{\qq}{\Map{\R}{A}{x}{e^{i x y} \func{f}{x} \cdot a}} - \func{\qq}{\Map{\R}{A}{x}{\func{f}{x} \cdot a}}} \\
& = \func{\qq}{\Map{\R}{A}{x}{\SqBr{\Br{\frac{e^{i x y} - 1}{y}} \func{f}{x}} \cdot a}},
\end{align*}
so
\begin{align*}
  & ~ \frac{1}{y} \cdot \SqBr{\FUNC{\func{V}{\gamma_{y}}}{\func{\qq}{f \odot a}} - \func{\qq}{f \odot a}} -
      \func{\qq}{\Map{\R}{A}{x}{i x \func{f}{x} \cdot a}} \\
= & ~ \underbrace{\func{\qq}{\Map{\R}{A}{x}{\SqBr{\Br{\frac{e^{i x y} - 1}{y} - i x} \func{f}{x}} \cdot a}}}_{\rightmoon}.
\end{align*}

Apply \autoref{The 2-Norm} to $ \rightmoon $ to find that
$$
     \Norm{\rightmoon}_{\L{\R}{A}{\alpha}}
\leq \Norm{\Map{\R}{\C}{x}{\Br{\frac{e^{i x y} - 1}{y} - i x} \func{f}{x}}}_{2,\R} \Norm{a}_{A}
=    \Norm{\Map{\R}{\C}{x}{\SqBr{\frac{\func{\gamma_{y}}{x} - 1}{y} - i x} \func{f}{x}}}_{2,\R} \Norm{a}_{A}.
$$
As
\begin{align*}
\forall x \in \R: \quad
       \Abs{\SqBr{\frac{\func{\gamma_{y}}{x} - 1}{y} - i x} \func{f}{x}}
& \leq \Abs{\frac{\func{\gamma_{y}}{x} - 1}{y} \func{f}{x}} + \Abs{i x \func{f}{x}} \\
& =    \frac{\Abs{\func{\gamma_{y}}{x} - 1}}{\Abs{y}} \Abs{\func{f}{x}} + \Abs{x \func{f}{x}} \\
& \leq \frac{\Abs{y} \Abs{x}}{\Abs{y}} \Abs{\func{f}{x}} + \Abs{x \func{f}{x}} \quad \Br{\text{by \autoref{A Basic Inequality}}} \\
& =    2 \Abs{x \func{f}{x}},
\end{align*}
and as we have, for all $ x \in \R $, that $ \ds \lim_{y \to 0} \SqBr{\frac{\func{\gamma_{y}}{x} - 1}{y} - i x} \func{f}{x} = 0 $, it follows readily from the LDCT that
$$
\lim_{y \to 0} \Norm{\Map{\R}{\C}{x}{\SqBr{\frac{\func{\gamma_{y}}{x} - 1}{y} - i x} \func{f}{x}}}_{2,\R} = 0.
$$
The Squeeze Theorem then yields $ \ds \lim_{y \to 0} \Norm{\rightmoon}_{\L{\R}{A}{\alpha}} = 0 $.

The foregoing argument reveals that
$$
  \lim_{y \to 0}
  \Norm{
       \frac{1}{y} \cdot \SqBr{\FUNC{\func{V}{\gamma_{y}}}{\func{\qq}{f \odot a}} - \func{\qq}{f \odot a}} - \func{\qq}{\Map{\R}{A}{x}{i x \func{f}{x} \cdot a}}
       }_{\L{\R}{A}{\alpha}}
= 0.
$$
Consequently,
$$
          \Span{\Set{\func{\qq}{f \odot a}}{f \in \HH \text{ and } a \in \Trip{\R}{A}{\alpha}^{\omega}}}
\subseteq \Dom{\Diff{y}{\FUNC{\func{\V}{\gamma_{y}}}{\bullet}} \bigg|_{y = 0}}
$$
and
$$
\forall f \in \HH, ~ \forall a \in \Trip{\R}{A}{\alpha}^{\omega}: \quad
\Diff{y}{\FUNC{\func{\V}{\gamma_{y}}}{\func{\qq}{f \odot a}}} \bigg|_{y = 0} = \func{\qq}{\Map{\R}{A}{x}{i x \func{f}{x} \cdot a}}.
$$
As $ \ds \PO = - i \cdot \Diff{y}{\FUNC{\func{\V}{\gamma_{y}}}{\bullet}} \bigg|_{y = 0} $ is a modular operator, we have $ \D \subseteq \Dom{\PO} $ and
\begin{align*}
\forall f \in \HH, ~ \forall a \in \Trip{\R}{A}{\alpha}^{\omega}, ~ \forall b \in A: \quad
    \func{\PO}{\FUNC{\func{\V}{\gamma_{y}}}{\func{\qq}{f \odot a}} \bullet b}
& = \func{\PO}{\FUNC{\func{\V}{\gamma_{y}}}{\func{\qq}{f \odot a}}} \bullet b \\
& = \SqBr{- i \cdot \func{\qq}{\Map{\R}{A}{x}{i x \func{f}{x} \cdot a}}} \bullet b \\
& = \func{\qq}{\Map{\R}{A}{x}{x \func{f}{x} \cdot a}} \bullet b \\
& = \func{\qq}{\Id_{\R} f \odot a} \bullet b.
\end{align*}
Finally, as $ \HH $ is closed under point-wise multiplication by $ \Id_{\R} $, we conclude that $ \Im{\PO}{\D} \subseteq \D $.
\end{proof}

\subsection{Analytic Vectors} 

We shall finally show that $ \D $ is a $ \SSet{\MO,\PO} $-analytic core for both $ \MO $ and $ \PO $; this explains --- concretely --- why every infinitesimal Schr\"{o}dinger representation is also an infinitesimal Heisenberg representation.


\begin{Prop}
$ \D $ is a $ \SSet{\MO,\PO} $-analytic core for both $ \MO $ and $ \PO $.
\end{Prop}

\begin{proof}
We shall divide the proof into several claims.

\textbf{Claim 1.} For all $ n \in \N_{0} $, $ f \in \HH $, and $ a \in \Trip{\R}{A}{\alpha}^{\omega} $, we have
\begin{equation} \label{Induction}
  \func{\MO^{n}}{\func{\qq}{f \odot a}}
= \Br{i \hbar}^{n} \cdot \SqBr{\sum_{m = 0}^{n} \Br{- 1}^{m} \binom{n}{m} \cdot \func{\qq}{f^{\Br{m}} \odot \func{\delta^{n - m}}{a}}};
\end{equation}
in particular,
$$
     \frac{1}{n!} \Norm{\func{\MO^{n}}{\func{\qq}{f \odot a}}}_{\L{\R}{A}{\alpha}}
\leq \hbar^{n} \sum_{m = 0}^{n} \frac{\Norm{f^{\Br{m}}}_{2,\R}}{m!} \frac{\Norm{\func{\delta^{n - m}}{a}}_{A}}{\Br{n - m}!}.
$$

\emph{Proof of Claim 1}. Fix arbitrary $ f \in \HH $ and $ a \in \Trip{\R}{A}{\alpha}^{\omega} $. Identity \eqref{Induction} is true for $ n = 0 $, so suppose that it holds for some $ n \in \N_{0} $; then
\begin{align*}
  & ~ \frac{1}{\Br{i \hbar}^{n + 1}} \cdot \func{\MO^{n + 1}}{\func{\qq}{f \odot a}} \\
= & ~ \frac{1}{\Br{i \hbar}^{n + 1}} \cdot \func{\MO}{\func{\MO^{n}}{\func{\qq}{f \odot a}}} \\
= & ~ \frac{1}{i \hbar} \cdot \func{\MO}{\sum_{m = 0}^{n} \Br{- 1}^{m} \binom{n}{m} \cdot \func{\qq}{f^{\Br{m}} \odot \func{\delta^{n - m}}{a}}} \quad
      \Br{\text{by the induction hypothesis}} \\
= & ~ \frac{1}{i \hbar} \cdot \SqBr{\sum_{m = 0}^{n} \Br{- 1}^{m} \binom{n}{m} \cdot \func{\MO}{\func{\qq}{f^{\Br{m}} \odot \func{\delta^{n - m}}{a}}}} \\
= & ~ \sum_{m = 0}^{n}
      \Br{- 1}^{m} \binom{n}{m} \cdot \func{\qq}{f^{\Br{m}} \odot \func{\delta^{n + 1 - m}}{a} - f^{\Br{m + 1}} \odot \func{\delta^{n - m}}{a}} \quad
      \Br{\text{by \autoref{An Explicit Description of the Momentum Operator}}} \\
= & ~ \sum_{m = 0}^{n} \Br{- 1}^{m} \binom{n}{m} \cdot \func{\qq}{f^{\Br{m}} \odot \func{\delta^{n + 1 - m}}{a}} -
      \sum_{m = 0}^{n} \Br{- 1}^{m} \binom{n}{m} \cdot \func{\qq}{f^{\Br{m + 1}} \odot \func{\delta^{n - m}}{a}} \\
= & ~ \sum_{m = 0}^{n} \Br{- 1}^{m} \binom{n}{m} \cdot \func{\qq}{f^{\Br{m}} \odot \func{\delta^{n + 1 - m}}{a}} -
      \sum_{m = 1}^{n + 1} \Br{- 1}^{m - 1} \binom{n}{m - 1} \cdot \func{\qq}{f^{\Br{m}} \odot \func{\delta^{n + 1 - m}}{a}} \\
= & ~ \func{\qq}{f \odot \func{\delta^{n + 1}}{a}} ~ + \\
  & ~ \sum_{m = 1}^{n} \Br{- 1}^{m} \binom{n}{m} \cdot \func{\qq}{f^{\Br{m}} \odot \func{\delta^{n + 1 - m}}{a}} +
      \sum_{m = 1}^{n} \Br{- 1}^{m} \binom{n}{m - 1} \cdot \func{\qq}{f^{\Br{m}} \odot \func{\delta^{n + 1 - m}}{a}} ~ + \\
  & ~ \Br{- 1}^{n + 1} \cdot \func{\qq}{f^{\Br{n + 1}} \odot a} \\
= & ~ \func{\qq}{f \odot \func{\delta^{n + 1}}{a}} +
      \sum_{m = 1}^{n} \Br{- 1}^{m} \SqBr{\binom{n}{m} + \binom{n}{m - 1}} \cdot \func{\qq}{f^{\Br{m}} \odot \func{\delta^{n + 1 - m}}{a}} +
      \Br{- 1}^{n + 1} \cdot \func{\qq}{f^{\Br{n + 1}} \odot a} \\
= & ~ \func{\qq}{f \odot \func{\delta^{n + 1}}{a}} +
      \sum_{m = 1}^{n} \Br{- 1}^{m} \binom{n + 1}{m} \cdot \func{\qq}{f^{\Br{m}} \odot \func{\delta^{n + 1 - m}}{a}} +
      \Br{- 1}^{n + 1} \cdot \func{\qq}{f^{\Br{n + 1}} \odot a} \\
  & ~ \Br{\text{by Pascal's Identity}} \\
= & ~ \sum_{m = 0}^{n + 1} \Br{- 1}^{m} \binom{n + 1}{m} \cdot \func{\qq}{f^{\Br{m}} \otimes \func{\delta^{n + 1 - m}}{a}}.
\end{align*}
Consequently, by the Principle of Mathematical Induction, Identity \eqref{Induction} is true for all $ n \in \N_{0} $.

From the foregoing discussion, it follows, for all $ n \in \N_{0} $, $ f \in \HH $, and $ a \in \Trip{\R}{A}{\alpha}^{\omega} $, that
\begin{align*}
       \frac{1}{n!} \Norm{\func{\MO^{n}}{\func{\qq}{f \odot a}}}_{\L{\R}{A}{\alpha}}
& =    \frac{1}{n!}
       \Norm{
            \Br{i \hbar}^{n} \cdot \SqBr{\sum_{m = 0}^{n} \Br{- 1}^{m} \binom{n}{m} \cdot \func{\qq}{f^{\Br{m}} \odot \func{\delta^{n - m}}{a}}}
            }_{\L{\R}{A}{\alpha}} \\
& \leq \frac{\hbar^{n}}{n!} \sum_{m = 0}^{n} \binom{n}{m} \Norm{\func{\qq}{f^{\Br{m}} \odot \func{\delta^{n - m}}{a}}}_{\L{\R}{A}{\alpha}} \\
& \leq \frac{\hbar^{n}}{n!} \sum_{m = 0}^{n} \binom{n}{m} \Norm{f^{\Br{m}}}_{2,\R} \Norm{\func{\delta^{n - m}}{a}}_{A} \quad
       \Br{\text{by \autoref{The 2-Norm}}} \\
& =    \hbar^{n} \sum_{m = 0}^{n} \frac{\Norm{f^{\Br{m}}}_{2,\R}}{m!} \frac{\Norm{\func{\delta^{n - m}}{a}}_{A}}{\Br{n - m}!}.
\end{align*}
Claim 1 is thereby proven. \hfill $ \triangle $
    
\textbf{Claim 2.} $ \D $ is $ \MO $-analytic.

\emph{Proof of Claim 2}. It suffices to prove, for $ f \in \HH $, $ a \in \Trip{\R}{A}{\alpha}^{\omega} $, and $ b \in A $, that the real power series
$$
t \mapsto \sum_{n = 0}^{\infty} \frac{\Norm{\func{\MO^{n}}{\func{\qq}{f \odot a} \bullet b}}_{\L{\R}{A}{\alpha}}}{n!} \cdot t^{n}
$$
has a positive radius of convergence.

Recall the following facts:
\begin{itemize}
\item
$ f $ is a $ D $-analytic vector in $ \Ll{2}{\R} $, so there exists an $ r \in \R_{> 0} $ such that $ \ds \sum_{m = 0}^{\infty} \frac{\Norm{f^{\Br{m}}}_{2,\R}}{m!} \cdot r^{m} < \infty $.

\item
$ a $ is an $ \alpha $-analytic vector in $ A $, so there exists an $ s \in \R_{> 0} $ such that $ \ds \sum_{n = 0}^{\infty} \frac{\Norm{\func{\delta^{n}}{a}}_{A}}{n!} \cdot s^{n} < \infty $.
\end{itemize}

For $ t \df \dfrac{\func{\min}{\SSet{r,s}}}{\hbar} > 0 $, observe that
\begin{align*}
     & ~ \sum_{n = 0}^{\infty} \frac{\Norm{\func{\MO^{n}}{\func{\qq}{f \odot a} \bullet b}}_{\L{\R}{A}{\alpha}}}{n!} \cdot t^{n} \\
=    & ~ \sum_{n = 0}^{\infty} \frac{\Norm{\func{\MO^{n}}{\func{\qq}{f \odot a}} \bullet b}_{\L{\R}{A}{\alpha}}}{n!} \cdot t^{n} \quad
         \Br{\text{as $ \MO $ is a modular operator}} \\
\leq & ~ \sum_{n = 0}^{\infty} \frac{\Norm{\func{\MO^{n}}{\func{\qq}{f \odot a}}}_{\L{\R}{A}{\alpha}} \Norm{b}_{A}}{n!} \cdot t^{n} \\
=    & ~ \Br{\sum_{n = 0}^{\infty} \frac{\Norm{\func{\MO^{n}}{\func{\qq}{f \odot a}}}_{\L{\R}{A}{\alpha}}}{n!} \cdot t^{n}} \Norm{b}_{A} \\
\leq & ~ \SqBr{
              \sum_{n = 0}^{\infty}
              \SqBr{\hbar^{n} \sum_{m = 0}^{n} \frac{\Norm{f^{\Br{m}}}_{2,\R}}{m!} \frac{\Norm{\func{\delta^{n - m}}{a}}_A}{\Br{n - m}!}} \cdot t^{m} t^{n - m}
              }
         \Norm{b}_{A} \quad \Br{\text{by Claim 1}} \\
=    & ~ \SqBr{
              \sum_{n = 0}^{\infty}
              \SqBr{\sum_{m = 0}^{n} \frac{\Norm{f^{\Br{m}}}_{2,\R}}{m!} \frac{\Norm{\func{\delta^{n - m}}{a}}_A}{\Br{n - m}!}} \cdot
              \Br{\hbar t}^{m} \Br{\hbar t}^{n - m}
              }
         \Norm{b}_{A} \\
=    & ~ \SqBr{\sum_{m = 0}^{\infty} \frac{\Norm{f^{\Br{m}}}_{2,\R}}{m!} \cdot \Br{\hbar t}^{m}}
         \SqBr{\sum_{n = 0}^{\infty} \frac{\Norm{\func{\delta^{n}}{a}}_{A}}{n!} \cdot \Br{\hbar t}^{n}}
         \Norm{b}_{A} \quad \Br{\text{by the Cauchy product for series}} \\
<    & ~ \infty.
\end{align*}
Hence, $ \D $ is $ \MO $-analytic. \hfill $ \triangle $

\textbf{Claim 3.} $ \D $ is $ \PO $-analytic.

\emph{Proof of Claim 3}. It suffices to prove, for $ f \in \HH $, $ a \in \Trip{\R}{A}{\alpha}^{\omega} $, and $ b \in A $, that the real power series
$$
        t
\mapsto \sum_{n = 0}^{\infty} \frac{\Norm{\func{\PO^{n}}{\func{\qq}{f \odot a} \bullet b}}_{\L{\R}{A}{\alpha}}}{n!} \cdot t^{n}
=       \sum_{n = 0}^{\infty} \frac{\Norm{\func{\qq}{\Id_{\R}^{n} f \odot a} \bullet b}_{\L{\R}{A}{\alpha}}}{n!} \cdot t^{n}
$$
has a positive radius of convergence.

Recall that $ f $ is an analytic vector in $ \Ll{2}{\R} $ for multiplication by $ \Id_{\R} $, so there exists an $ r \in \R_{> 0} $ such that $ \ds \sum_{n = 0}^{\infty} \frac{\Norm{\Id_{\R}^{n} f}_{2,\R}}{n!} \cdot r^{n} < \infty $. Hence,
\begin{align*}
       \sum_{n = 0}^{\infty} \frac{\Norm{\func{\PO^{n}}{\func{\qq}{f \odot a} \bullet b}}_{\L{\R}{A}{\alpha}}}{n!} \cdot r^{n}
& =    \sum_{n = 0}^{\infty} \frac{\Norm{\func{\qq}{\Id_{\R}^{n} f \odot a} \bullet b}_{\L{\R}{A}{\alpha}}}{n!} \cdot r^{n} \\
& \leq \sum_{n = 0}^{\infty} \frac{\Norm{\func{\qq}{\Id_{\R}^{n} f \odot a}}_{\L{\R}{A}{\alpha}} \Norm{b}_{A}}{n!} \cdot r^{n} \\
& =    \SqBr{\sum_{n = 0}^{\infty} \frac{\Norm{\func{\qq}{\Id_{\R}^{n} f \odot a}}_{\L{\R}{A}{\alpha}}}{n!} \cdot r^{n}} \Norm{b}_{A} \\
& \leq \SqBr{\sum_{n = 0}^{\infty} \frac{\Norm{\Id_{\R} f}_{2,\R} \Norm{a}_{A}}{n!} \cdot r^{n}} \Norm{b}_{A} \quad \Br{\text{by \autoref{The 2-Norm}}} \\
& =    \Br{\sum_{n = 0}^{\infty} \frac{\Norm{\Id_{\R} f}_{2,\R}}{n!} \cdot r^{n}} \Norm{a}_{A} \Norm{b}_{A} \\
& <    \infty.
\end{align*}
Hence, $ \D $ is indeed $ \PO $-analytic. \hfill $ \triangle $

\textbf{Claim 4.} $ \D $ is a core for both $ \MO $ and $ \PO $.

\emph{Proof of Claim 4}. As $ i \cdot \MO $ and $ i \cdot \PO $ are infinitesimal generators of $ C_{0} $-groups of isometries on $ \L{\R}{A}{\alpha} $ viewed as a Banach space, it follows from Claims 2 and 3, \autoref{An Explicit Description of the Momentum Operator}, \autoref{An Explicit Description of the Position Operator}, and \autoref{Contractivity and Invariance with Respect to a Contractive C0-Group's Infinitesimal Generator} that $ \D $ is a core for both $ \MO $ and $ \PO $. \hfill $ \triangle $

The theorem is therefore proven.
\end{proof}


\section*{Acknowledgements}

L.I. was supported in part by the Army Research Laboratory under Grant/Contract Number W911NF-25-1-0052. Additional support was provided by the National Science Foundation Grant Number DMS-2603107.


\bibliographystyle{plain}
\bibliography{References}

@article{Amyari-Chakoshi,
 title   = {{A Generalization of Stone's Theorem in Hilbert $ C^{\ast} $-Modules}},
 author  = {M. Amyari and M. Chakoshi},
 journal = {J. Korean Soc. Math. Educ. Ser. B Pure Appl. Math.},
 volume  = {18},
 number  = {1},
 pages   = {31--39},
 year    = {2011},
 doi     = {10.7468/jksmeb.2011.18.1.031}
}

@book{Bratteli-Robinson,
 title     = {{Operator Algebras and Quantum Statistical Mechanics 1: $ C^{\ast} $- and $ W^{\ast} $-Algebras. Symmetry Groups. Decomposition of States}},
 author    = {O. Bratteli and D. W. Robinson},
 series    = {Texts Monogr. Phys.},
 year      = {1987},
 publisher = {Springer-Verlag, New York}
}

@book{Conway,
 title     = {{A Course in Functional Analysis}},
 author    = {J. B. Conway},
 series    = {Grad. Texts in Math.},
 volume    = {96},
 year      = {1990},
 publisher = {Springer-Verlag, New York}
}

@article{Gimperlein-Krotz-Schlichtkrull,
 title   = {{Analytic Representation Theory of Lie Groups: General Theory and Analytic Globalizations of Harish-Chandra Modules}},
 author  = {H. Gimperlein and B. Kr\"{o}tz and H. Schlichtkrull},
 journal = {Compos. Math.},
 volume  = {147},
 number  = {5},
 pages   = {1581--1607},
 year    = {2011},
 doi     = {https://doi.org/10.1112/S0010437X11005392}
}

@article{Huang-Ismert,
 title   = {{The Covariant Stone-von Neumann Theorem for Actions of Abelian Groups on $ C^{\ast} $-Algebras of Compact Operators}},
 author  = {L. Huang and L. Ismert},
 journal = {Commun. Math. Phys.},
 volume  = {378},
 number  = {1},
 pages   = {117--147},
 year    = {2020}
}

@book{Lance,
 title     = {{Hilbert $ C^{\ast} $-Modules: A Toolkit for Operator Algebraists}},
 author    = {E. C. Lance},
 series    = {London Math. Soc. Lecture Note Ser.},
 volume    = {210},
 year      = {1995},
 publisher = {Cambridge University Press, Cambridge}
}

@article{Nelson,
 title   = {{Analytic Vectors}},
 author  = {E. Nelson},
 journal = {Ann. of Math.},
 volume  = {70},
 number  = {3},
 pages   = {572--615},
 year    = {1959},
 doi     = {https://doi.org/10.2307/1970331}
}

@book{Pedersen,
 title     = {{Analysis Now}},
 author    = {G. Pedersen},
 series    = {Grad. Texts in Math.},
 volume    = {118},
 year      = {1989},
 publisher = {Springer-Verlag, New York}
}

@book{Rudin,
 title     = {{Functional Analysis}},
 author    = {W. Rudin},
 series    = {Internat. Ser. Pure Appl. Math.},
 year      = {1991},
 publisher = {McGraw-Hill, Inc., New York}
}

@article{Rusinek,
 title   = {{Analytic Vectors and Integrability of Lie Algebra Representations}},
 author  = {J. Rusinek},
 journal = {J. Funct. Anal.},
 volume  = {74},
 number  = {1},
 pages   = {10--23},
 year    = {1987},
 doi     = {https://doi.org/10.1016/0022-1236(87)90036-X}
}

@article{Schlieter-Werner,
 title   = {{Unbounded Multipliers on Operator Spaces}},
 author  = {H. Schlieter and W. Werner},
 journal = {Proc. Amer. Math. Soc.},
 volume  = {141},
 number  = {5},
 pages   = {1719--1733},
 year    = {2013},
 doi     = {https://doi.org/10.1090/S0002-9939-2012-11478-9}
}

@book{Taylor,
 title     = {{Noncommutative Harmonic Analysis}},
 author    = {M. Taylor},
 series    = {Math. Surveys Monogr.},
 volume    = {22},
 year      = {1986},
 publisher = {American Mathematical Society, Providence, RI}
}

@article{VonNeumann,
 title   = {{\"{U}ber einen Satz von Herrn M. H. Stone}},
 author  = {J. von Neumann},
 journal = {Ann. Math.},
 volume  = {33},
 number  = {3},
 pages   = {567--573},
 year    = {1932}
}

@article{Woronowicz,
 title   = {{Unbounded Elements Affiliated with $ C^{\ast} $-Algebras and Noncompact Quantum Groups}},
 author  = {S. L. Woronowicz},
 journal = {Commun. Math. Phys.},
 volume  = {136},
 number  = {2},
 pages   = {399--432},
 year    = {1991}
}


\end{document}